\documentclass[%
reprint,
amsmath,
amssymb,
aps,
prl,
nofootinbib,
superscriptaddress,
floatfix]{revtex4-1}

\usepackage[pdftex]{graphicx}
\usepackage{hyperref}

\usepackage{bm}
\usepackage{braket}
\usepackage{mathtools}
\usepackage{xspace}
\usepackage{array}
\usepackage{multirow}
\usepackage{comment}

\makeatletter

\let\MYcaption\@makecaption
\makeatother

\usepackage{subcaption}
\makeatletter
\let\@makecaption\MYcaption
\makeatother

\usepackage{color}
\usepackage[normalem]{ulem} 

\newcommand{\bk}{\bm{k}}
\newcommand{\bdg}{\mathrm{BdG}}
\newcommand{\ztwo}{\mathbb{Z}_2}
\newcommand{\wien}{\textsc{wien}2k\xspace}

\begin{document}

\title{Topological crystalline superconductivity in locally noncentrosymmetric CeRh$_2$As$_2$}

\author{Kosuke Nogaki}
\email[]{nogaki.kosuke.83v@st.kyoto-u.ac.jp}
\affiliation{%
  Department of Physics, Kyoto University, Kyoto 606-8502, Japan
}%

\author{Akito Daido}
\affiliation{%
  Department of Physics, Kyoto University, Kyoto 606-8502, Japan
}%

\author{Jun Ishizuka}
\affiliation{%
  Department of Physics, Kyoto University, Kyoto 606-8502, Japan
}%

\author{Youichi Yanase}
\affiliation{%
  Department of Physics, Kyoto University, Kyoto 606-8502, Japan
}%
\affiliation{%
  Institute for Molecular Science, Okazaki 444-8585, Japan
}%

\date{\today}

\begin{abstract}
Recent discovery of superconductivity in CeRh$_2$As$_2$ clarified an unusual $H$-$T$ phase diagram with two superconducting phases [Khim {\it et al.} arXiv:2101.09522]. The experimental observation has been interpreted based on the even-odd parity transition characteristic of locally noncentrosymmetric superconductors. 
Indeed, the inversion symmetry is locally broken at the Ce site, and CeRh$_2$As$_2$ molds a new class of exotic superconductors.
The low-temperature and high-field superconducting phase is a candidate for the odd-parity pair-density-wave state,  
suggesting a possibility of topological superconductivity as spin-triplet superconductors are. 
In this paper, we first derive the formula expressing the $\mathbb{Z}_2$ invariant of glide symmetric and time-reversal symmetry broken superconductors by the number of Fermi surfaces on a glide invariant line.
Next, we conduct a first-principles calculation for the electronic structure of CeRh$_2$As$_2$.
Combining the results, 
we show that the field-induced odd-parity superconducting phase of CeRh$_2$As$_2$ is a platform of topological crystalline superconductivity protected by the nonsymmorphic glide symmetry and accompanied by boundary Majorana fermions.
\end{abstract}

\maketitle

\textit{Introduction.} --- 
For decades, symmetry breaking has been the most important concept to describe various phases of quantum matter such as magnetism, density-wave, and superconductivity as well as critical phenomena in condensed matter physics~\cite{Landau2013statistical}.
Recently, topological science has shed light on new aspects 
\cite{Qi2011,Tanaka2012,Sato2016,Sato2017}.
Topological phase transitions without symmetry breaking have been uncovered, and topological insulators/superconductors are widely recognized as intriguing phases of matter. 
One of the exotic phenomena in topological materials is the appearance of gapless modes at boundaries and defects of the systems, though the topology is determined by only the bulk information.
In particular, topological superconductors host Majorana fermions, which have been proposed for topological fault-tolerant quantum computation~\cite{Kitaev2001,Nayak2008}. Such unique properties attractive from the viewpoints of basic and applied science triggered tremendous efforts for searching topological superconductivity~\cite{Fu2008,Sato2009,Sau2010,Lutchyn2010,Oreg2010,Alicea2010,Qi2010,Chung2011,Mourik2012,Das2012,Deng2012,Wang2012,Nadj-Perge2014,Xu2014,Xu2015,Wang2015,Sun2016,He2017,Menard2017,Zhang2018,Wang2018,Machida2019,Wang2020,Fu2010,Sato2010,Hosuer2011,Fu2014,Hosuer2014,Kobayashi2015,Wang2015_2,Xu2016,Pan2016,Matano2016,Yonezawa2017,Zhang2018_2,Yoshida2016,Daido2016,Can2021}. 
However, the realization of topological superconductivity is still under intensive debate.

By the pioneering research~\cite{Schnyder2008,Kitaev2009,Ryu2010}, the topological phases were classified based on the local symmetries, namely the time-reversal, particle-hole, and chiral symmetries.
It has later been recognized that the symmetries unique to solids enrich the topological properties of materials.
This idea led to the concept of topological crystalline insulators/superconductors (TCIs/TCSCs)~\cite{Fu2011,Zhang2013,Chiu2013,Morimoto2013,Shiozaki2014,Chiu2014,Ueno2013,Tsutsumi2013,Yoshida2015,Fang2015,Shiozaki2015,Shiozaki2016,Shapourian2018,Yanase2017,Daido2019,Ono2019,Ono2020,Ono2020_2,Skurativska2020,Geier2020,Shiozaki2019,Ahn2020}.
However, major candidates of the TCSC are odd-parity superconductors, as are the case for usual topological superconductors~\cite{Tanaka2012,Sato2016,Sato2017}. Unfortunately, the odd-parity superconductors are usually spin-triplet superconductors, which are rarely known in nature. 
On the other hand, superconductivity with exotic symmetry beyond the paradigm of standard classification theory~\cite{Sigrist-Ueda} is recently attracting attention, and odd-parity superconductivity due to an ordinary spin-singlet pairing has been reported in CeRh$_2$As$_2$~\cite{Khim2021}. 
This discovery may open a new route to realize the topological superconductivity.

\begin{figure}[tbp]
 \begin{center}
    \includegraphics[keepaspectratio, scale=0.19]{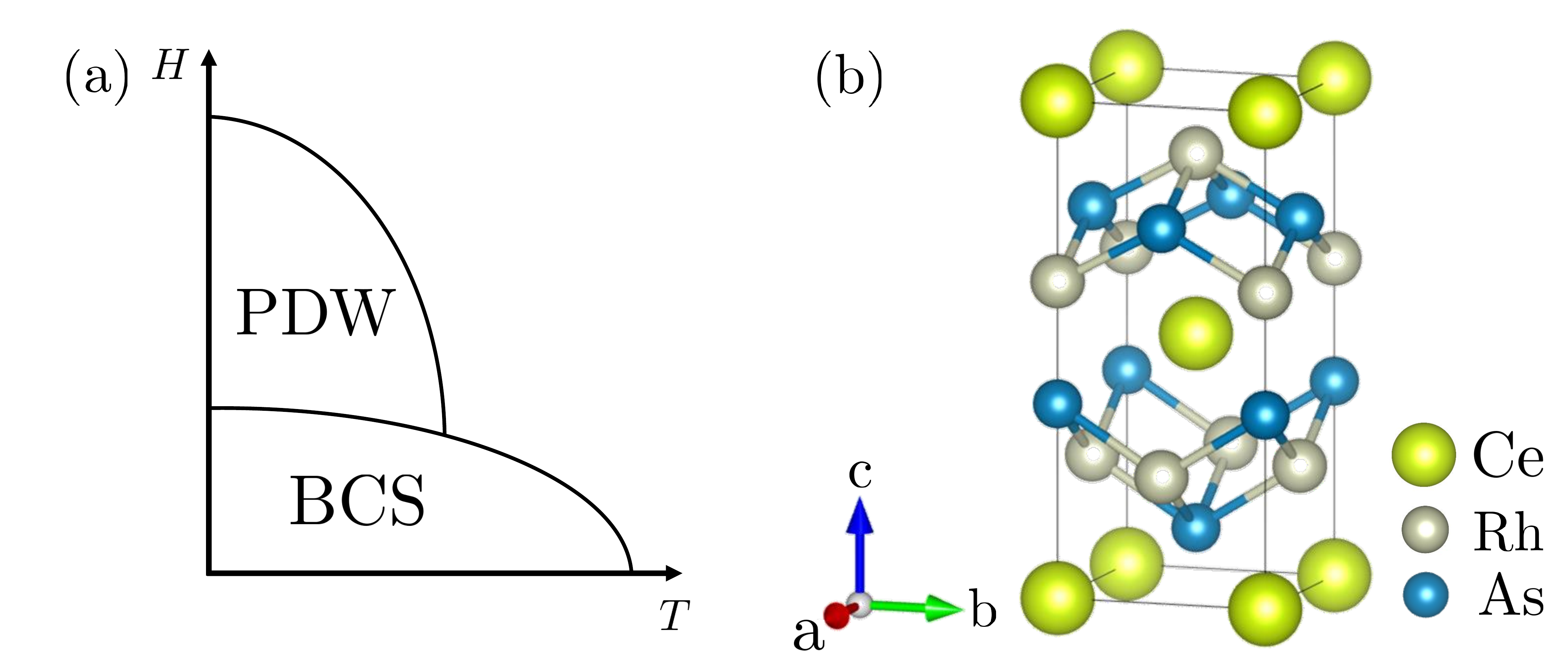}
  \end{center}
  \caption{(a) Schematic $H$-$T$ phase diagram of CeRh$_2$As$_2$~\cite{Khim2021}.
  The BCS state is assumed to be an even-parity superconducting phase, and the high-field phase is supposed to be the PDW state, the main topic in this Letter.
  (b) Crystal structure of CeRh$_2$As$_2$ generated by {\sc Vesta}~\cite{Momma2011}. 
  }
  \label{fig:crys_struc}
\end{figure}

This work was triggered by a recent experimental report of an unusual superconducting $H$-$T$ phase diagram in a newly discovered heavy-fermion superconductor, CeRh$_2$As$_2$~\cite{Khim2021}.
A high upper critical field much beyond the Pauli-Clogston-Chandrasekhar limit and the phase transition between two superconducting phases have been observed [Fig.~\ref{fig:crys_struc}(a)]. Thus, CeRh$_2$As$_2$ is a fascinating platform of multiple superconducting phases. 
Different from the previous examples, UPt$_3$~\cite{Joynt_review} and UTe$_2$~\cite{Braithwaite_UTe2_2019,Ran_UTe2_pressure,Aoki_UTe2_2020,Ishizuka2020}, the phase diagram was attributed to the locally noncentrosymmetric crystal structure~\cite{Khim2021} in accordance with a theoretical proposal~\cite{Yoshida2012}. 
Soon after the experimental report, theoretical works along this line were conducted~\cite{Schertenleib2021,moeckli2021}.

Noncentrosymmetric superconductivity in globally inversion asymmetric systems have been investigated for several decades 
\cite{Edelstein1989,Edelstein1995,Sato2009,Sau2010,Lutchyn2010,Oreg2010,Alicea2010,Bauer2004,Agterberg2007,Bauer2012,Smidman_2017,Saito2016,Wakatsuki2018,Ando2020,Nogaki2020,Daido2016,Yoshida2016}.
The concept was recently extended to locally noncentrosymmetric superconductivity, and unique superconducting phenomena have been uncovered~\cite{Fischer2011,Maruyama2012,Maruyama2013,Yoshida2012,Yoshida2013,Yoshida2014,Yoshida2015,Shimozawa_2016,Mockli2018}.
The $H$-$T$ phase diagram of CeRh$_2$As$_2$ is consistent with the prediction based on a
two-sublattice Rashba model~\cite{Yoshida2012,Mockli2018}.
The similarity of phase diagrams between the experiment~\cite{Khim2021} and theory~\cite{Yoshida2012,Mockli2018} suggests that the local inversion symmetry breaking plays an essential role in CeRh$_2$As$_2$, and the superconducting phase in the high magnetic field region is the pair-density-wave (PDW) state.  
In the PDW state, the superconducting gap function changes sign depending on Ce layers~\cite{supplement}. The most crucial property of this phase is the odd-parity superconductivity in spite of dominantly spin-singlet pairing. The odd parity is owing to the sign change of the gap function between the sublattices related by inversion symmetry~\cite{Yoshida2012}. Thus, the locally noncentrosymmetric crystal is a platform of odd-parity superconductivity without requiring rare spin-triplet pairing. It may realize in CeRh$_2$As$_2$, making a candidate of the topological superconductor, different from potential spin-triplet superconductors, UPt$_3$~\cite{Tsutsumi2013,Yanase2017}, UCoGe~\cite{Daido2019}, and UTe$_2$~\cite{Ishizuka2019}.

\textit{Crystal structure and glide symmetry.} --- 
CeRh$_2$As$_2$ crystallizes in the centrosymmetric tetragonal CaBe$_2$Ge$_2$-type structure~\cite{Khim2021}
with stacking Ce layers and Rh$_2$As$_2$ layers [Fig.~\ref{fig:crys_struc}(b)]. Importantly, Rh$_2$As$_2$ layers at the top and bottom of the Ce layer have different compositions.  
Therefore, the inversion symmetry is locally broken at the Ce sites, although the global inversion center exists in the middle of the two Ce sites in the unit cell.
The space group is $P4/nmm$ (No.129) including one glide reflection and three screw rotations.
Since the magnetic field parallel to the $c$-axis destroys the three screw symmetries, only the glide symmetry is respected in the field-induced phase, and we hereafter focus on the glide operation $\hat{G}$ which is $\{ M_z | \bm{a}/2+\bm{b}/2 \}$ in Seitz notation.
Here, $\bm{a}$ and $\bm{b}$ are the lattice vectors along the $a$ and $b$ axes, respectively.

In the Brillouin zone, the glide symmetry is preserved in the two glide-invariant planes, $k_z=0,\pi$,
and we can divide Hilbert space into each Bloch state and glide sector,
\begin{align}
  \mathbb{V}_{k_z=0,\pi} = \bigoplus_{\bk \in \mathrm{BZ}_{k_z=0,\pi}} \mathbb{V}^{\mathfrak{g}^+}_{\bk} \oplus \mathbb{V}^{\mathfrak{g}^-}_{\bk}.
\end{align}
Here, $\mathbb{V}_{\bk}$ is Hilbert space consisting of Bloch states labeled by $\bk$, and $\mathfrak{g}^\pm$ are eigenvalues of the glide operation.
Because the Hamiltonian preserves the glide symmetry, $[\hat{H},\hat{G}]=0$, we can label eigenstates of the Hamiltonian by glide eigenvalues as $\ket{\bk^{\mathfrak{g}^{\pm}}} \in \mathbb{V}^{\mathfrak{g}^{\pm}}_{\bk}$.
To evaluate the eigenvalues in the spinful case, we focus on the following relation:
\begin{equation}
  \hat{G}^2 = \{ -E | \bm{a}+\bm{b} \}.
  \label{eq:double_glide}
\end{equation}
Owing to Eq.~\eqref{eq:double_glide}, 
any state $\ket{\bm{k}}$ in $\mathbb{V}_{k_z=0,\pi}$ satisfies
\begin{equation}
  \hat{G}^2 \ket{\bk} = -e^{-i(k_x+k_y)}\ket{\bk},
\end{equation}
and we conclude that the glide eigenvalues are
\begin{align}
  \mathfrak{g}^\pm = \pm i e^{-i(k_x+k_y)/2}.
  \label{eq:eigen_value}
\end{align}

\textit{Symmetry of superconductivity.} --- 
Superconductivity is classified by the point group $D_{4h}$, which has
8 one-dimensional (1D) irreducible representations and 2 two-dimensional (2D) ones.
Many Ce-based heavy-fermion systems undergo spin-singlet superconductivity, which is also expected in CeRh$_2$As$_2$ at $H=0$. 
Thus, the low-field superconducting phase is supposed to be even-parity, either of $A_{1g}$, $A_{2g}$, $B_{1g}$, $B_{2g}$, or $E_g$ state.
Accordingly, the high-field phase is an odd-parity $A_{1u}$, $A_{2u}$, $B_{1u}$, $B_{2u}$, or $E_{u}$ state when the PDW state is assumed. 
Hereafter, we focus on 1D odd-parity representations because the $E_u$ state is unlikely in CeRh$_2$As$_2$~\cite{Khim2021}.

\textit{$\mathbb{Z}_2$ invariants.} --- 
Let us discuss the $\mathbb{Z}_2$ topological invariants protected by the glide symmetry.
By definition, the PDW state is glide-odd superconductivity in which the superconducting gap function obeys the following relation,
\begin{align}
  \mathcal{G}(\bk) \Delta(\bm{k}) \mathcal{G}^\top(-\bk) &= -\Delta(M_z\bm{k}).
  \label{eq:glide_odd_delta}
\end{align}
Here, $\mathcal{G}(\bk)$ is the representation matrix of the glide operation $\hat{G}$ in the Hilbert space $\mathbb{V}_{\bk}$, $M_z\bm{k}\equiv(k_x,k_y,-k_z)$, and $\Delta(\bm{k})$ represents the gap function.

In the following, we focus on the glide invariant planes in the Brilloiun zone, $k_z=0,\pi$. 
From Eq.~\eqref{eq:glide_odd_delta}, the particle-hole operation $\hat{C}$ and glide operation $\hat{G}$ anti-commutes, $\{\hat{C}, \hat{G}\}=0$, and from Eq.~\eqref{eq:eigen_value}, the glide eigenvalues become pure imaginary $\pm i$ in the restricted Hilbert space on $k_x+k_y=0$.
Combining these results, we show that the particle-hole symmetry is closed in each glide sector,
\begin{align}
   \hat{G}\hat{C}\ket{\bk^{\mathfrak{g}^{\pm}}} = -\hat{C}(\pm i)\ket{\bk^{\mathfrak{g}^{\pm}}} = (\pm i)\hat{C}\ket{\bk^{\mathfrak{g}^{\pm}}}.
\end{align}
As the time-reversal symmetry is broken under the magnetic field, 
each glide sector on the line is classified into 1D class D superconductivity. 
This class is specified by the topological invariant $\mathbb{Z}_2$, which is given by the integral of the Berry connection as introduced in Ref.~\onlinecite{Shiozaki2016},
\begin{equation}
 \nu^{\mathfrak{g}^\pm} = \frac{1}{\pi}\int^{\Gamma_2}_{\Gamma_1} d\,k_i \, \mathcal{A}^{\mathfrak{g}^\pm}_i(\bk) 
  \hspace{5mm}({\rm mod} \,\,\, 2),
  \label{eq:z_2inv}
\end{equation}
where $\mathcal{A}_i^{\mathfrak{g}^\pm}(\bk)$ are $i$-th components of the Berry connection of each glide sector given by $i\sum_{\mathrm{occ.}} \braket{\psi_{\bk}^{\mathfrak{g}^\pm}|\partial_{k_i}|\psi_{\bk}^{\mathfrak{g}^\pm}}$. 
Here $\ket{\psi_{\bk}^{\mathfrak{g}^\pm}}$ are the occupied eigenstates of Bogoliubov–de Gennes (BdG) Hamiltonian in each glide sector with the glide eigenvalue $\mathfrak{g}^\pm$.
The time-reversal invariant momenta (TRIM) are $\Gamma_1=(0,0,0)$ ($\Gamma$ point) or $(0,0,\pi)$ ($Z$ point) and $\Gamma_2 = (\pi,-\pi,0)$ ($M$ point) or $(\pi,-\pi,\pi)$ ($A$ point), depending on $k_z=0$ or $\pi$.

In addition to the glide $\mathbb{Z}_2$ invariants $\nu^{\mathfrak{g}^\pm}$, we can define the Chern number $C$ as well as two Zak phases $\gamma$ and $\gamma'$ on the lines $k_x+k_y=\pi$ and $0$, respectively. However, they are not independent of each other.
Actually, $\gamma'=\nu^{\mathfrak{g}^+}+\nu^{\mathfrak{g}^-}$ holds, and later we will show $\gamma=0$.
Thus, the topology of the PDW state for $k_z=0,\pi$ can be characterized by identifying $\nu^{\mathfrak{g}^\pm}$ and $C$ (denoted by $\nu^{\mathfrak{g}^\pm}_{0,\pi}$ and $C_{0,\pi}$ for each $k_z$).
Furthermore, the Chern number satisfies the relation, 
\begin{align}
    C &= \gamma' - \gamma \hspace{11mm} ({\rm mod} \,\,\, 2), \\            &=\nu^{\mathfrak{g}^+} + \nu^{\mathfrak{g}^-}  \hspace{5mm} ({\rm mod} \,\,\, 2).
\end{align}
Therefore, only one of $\nu^{\mathfrak{g}^\pm}$ is a strong topological index in accordance with the $K$-theory classification for the strong indices $\mathbb{Z}\oplus\ztwo$~\cite{Shiozaki2016}. 
In this paper, we discuss $\nu^{\mathfrak{g}^\pm}$ and $C$ modulo two in CeRh$_2$As$_2$, since they do not rely on the details of the order parameter as we see below.
They are determined only by the topology of the Fermi surfaces, and do not require the full calculation of the symmetry indicators~\cite{Ono2019,Ono2020,Ono2020_2,Skurativska2020,Geier2020,Shiozaki2019,Ahn2020}.

Next, to simplify the expression \eqref{eq:z_2inv}, we pay attention to the following relation:
\begin{equation}
  \hat{I} \hat{G} = \hat{G} \hat{I} \{ E | \bm{a}+\bm{b} \},
\end{equation}
where $\hat{I}$ is the space inversion operation. 
As $\{ E | \bm{a}+\bm{b} \}$ is reduced to merely a phase factor $e^{-i(k_x+k_y)}$, $\hat{I}$ and $\hat{G}$ commute in the Hilbert space $\mathbb{V}_{k_x+k_y=0, k_z=0,\pi}$.
Therefore, the space inversion parity is well-defined in the glide sector, and
each glide sector corresponds to a 1D class D odd-parity superconductor. 
We can use the Fermi-surface formula for odd-parity superconductors~\cite{Sato2010,Fu2010}, by which the $\mathbb{Z}_2$ invariant is evaluated based on the topology of Fermi surfaces.
As a consequence, we finally get the conclusion that $\nu^{\mathfrak{g}^\pm}$ is nontrivial (trivial) when the number of Fermi surfaces between $\Gamma_1$ and $\Gamma_2$ is odd (even):
\begin{align}
  \nu^{\mathfrak{g}^\pm}_0 = \#{\rm FS}_{\Gamma \rightarrow M}^\pm \hspace{5mm} ({\rm mod} \,\,\, 2), 
\label{eq:FS_formula}
  \\
  \nu^{\mathfrak{g}^\pm}_{\pi} = \#{\rm FS}_{Z \rightarrow A}^\pm \hspace{5mm} ({\rm mod} \,\,\, 2).
\label{eq:FS_formula2}
\end{align}

In a similar way, we can relate $\gamma$ to the number of Fermi surfaces.
Because the BdG Hamiltonian on the line $k_x+k_y=\pi$ can be regarded as a 1D odd-parity superconductor, 
the Fermi surface formula~\cite{Fu2010,Sato2010} reads
\begin{align}
   \gamma_{0,\pi} = \#{\rm FS}_{\Gamma_1 \rightarrow \Gamma_2} \hspace{5mm} ({\rm mod} \,\,\, 2). 
\end{align}
Here, $\Gamma_1=X,R$ and $\Gamma_2=X',R'$ for $k_z=0,\pi$, respectively.
The four-fold rotation symmetry $C^z_4$ forces the occupation numbers of electrons at
$\Gamma_1$ and $\Gamma_2$ equal.
Therefore, the electron bands must cross the Fermi level even times between $\Gamma_1$ and $\Gamma_2$, leading to $\gamma_{0,\pi}=0$.

Below, we conduct first-principles calculations for CeRh$_2$As$_2$ and evaluate the $\mathbb{Z}_2$ invariants. The first-principles calculations are carried out at the zero magnetic field, where all the electronic bands are doubly degenerate, $\ket{\bk^{\mathfrak{g}^{\pm}}}$ and $\hat{I}\hat{\Theta}\ket{\bk^{\mathfrak{g}^{\pm}}}$, because of the inversion symmetry $\hat{I}$ and time-reversal symmetry $\hat{\Theta}$. 
The relation:
\begin{align}
  \hat{G}\hat{I}\hat{\Theta}\ket{\bk^{\mathfrak{g}^{\pm}}} &= \hat{I}\hat{\Theta}\hat{G}\ket{\bk^{\mathfrak{g}^{\pm}}} 
  = \mathfrak{g}^\mp \hat{I}\hat{\Theta}\ket{\bk^{\mathfrak{g}^{\pm}}},
\end{align}
reveals that the degenerate states belong to different glide sectors. 
Consequently, the numbers of Fermi surfaces are equivalent between the two glide sectors and coincides with the number of spinful bands crossing the Fermi level, $\#{\rm FS}_{\Gamma_1 \rightarrow \Gamma_2}^+ = \#{\rm FS}_{\Gamma_1 \rightarrow \Gamma_2}^- \equiv \#{\rm FS}_{\Gamma_1 \rightarrow \Gamma_2}$. 
The magnetic field does not alter the $\ztwo$ invariants and Chern number modulo two unless it causes the Lifshitz transition. We will discuss the effect of possible Lifshitz transitions later.

\begin{figure}[htbp]
 \begin{center}
    \includegraphics[keepaspectratio, scale=0.25]{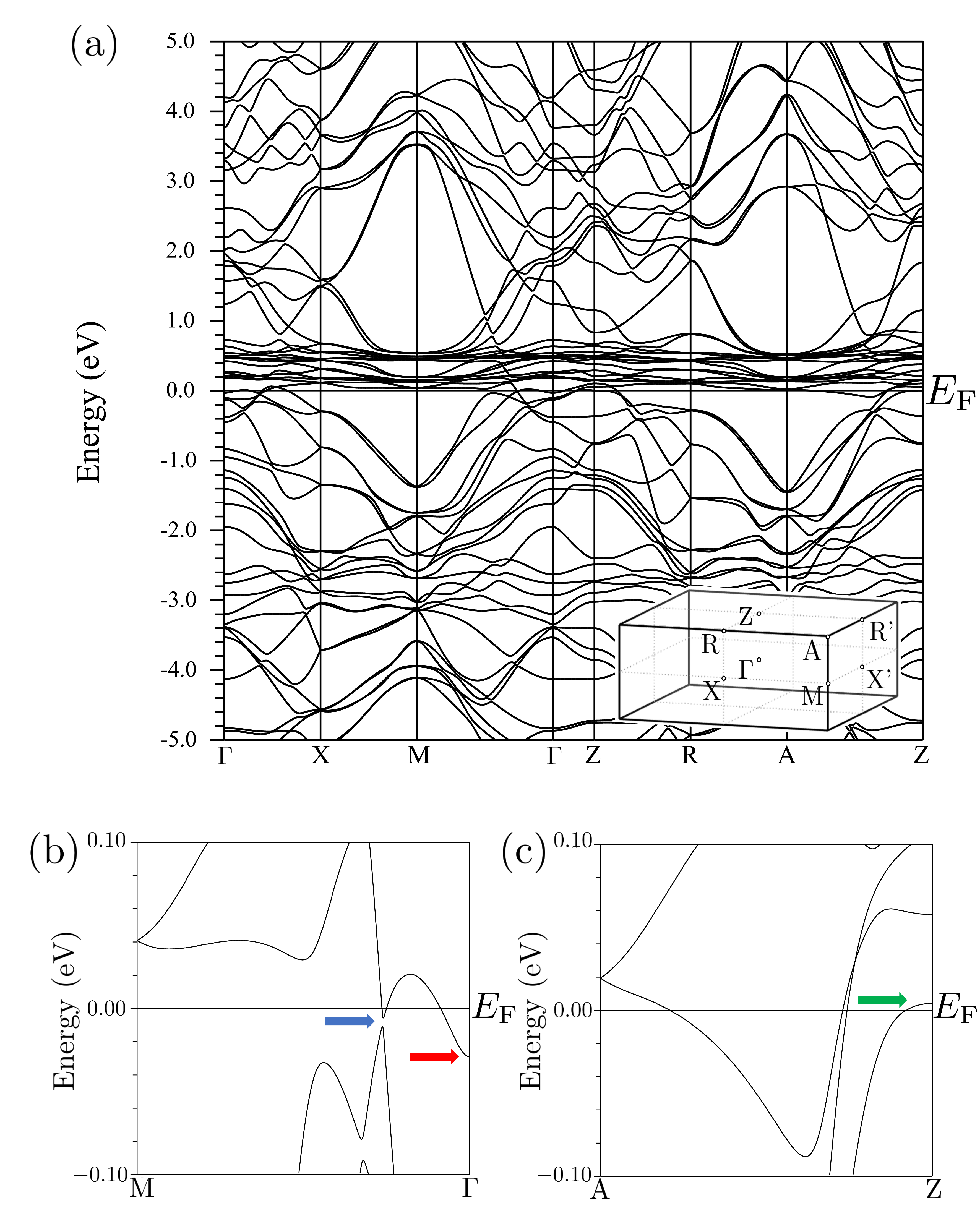}
  \end{center}
  \caption{Electronic band structure of CeRh$_2$As$_2$ with spin-orbit coupling calculated by \wien.
  (a) The whole bands along symmetric lines. 
  The heavy bands of Ce 4$f$ electrons are seen near the Fermi level. 
  (b) Enlarged view along the $M$-$\Gamma$ line. The bands cross the Fermi level three times.
  (c) Enlarged view along the $A$-$Z$ line. The bands cross the Fermi level four times.}
  \label{fig:CeRh2As2_band}
\end{figure}

\begin{figure}[htbp]
 \begin{center}
    \includegraphics[keepaspectratio, scale=0.26]{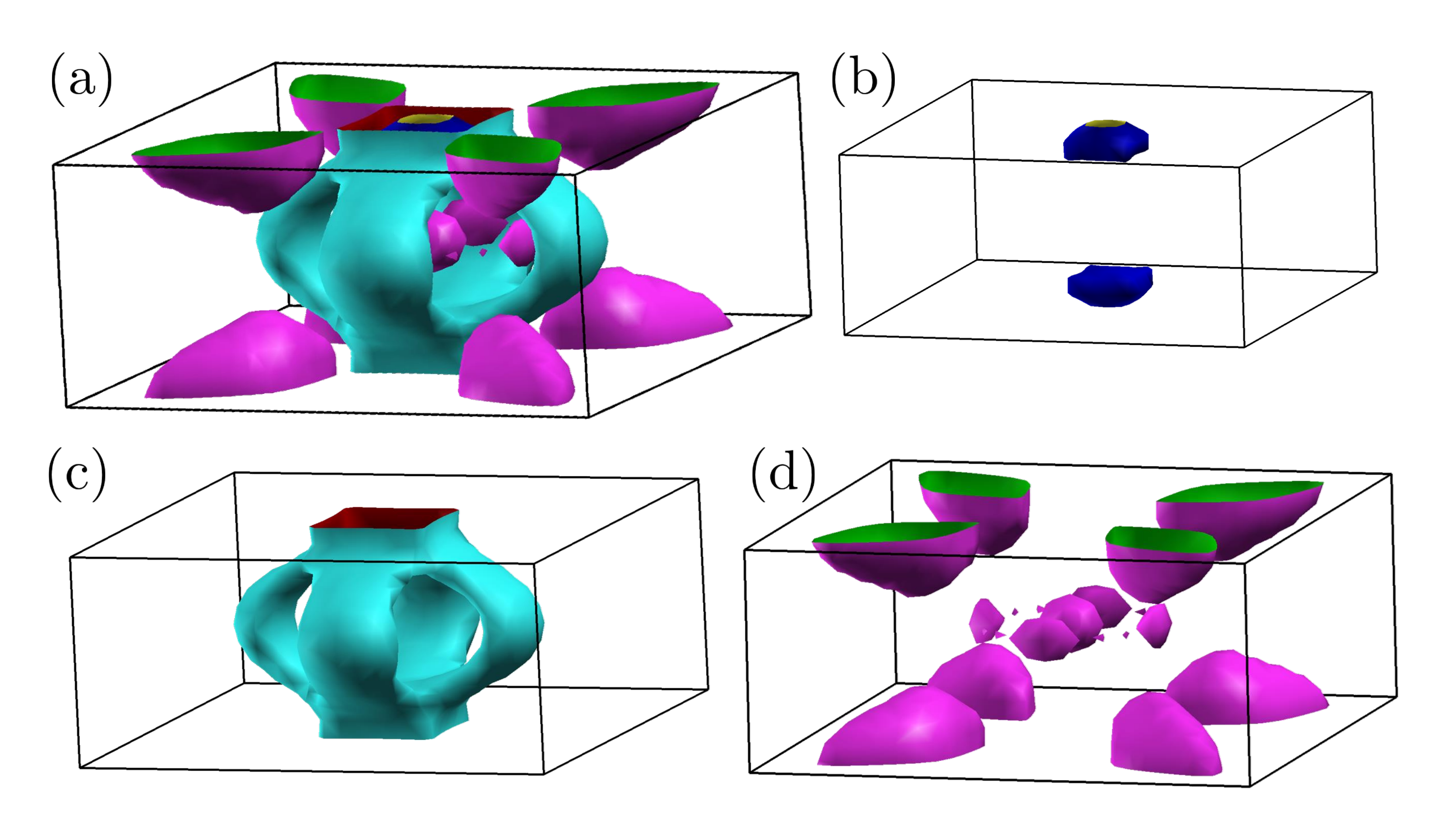}
  \end{center}
  \caption{(a) Fermi surfaces of CeRh$_2$As$_2$. Each Fermi surface is shown in (b)–(d). The Fermi surfaces other than the pockets around the $A$ point in (d) are mainly constituted by Ce 4$f$ electrons. 
  The figures are generated by {\sc XCrySDen} software~\cite{Kokaji1999}.}
  \label{fig:CeRh2As2_fs}
\end{figure}

\textit{Band structure of CeRh$_2$As$_2$.} --- 
We carry out the density functional theory (DFT) for electronic structure calculations using the \wien package~\cite{Blaha2019}. 
The crystallographic parameters are experimentally obtained values~\cite{Khim2021}.
We employ the full-potential linearized augmented plane wave+local orbitals method within the generalized gradient approximation (GGA) with the spin-orbit coupling. 
Details of our band calculations are given in the Supplemental Materials~\cite{supplement}.
Figure~\ref{fig:CeRh2As2_band}(a) shows the band structure of CeRh$_2$As$_2$, revealing the heavy bands of Ce 4$f$ electrons near the Fermi level. The main contribution to the density of states at the Fermi level comes from the Ce 4$f$ orbitals~\cite{supplement}. 
We show the Fermi surfaces in Fig.~\ref{fig:CeRh2As2_fs}.
The Ce 4$f$ electrons with hybridization to Rh 4$d_{x^2-y^2}$ electrons mainly constitute the Fermi surface except for the pink pockets near the $A$ point [Fig.~\ref{fig:CeRh2As2_fs}(d)].
The largest Fermi surface [Fig.~\ref{fig:CeRh2As2_fs}(c)] is quasi-two dimensional, supporting the field-induced PDW state~\cite{Yoshida2012}.

Before moving to discussions about $\mathbb{Z}_2$ topological invariants, we compare the result with the band structure calculated by {\sc Vasp} and {\sc Quantum Espresso} software in which the pseudopotential method was adopted~\cite{Ptok2021}.
The authors claim that the heavy bands of Ce 4$f$ electrons are away from the Fermi level, and therefore, 4$f$ electrons only weakly influence the system properties. 
On the other hand, our result shows heavy-fermion bands crossing the Fermi level. 
In the experiments, the Kondo effect was observed~\cite{Khim2021,Ishida2021}, 
and the electronic specific heat coefficient $\gamma$ is about $1000\,\mathrm{mJ/mol\,K^2}$, which support the presence of heavy-fermion bands in CeRh$_2$As$_2$.
Therefore, our result seems to be consistent with experiments. 
Hereafter, we proceed discussions based on our result. 
The formulas~\eqref{eq:FS_formula} and \eqref{eq:FS_formula2} of $\ztwo$ invariants are universal, and we can judge CeRh$_2$As$_2$ is TCSC or not, when the topology of Fermi surfaces is determined by experiments.

\textit{TCSC in CeRh$_2$As$_2$} --- 
We now show the $\ztwo$ invariants.
First, we discuss the topology on a glide-invariant plane, $k_z=0$. 
Focusing on the $\Gamma$-$M$ line [Fig.~\ref{fig:CeRh2As2_band}(b)], we notice that bands cross the Fermi level three times.  Therefore, both  $\nu_{0}^{\mathfrak{g}^\pm}$ are nontrivial, revealing the TCSC.
The TCSC is robust against the Zeeman splitting. 
In CeRh$_2$As$_2$ the PDW state is observed under the magnetic field between 4\,T and 14\,T~\cite{Khim2021}, in which the Zeeman splitting energy is estimated as $2\sim8\,\mathrm{meV}$ when we assume a probably-overestimated g-factor, $g=10$.
At the $\Gamma$ point (red arrow), the conduction band is approximately $30 \,\mathrm{meV}$ below the Fermi level, and thus, the occupation number does not change. 
Although the crossing points indicated by the blue arrow may be lifted, it does not influence the $\ztwo$ invariants according to the formula~\eqref{eq:FS_formula}. 
Therefore, the number of Fermi surfaces modulo two remains the same under the magnetic field, and the $\ztwo$ invariants $\nu_{0}^{\mathfrak{g}^\pm}$ are nontrivial in the PDW state.

Next, we discuss the other glide-invariant plane $k_z=\pi$. 
On the $Z$-$A$ line [Fig.~\ref{fig:CeRh2As2_band}(c)], bands cross the Fermi level four times, and accordingly, $\nu_{\pi}^{\mathfrak{g}^\pm}$ are trivial.
However, a hole band (green arrow) is shallow, and the energy at the Z point is estimated to be $\sim4\,\mathrm{meV}$.  
Therefore, the Lifshitz transition may occur under the magnetic field, and the $\ztwo$ invariants may become nontrivial. 
The $\ztwo$ topological invariants of CeRh$_2$As$_2$ are summarized in Table~\ref{tab:z2_invariant} based on the band structure calculation.
We assume that one of the Zeeman split bands undergoes the Lifshitz transition at the $Z$ point in the high-field region. 
Then, the Chern number as well as a $\ztwo$ invariant are nontrivial. Because the Chern number is well-defined on any constant $k_z$ plane, the difference in $C_0$ and $C_\pi$ indicates a Weyl superconducting state as in UPt$_3$~\cite{Yanase2016}. 
\begin{table}[htbp]
\caption{$\ztwo$ topological invariants 
and Chern number
of CeRh$_2$As$_2$ in the PDW state based on the first-principle calculation. 
The possibility of a Weyl superconducting state is determined. 
In the high-field region, one of the Zeeman split bands is supposed to cause the Lifshitz transition at the $Z$ point.}
\label{tab:z2_invariant}
\begin{tabular}{cccc}
\hline
 & ($\nu_{0}^{\mathfrak{g}^+}$, $\nu_{0}^{\mathfrak{g}^-}$, $C_0$)  & ($\nu_{\pi}^{\mathfrak{g}^+}$, $\nu_{\pi}^{\mathfrak{g}^-}$, $C_\pi$) & Weyl SC \\ \hline
Low field & $(1,1, \text{even})$ & $(0,0, \text{even})$ & $\times$ \\ 
High field & $(1,1, \text{even})$ & $(1,0, \text{odd})$ or $(0,1, \text{odd})$ & $\circ$ \\ 
\hline
\end{tabular}
\end{table}

From these results, CeRh$_2$As$_2$ is shown to be a platform of TCSC.
We would like to stress that the glide $\ztwo$ invariants are nontrivial at least on the $k_z=0$ plane.
Therefore, boundary Majorana states appear at the $(\bar{1}10)$ surface preserving the glide symmetry, which is generated by $\bm{a}+\bm{b}$ and $\bm{c}$.

\begin{figure}[htbp]
 \begin{center}
    \includegraphics[keepaspectratio, scale=0.27]{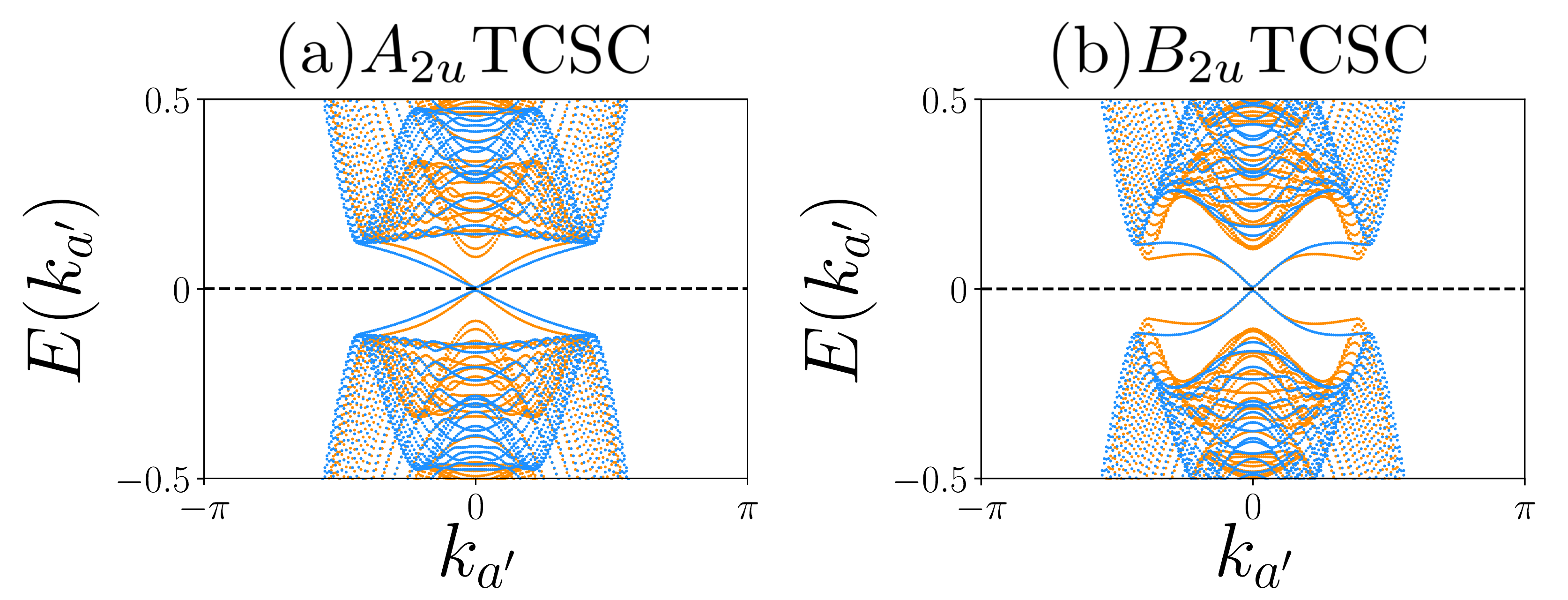}
  \end{center}
  \caption{$(\bar{1}10)$ and $(1\bar{1}0)$ surface states of (a) $A_{1u}$ and (b) $B_{1u}$ superconducting states at one of the glide-invariant planes, $k_z=0$.
  The orange (blue) color represents the glide-even (odd) sector.
  The axis $k_{a'}$ corresponds to $k_x+k_y$, and Majorana surface states appear at $k_{a'}=0$.}
  \label{fig:CeRh2As2_edge}
\end{figure}

\textit{Model study.} --- 
To demonstrate the emergence of Majorana surface states, we conduct numerical calculations based on a tight-binding model for a Ce 4$f$-orbital. 
We construct the tight-binding model and set parameters~\cite{supplement} so that $\#{\rm FS}_{\Gamma_1 \rightarrow \Gamma_2}^{\pm}$ are odd (even) at $k_z=0$, $(k_z=\pi)$, consistent with the first-principles calculation.
In the spectrum for $k_z=0$ (Fig.~\ref{fig:CeRh2As2_edge}), we recognize the stable Majorana surface states for both $A_{2u}$ representation ($s+p$-wave state) and $B_{2u}$ representation ($d_{x^2-y^2}+p$-wave state).
We also verified the Majorana states for the $A_{1u}$ and $B_{1u}$ representations, while the surface states are gapped on the other glide-invariant plane $k_z=\pi$ (See Supplemental Materials~\cite{supplement}). All the results are consistent with our analysis of topological invariants discussed above.

\textit{Summary and conclusion.} --- 
In this Letter, we theoretically investigated the electronic structure and topological superconductivity in a recently-discovered heavy-fermion superconductor CeRh$_2$As$_2$.
First, using the group theory, we clarified the algebra of symmetry operations in the Bloch representation and decomposed the Hilbert space on the glide-invariant planes $k_z=0,\pi$ into the glide sectors. 
Supposing the PDW state in the high-field superconducting phase as proposed, 
based on these results, we derived the Fermi-surface formula of $\ztwo$ invariants specifying the TCSC protected by the nonsymmorphic glide symmetry.
Second, we conducted the first-principles calculation for the electronic structure of CeRh$_2$As$_2$. 
Evaluating the $\ztwo$ invariants, we found the TCSC due to heavy-fermion bands of Ce 4$f$ electrons hybridizing with conduction electrons. 
The emergence of Majorana fermions at the surface preserving the glide symmetry was demonstrated based on the tight-binding model. This work proposes CeRh$_2$As$_2$ as a new class of topological superconductors in the following two senses. 
(1) The topological superconductivity requires neither spin-triplet pairing nor topological band structures. 
(2) The topological structure is protected by nonsymmorphic symmetry, which does not have a counterpart in continuous systems.

\begin{acknowledgments}
The authors are grateful to S.~Sumita, H.~Watanabe, T.~Kitamura, S. Khim, C. Geibel, and K. Ishida for fruitful discussions.
We thank S. Khim and C. Geibel for providing the parameters of the crystal structure
before the submission of their paper.
Some figures in this work were created by using {\sc Vesta}~\cite{Momma2011} and {\sc XCrySDen}~\cite{Kokaji1999}.
This work was supported by JSPS KAKENHI (Grants No. JP18H05227, No. JP18H01178, and No. 20H05159), SPIRITS 2020 of Kyoto University and Research Grants, 2020 of WISE Program, MEXT.
\end{acknowledgments}

%


\clearpage

\renewcommand{\bibnumfmt}[1]{[S#1]}
\renewcommand{\citenumfont}[1]{S#1}
\renewcommand{\thesection}{S\arabic{section}}
\renewcommand{\theequation}{S\arabic{equation}}
\setcounter{equation}{0}
\renewcommand{\thefigure}{S\arabic{figure}}
\setcounter{figure}{0}
\renewcommand{\thetable}{S\arabic{table}}
\setcounter{table}{0}
\makeatletter
\c@secnumdepth = 2
\makeatother

\onecolumngrid

\begin{center}
 {\large \textmd{Supplemental Materials:} \\[0.3em]
 {\bfseries Topological crystalline superconductivity in locally noncentrosymmetric CeRh$_2$As$_2$}}
 
\end{center}

\setcounter{page}{1}

\section{Symmetry and notations in superconducting state}
We begin with clarifying notations for general space group operation $\hat{g}$ in the normal and superconducting states.
In the main text, the special case $\hat{g}=\hat{G}$ (the glide operation) is considered.
In Seitz notation, $\hat{g}$ is given by $\{p|\bm{\tau}\}$, in which $p$ and $\bm{\tau}$ are point group operation and translation. It acts on real-space coordinates as:
\begin{equation}
    \hat{g}\bm{x} = p\bm{x}+\bm{\tau}.
\end{equation}
In the crystals, the relation $\bm{x} = \bm{R} + \bm{r}_n$ holds, where $\bm{R}$ and $\bm{r}_n$ are the center of a unit cell and relative position of the $n$-th sublattice within a unit cell.
For electrons with internal degrees of freedom labeled by $l$,
\begin{equation}
\hat{g} c_{l}^{\dagger}\left(\boldsymbol{R}+\boldsymbol{r}_{n}\right) \hat{g}^{-1}=c_{l^{\prime}}^{\dagger}\left(p \boldsymbol{R}+\Delta \boldsymbol{R}_{n}^{g}+\boldsymbol{r}_{n^{\prime}}\right) \mathcal{D}^{\mathrm{SL}}_{n^{\prime} n}(p) \mathcal{D}^{\mathrm{int}}_{l^{\prime} l}(p),
\end{equation}
where $\mathcal{D}^{\mathrm{SL}}_{n'n}(p) = \delta_{n',\hat{g}(n)}$, $\mathcal{D}^{\mathrm{int}}_{l'l}(p)$ is a representation matrix associated with internal degrees of freedom, and $\Delta\bm{R}^g_n$ represents the displacement of the unit cells before and after the symmetry operation $\hat{g}$.
The relation: $p\boldsymbol{r}_{n} = \Delta \boldsymbol{R}_{n}^{g}+\boldsymbol{r}_{n}^{\prime}$ must be held.
In the periodic boundary conditions, we conduct Fourier transform:
\begin{equation}
c_{\bk l n}^{\dagger} \equiv \frac{1}{\sqrt{V}} \sum_{\bm{R}} e^{i \bk \cdot \bm{R}} c_{l}^{\dagger}\left(\bm{R}+\bm{r}_{n}\right),
\end{equation}
in which the basis is periodic in the Brillouin zone,
\begin{equation}
    c^{\dagger}_{\bk ln} = c^{\dagger}_{\bk+\bm{G}ln},
\end{equation}
with any reciprocal lattice vector $\bm{G}$.
Assuming that $\hat{T}_{\bm{a}}$ is a primitive lattice translation operator, we have
\begin{align}
    \hat{T}_{\bm{a}}c^{\dagger}_{\bk ln}\hat{T}_{\bm{a}}^\dagger &= \frac{1}{\sqrt{V}} \sum_{\bm{R}} e^{i \bk \cdot \bm{R}} c_{l}^{\dagger}\left(\bm{R}+\bm{a}+\bm{r}_{n}\right) \\
    &= \frac{1}{\sqrt{V}} e^{-\bk \cdot \bm{a}} \sum_{\bm{R}} e^{i \bk \cdot (\bm{R}+\bm{a})} c_{l}^{\dagger}\left(\bm{R}+\bm{a}+\bm{r}_{n}\right) \\
    &= e^{-\bk \cdot \bm{a}} c^{\dagger}_{\bk ln}.
\end{align}
Therefore, $\bk$ labels eigenvalues of the primitive lattice translation, and from the fact that $[\hat{H},\hat{T}_{\bm{a}}]=0$, we can divide the Hilbert space into each $\bk$ sector,
\begin{equation}
    \mathbb{V} = \bigoplus_{\bk} \mathbb{V}_{\bk}.
\end{equation}
This is nothing but the Bloch theorem.
The dimension of $\mathbb{V}_{\bk}$ is identical with the number of degrees of freedom such as spin, orbital, and sublattice.
In the space $\mathbb{V}_{\bk}$, we can represent the symmetry operation $\hat{g}$,
\begin{align}
    \hat{g}c^{\dagger}_{\bk,\alpha}\hat{g}^{-1} &=\frac{1}{\sqrt{V}} \sum_{\bm{R}}e^{i \bk \cdot \bm{R}}c_{l^{\prime}}^{\dagger}\left(p \boldsymbol{R}+\Delta \boldsymbol{R}_{n}^{g}+\boldsymbol{r}_{n^{\prime}}\right) \mathcal{D}^{\mathrm{SL}}_{n^{\prime} n}(p) \mathcal{D}^{\mathrm{int}}_{l^{\prime} l}(p) \\
    &=\frac{1}{\sqrt{V}} \sum_{\bm{R}'=p\bm{R}+\Delta\bm{R}^g_n}e^{i \bk \cdot (p^{-1}(\bm{R}'-\Delta\bm{R}^g_n))}c_{l^{\prime}}^{\dagger}\left(\bm{R}'+\boldsymbol{r}_{n^{\prime}}\right) \mathcal{D}^{\mathrm{SL}}_{n^{\prime} n}(p) \mathcal{D}^{\mathrm{int}}_{l^{\prime} l}(p) \\
    &=\frac{1}{\sqrt{V}} e^{-i p\bk \cdot \Delta\bm{R}^g_n} \sum_{\bm{R}'}e^{i p\bk \cdot \bm{R}'}c_{l^{\prime}}^{\dagger}\left(\bm{R}'+\boldsymbol{r}_{n^{\prime}}\right) \mathcal{D}^{\mathrm{SL}}_{n^{\prime} n}(p) \mathcal{D}^{\mathrm{int}}_{l^{\prime} l}(p) \\
    &= c^{\dagger}_{p\bk \,l'n'} e^{-i p\bk \cdot \Delta\bm{R}^g_n} \mathcal{D}^{\mathrm{SL}}_{n^{\prime} n}(p) \mathcal{D}^{\mathrm{int}}_{l^{\prime} l}(p) \\ 
    &= c^{\dagger}_{p\bk,\beta} \mathcal{D}^g(\bk)_{\beta\alpha},
    \label{eq:rep_g}
\end{align}
where $\alpha,\beta$ represent total internal degrees of freedom of electrons, $\alpha=(l,n)$, $\beta=(l',n')$, and $\mathcal{D}^g(\bk)$ is the representation matrix of $\hat{g}$. From Eq.~\eqref{eq:rep_g}, $\mathcal{D}^g(\bk)_{\beta\alpha}= e^{-i p\bk \cdot \Delta\bm{R}^g_n} \mathcal{D}^{\mathrm{SL}}_{n^{\prime} n}(p) \mathcal{D}^{\mathrm{int}}_{l^{\prime} l}(p)$.
When the Hamiltonian $\hat{H}=\sum_{\bm{k},\alpha,\beta}c^\dagger_{\bm{k}\alpha}[\mathcal{H}(\bm{k})]_{\alpha\beta}c_{\bm{k}\beta}$ preserves the symmetry $\hat{g}$, $[\hat{H},\hat{g}]=0$, the following relation is satisfied
\begin{equation}
    \mathcal{D}^g(\bk) \mathcal{H}(\bk) (\mathcal{D}^g(\bk))^{\dagger} = \mathcal{H}(p\bk).
\end{equation}

Though superconducting gap function may break the space group symmetry $\hat{g}$, the symmetry in the normal state is preserved in combination with the $U(1)$ symmetry, at least when the superconductivity belongs to a  one-dimensional irreducible representation.
The representation matrix is given by
\begin{align}
  \mathcal{D}^g_{\mathrm{BdG}}(\bk) = \left(
  \begin{array}{cc}
    \mathcal{D}^g(\bk) & 0 \\
    0 & \pm\left(\mathcal{D}^g(-\bk)\right)^*  \\
  \end{array}
  \right)_\tau,
\end{align}
where the positive (negative) sign corresponds to $\hat{g}$-even ($\hat{g}$-odd) superconductivity, as
\begin{equation}
    \mathcal{D}^g(\bk) \Delta(\bk) (\mathcal{D}^g(-\bk))^{\top} = \pm \Delta(p\bk),
\end{equation}
and we take Nambu basis as $\psi(\bk) = (c_{\bk},c^{\dagger}_{-\bk})^\top$.
In the Nambu space, the Bogoliubov–de Gennes (BdG) Hamiltonian takes the form:
\begin{align}
    \hat{H}_{\bdg} &= \frac{1}{2} \sum_{\bk} \psi^{\dagger}(\bk) \mathcal{H}_{\mathrm{BdG}}(\bk) \psi(\bk), \\
    \mathcal{H}_{\mathrm{BdG}}(\bk) &= \left(
  \begin{array}{cc}
    \mathcal{H}(\bk) & \Delta(\bk)\\
    \Delta^\dagger(\bk) & -\mathcal{H}^{\top}(-\bk) \\
  \end{array}
  \right)_\tau,
\end{align}
where we represent the index of Nambu space as $\tau$.
In the superconducting state, the particle-hole symmetry is always preserved,
\begin{align}
    \mathcal{C} \mathcal{H}_{\mathrm{BdG}}(\bk)\mathcal{C}^{-1} &= - \mathcal{H}_{\mathrm{BdG}}(-\bk), \\
    \mathcal{C} &= \left(
  \begin{array}{cc}
    0 & 1 \\
    1 & 0 \\
  \end{array}
  \right)_\tau \mathcal{K},
\end{align}
with $\mathcal{K}$ being the complex conjugation.
From direct calculations, we confirm the following relation:
\begin{equation}
    \mathcal{C}\mathcal{D}^g_{\mathrm{BdG}}(\bk) = \pm \mathcal{D}^g_{\mathrm{BdG}}(-\bk)\mathcal{C}.
    \label{eq:nambu_PH}
\end{equation}
In the main text, we adopt the notation $\{\hat{g},\hat{C}\}=0$ in the sense of Eq.~\eqref{eq:nambu_PH} with the negative sign.

\section{Details of band calculation}
We perform the DFT band structure calculation for CeRh$_2$As$_2$ using the \wien code~\cite{Supp_Blaha2019}.
The crystal structure of CeRh$_2$As$_2$ is characterized by the space group $P4/nmm$ (No.129).
The crystallographic parameters have been experimentally obtained as shown in Table~\ref{tab:struct}~\cite{Supp_Khim2021}.
The maximum reciprocal lattice vector $K_{\mathrm{max}}$ is $R_{\mathrm{MT}}K_{\mathrm{max}}=8.0$, and 
the muffin-tin radii $R_\mathrm{MT}$ of 2.50, 2.42, and 2.15 a.u. are chosen for Ce, Rh, and As atoms, respectively. $13\times13\times5$ $k$-points sampling is adopted for the self-consistent calculation. We confirmed that $18\times18\times8$ $k$-points sampling gives almost the same results. 

Because the experimental values of the crystallographic parameters are different from those obtained by a lattice optimizing calculation~\cite{Supp_Ptok2021}, we calculated the electronic band structure using \wien for the latter parameters and obtained nearly the same results as in the main text. Thus, the discrepancy of the Ce 4$f$-electron level does not result from the lattice parameters. 

\begin{figure}[tbp]
 \begin{center}
    \includegraphics[keepaspectratio, scale=0.3]{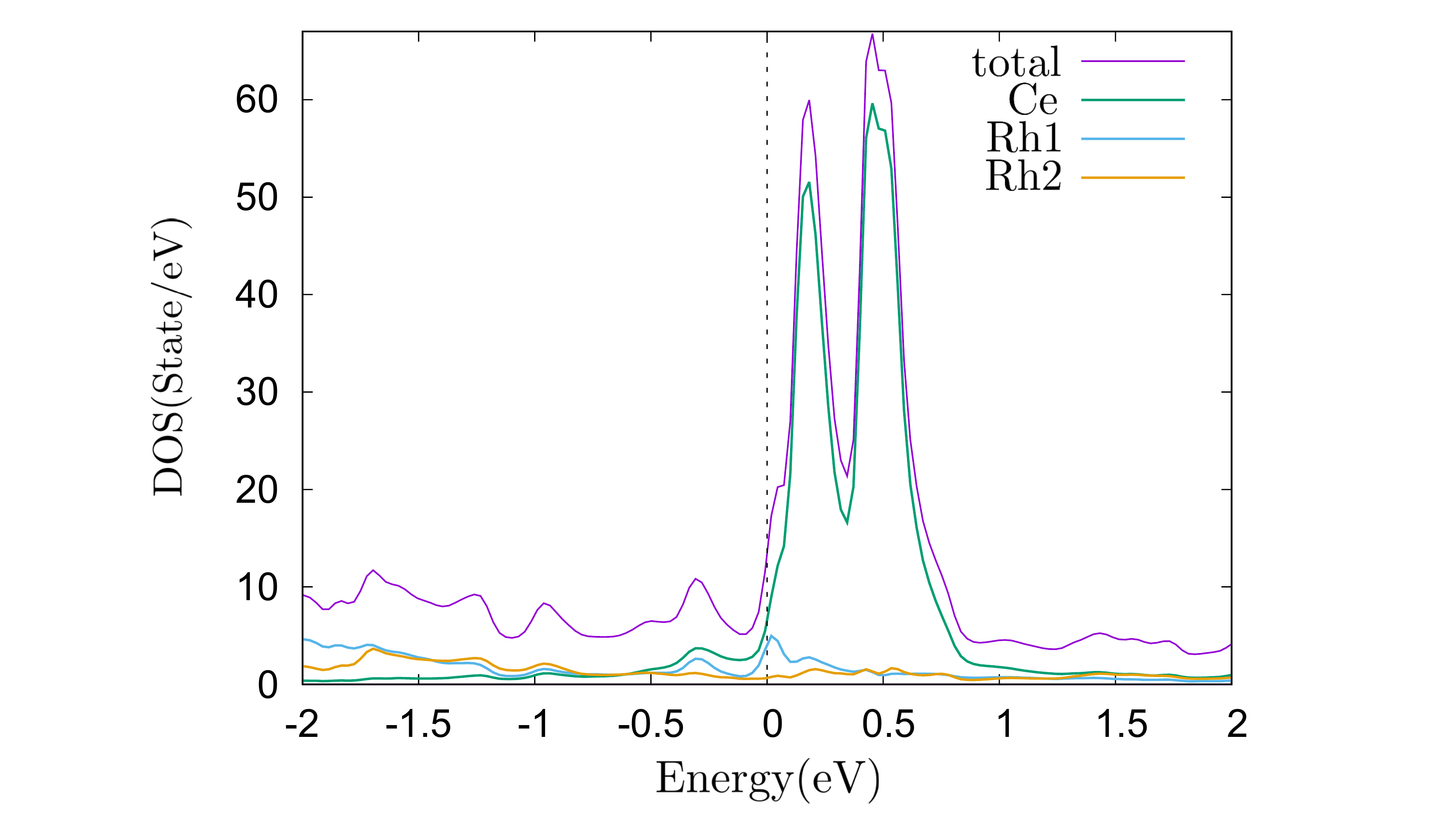}
  \end{center}
  \caption{Total density of states and partial density of states of Ce, Rh1, and Rh2 sites.
  }
  \label{fig:dos}
\end{figure}

The partial density of states of Ce and Rh atoms as well as the total density of states are shown in Fig.~\ref{fig:dos}. We see that the main contribution near the Fermi level comes from the
Ce 4$f$-orbital, consistent with the heavy-fermion behaviors observed in CeRh$_2$As$_2$~\cite{Supp_Khim2021,Supp_Ishida2021}.

\begin{table}[htbp]
\caption{Atomic coordinates of CeRh$_2$As$_2$~\cite{Supp_Khim2021}. Lattice constants are $a=b=4.2801$ \AA \hspace{1mm} and $c=9.8616$ \AA.}
\label{tab:struct}
\begin{tabular*}{0.3\columnwidth}{@{\extracolsep{\fill}}llll}
\hline\hline
Atom & $x$ & $y$ & $z$ \\ \hline
Ce ($2c$) & $0.25$ & $0.25$ & $0.25469$ \\
Rh1 ($2a$) & $0.75$ & $0.25$ & $0$ \\
Rh2 ($2c$) & $0.25$ & $0.25$ & $0.61742$ \\
As1 ($2b$) & $0.75$ & $0.25$ & $0.5$ \\
As2 ($2c$) & $0.25$ & $0.25$ & $0.86407$ \\
\hline\hline 
\end{tabular*}
\end{table}

\section{Tight-binding Hamiltonian for CeRh$_2$As$_2$}
In this section, we provide a single-orbital tight-binding model for CeRh$_2$As$_2$. 
Superconducting gap functions in the $A_{1u}$, $A_{2u}$, $B_{1u}$, and $B_{2u}$ representations are also shown.
The model is constructed by taking only the Ce atoms into account. We adopt the following model for calculating topological surface states.

\subsection{Normal-part Hamiltonian and model parameters}
\begin{figure}[htbp]
 \begin{center}
    \includegraphics[keepaspectratio, scale=0.4]{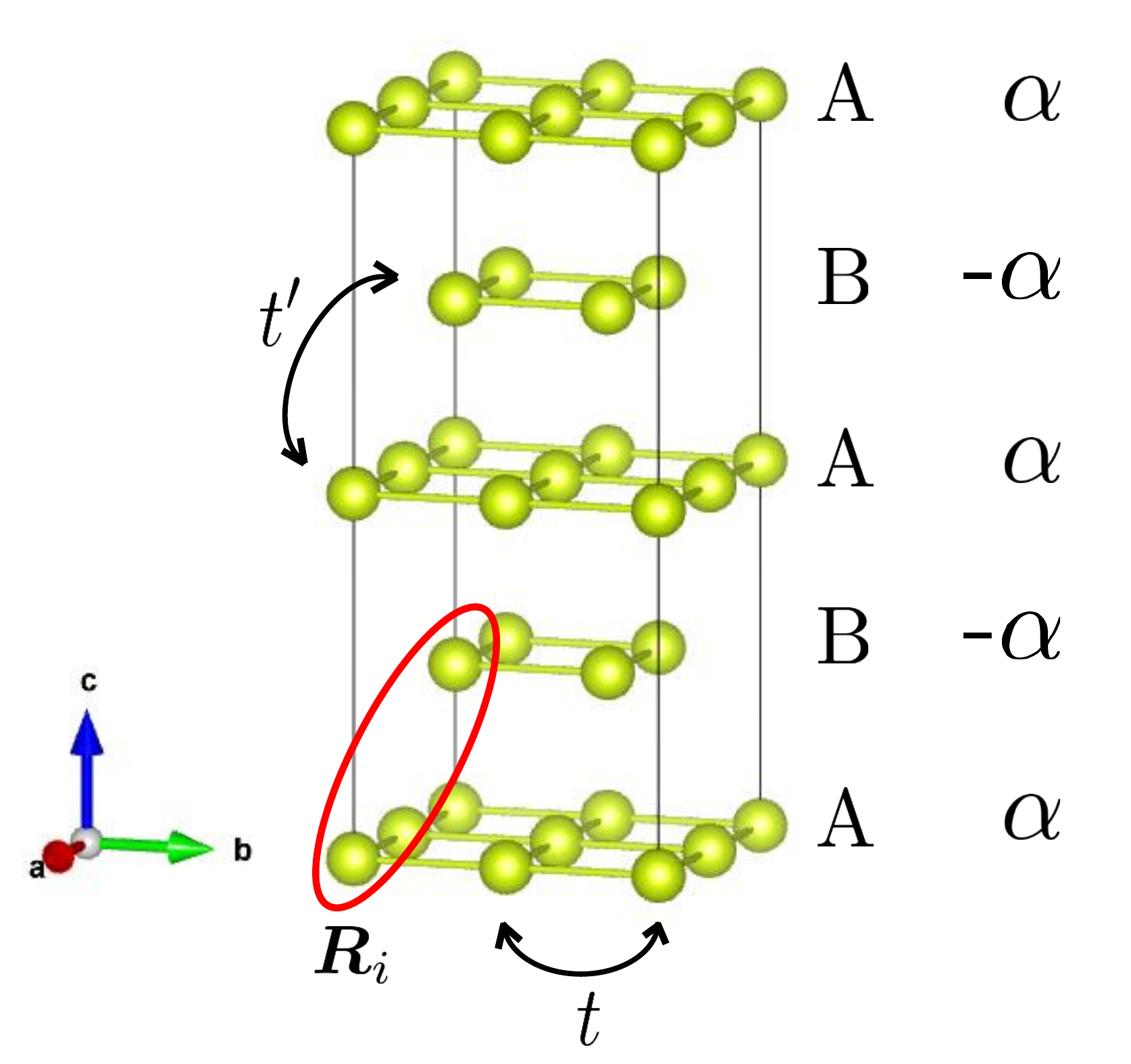}
  \end{center}
  \caption{Structure of the tight-binding model.
  The Ce atoms of CeRh$_2$As$_2$ form the body-centered tetragonal lattice. 
  The environmental Rh$_2$As$_2$ layers break the local inversion symmetry at the Ce sites. 
  The local inversion symmetry breaking introduces the A and B sublattice structure. 
  The Rashba-type antisymmetric spin-orbit coupling on each sublattice affects the electronic structure, and the sign of the spin-orbit coupling is opposite between the A and B sublattices.
  The red circle represents a unit cell, and $t\,(t')$ represents intra- (inter-)sublattice hopping.
  We set the A sublattice as a center of a unit cell.
  }
  \label{fig:onlyCe}
\end{figure}

First, we show the spin-independent part of the Hamiltonian:
\begin{align}
  \label{eq:intra_sub_hop}
  \hat{H}_{\mathrm{kin}} &= -t \sum_{i,s,\sigma} \left(c^\dagger_{i+a,s,\sigma}c_{i,s,\sigma} + c^\dagger_{i+b,s,\sigma}c_{i,s,\sigma} + h.c \right) \\
  \label{eq:inter_sub_hop}
  & \ \ \ - t' \sum_{i,s} \Bigl(c^\dagger_{i,s,A}c_{i,s,B} + c^\dagger_{i+a,s,A}c_{i,s,B} + c^\dagger_{i+b,s,A}c_{i,s,B} + c^\dagger_{i+a+b,s,A}c_{i,s,B} \notag \\
  & \ \ \ \ \ \ \ \ \ \ \ \, + c^\dagger_{i+c,s,A}c_{i,s,B} + c^\dagger_{i+a+c,s,A}c_{i,s,B} + c^\dagger_{i+b+c,s,A}c_{i,s,B} + c^\dagger_{i+a+b+c,s,A}c_{i,s,B} + h.c. \, \Bigr) \\ 
  \label{eq:kin_wave_hop}
  & = \sum_{\bk,s} \left[-2t(\cos k_x a + \cos k_y a) \sigma_0 -t'(1+e^{-ik_x a})(1+e^{-ik_y a})(1+e^{-ik_z c}) \sigma_x\right]_{\sigma \sigma'} c^\dagger_{\bm{k},s,\sigma}c_{\bm{k},s,\sigma'}.
\end{align}
where we take primitive translation vectors as $\bm{a}=(a,0,0)$, $\bm{b}=(0,a,0)$, and $\bm{c}=(0,0,c)$.
Here, $c_{i,s,\sigma}$ $(c^\dagger_{i,s,\sigma})$ are annihilation (creation) operators for the Ce 4$f$ electron with the lattice point $\bm{R}_i$, spin $s$, and sublattice $\sigma$. Equations~\eqref{eq:intra_sub_hop} and \eqref{eq:inter_sub_hop} represent the intra-sublattice hopping and inter-sublattice hopping, respectively, with $t$ and $t'$ being intra- and inter-sublattice hopping integrals (Fig.~\ref{fig:onlyCe}).
To obtain Eq.~\eqref{eq:kin_wave_hop}, we conducted Fourier transform in which Bloch basis is periodic in the Brillouin zone,
\begin{align}
  c_{i,s,A} & = \frac{1}{\sqrt{N}} \sum_{\bm{k}} c_{\bm{k}, s, A} e^{i\bm{k} \cdot \bm{R}_i}, \\
  c_{i,s,B} & = \frac{1}{\sqrt{N}} \sum_{\bm{k}} c_{\bm{k}, s, B} e^{i\bm{k} \cdot \bm{R}_i}.
\end{align}

Next, we show the spin-dependent part of the Hamiltonian:
\begin{align}
\hat{H}_{\mathrm{ASOC}} &= \sum_{s,s',\sigma}\alpha_\sigma \bm{g}(\bm{k})\cdot c^\dagger_{\bm{k},s,\sigma}\bm{s}_{s,s'}c_{\bm{k},s',\sigma}, \\
  \hat{H}_{\mathrm{Zeeman}} &= \sum_{s,s',\sigma} -\mu_B \bm{H} \cdot c^\dagger_{\bm{k},s,\sigma}\bm{s}_{s,s'}c_{\bm{k},s',\sigma}.
\end{align}
$\hat{H}_{\mathrm{ASOC}}$ represents the antisymmetric spin-orbit coupling (ASOC) depending on the sublattice, that is, $\alpha_A= -\alpha_B=\alpha$.
Since the site symmetry of the Ce site is $C_{4v}$, the g-vector has a Rashba-type form, $\bm{g}(\bk) = (-2t\sin k_y a,2t\sin k_x a,0)$.
The inter-sublattice ASOC is forbidden due to the presence of inversion center between A and B sublattices.
$\hat{H}_{\mathrm{Zeeman}}$ represents the Zeeman coupling, and hereafter we assume $\bm{H}=(0,0,H)$.

Taking the basis as $\phi(\bm{k}) = \left(c_{\bk\uparrow A},c_{\bk\downarrow A},c_{\bk\uparrow B},c_{\bk\downarrow B}\right)^\top$, the representation matrix of total Hamiltonian takes the form,
\begin{align}
  \hat{H} &= \hat{H}_{\mathrm{kin}} + \hat{H}_{\mathrm{ASOC}} + \hat{H}_{\mathrm{Zeeman}} = \sum_{\bm{k}} \phi^\dagger(\bm{k}) \mathcal{H}(\bm{k}) \phi(\bm{k}), \\
  \mathcal{H}(\bm{k}) &= \left(
  \begin{array}{cc}
    -2t(\cos k_x a + \cos k_y a)s_0 + (\alpha \bm{g}(\bm{k})-\mu_B \bm{H})\cdot \bm{s} & -t'(1+e^{-ik_x a})(1+e^{-ik_y a})(1+e^{-ik_z c})s_0 \\
    -t'(1+e^{ik_x a})(1+e^{ik_y a})(1+e^{ik_z c})s_0 & -2t(\cos k_x a + \cos k_y a)s_0 + (-\alpha \bm{g}(\bm{k})-\mu_B \bm{H}) \cdot \bm{s} \\
  \end{array}
  \right)_\sigma.
\end{align}

The glide symmetry playing the key role in this Letter is 
\begin{equation}
    \hat{G} = \{M_z|\bm{a}/2+\bm{b}/2\}.
\end{equation}
Acting on the one-particle basis, $\hat{G}$ yields the following relationship:
\begin{align}
  \hat{G} \ket{\bm{R},\uparrow,A} &= i\ket{R_x,R_y,-R_z,\uparrow,B}, \\
  \hat{G} \ket{\bm{R},\downarrow,A} &= -i\ket{R_x,R_y,-R_z,\downarrow,B}, \\
  \hat{G} \ket{\bm{R},\uparrow,B} &= i\ket{R_x+a,R_y+a,-R_z,\uparrow,A}, \\
  \hat{G} \ket{\bm{R},\downarrow,B} &= -i\ket{R_x+a,R_y+a,-R_z,\downarrow,A}.
\end{align}
Therefore, we represent $\hat{G}$ in this basis as,
\begin{align}
  \hat{G}\phi^\dagger(\bm{k})\hat{G}^{-1} &= \phi^\dagger(k_x,k_y,-k_z)\left(
  \begin{array}{cccc}
    0 & 0 & i e^{-i(k_x+k_y)a} & 0 \\
    0 & 0 & 0 & -i e^{-i(k_x+k_y)a} \\
    i & 0 & 0 & 0 \\
    0 & -i & 0 & 0 \\
  \end{array}
  \right) \\
  &= \phi^\dagger(k_x,k_y,-k_z)\left(
  \begin{array}{cc}
    0 & e^{-i(k_x+k_y)a} \\
    1 & 0 \\
  \end{array}
  \right)_\sigma \otimes (is_z) \\
  &=\phi^\dagger(k_x,k_y,-k_z) \mathcal{G}(\bk).
\end{align}
On the glide-invariant planes, $k_z=0$ or $\pi$, we can block diagonalize the Hamiltonian with the basis diagonalizing the representation matrix of glide symmetry $\mathcal{G}(\bk)$,
\begin{align}
  U^\dagger \mathcal{H}(\bk) U &=\left(
  \begin{array}{cc}
    \mathcal{H}^{\mathfrak{g}^+}(\bk) & 0  \\
    0 & \mathcal{H}^{\mathfrak{g}^-}(\bk)  \\
  \end{array}
  \right), \\
  \mathcal{H}^{\mathfrak{g}^{\pm}}(\bk) &= -2t(\cos k_x + \cos k_y)s_0 - \alpha \bm{g}(\bm{k}) \cdot \bm{s} -\left(\mu_B H \pm 4t'\cos \frac{k_x}{2} \cos \frac{k_y}{2} \, \delta_{k_z,0}\right)s_z.
\end{align}
Here, $\mathcal{H}^{\mathfrak{g}^{\pm}}(\bk)$ are the glide sector Hamiltonian. 
The unitary matrix $U$ is obtained as,
\begin{align}
  U &= \frac{1}{\sqrt{2}}\left(
  \begin{array}{cccc}
    e^{-i(k_x+k_y)a/2} & 0 & -e^{-i(k_x+k_y)a/2} & 0 \\
    0 & -e^{-i(k_x+k_y)a/2} & 0 &e^{-i(k_x+k_y)a/2} \\
    1 & 0 & 1 & 0 \\
    0 & 1 & 0 & 1 
  \end{array}
  \right)  \\
  &= \frac{1}{\sqrt{2}}\left(
  \begin{array}{cc}
    e^{-i(k_x+k_y)a/2} s_z & -e^{-i(k_x+k_y)a/2} s_z \\
    s_0 & s_0 
  \end{array}
  \right).
\end{align}

\begin{figure}[htbp]
 \begin{center}
    \includegraphics[keepaspectratio, scale=0.4]{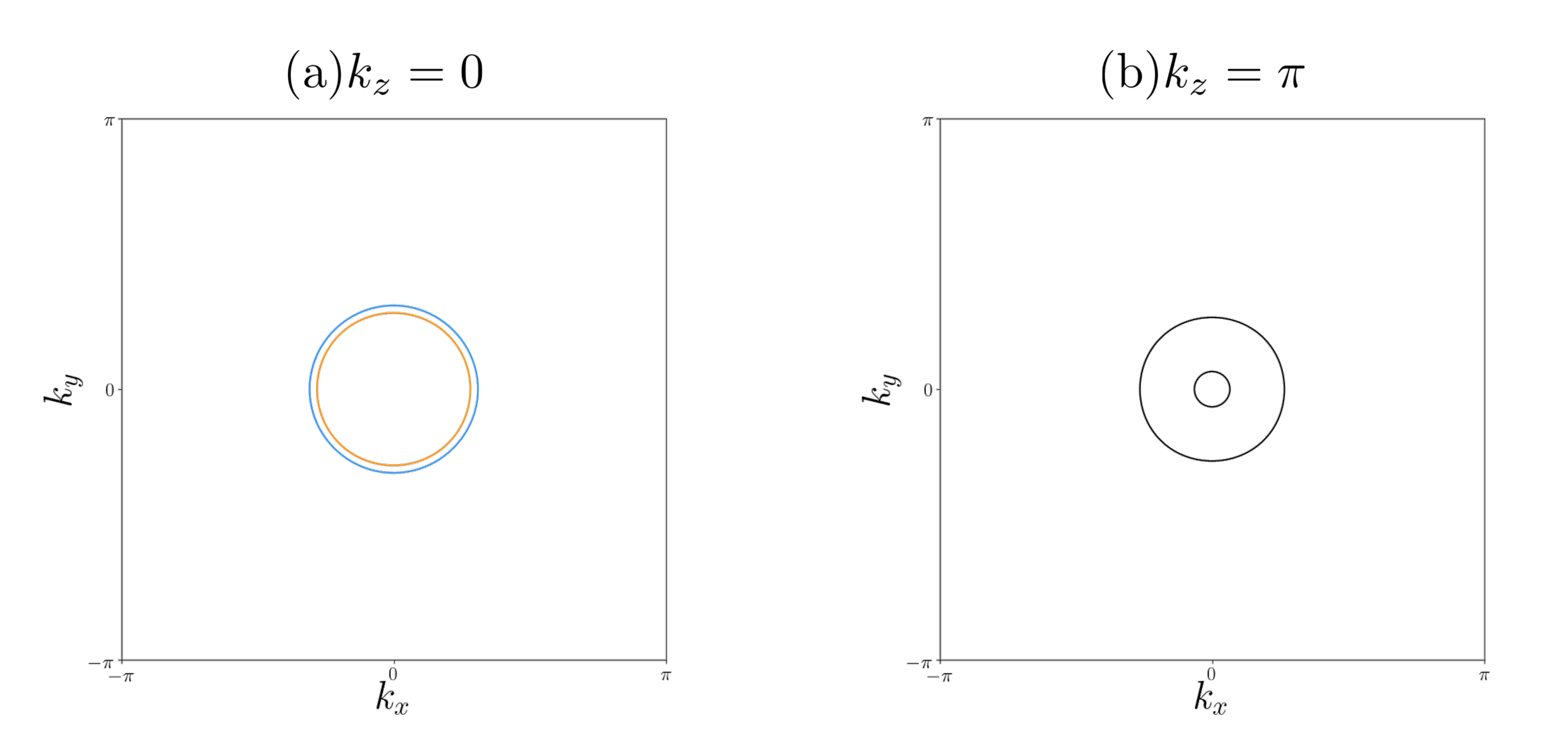}
  \end{center}
  \caption{Fermi surfaces of the tight-binding model on (a) $k_z=0$ and (b) $k_z=\pi$. 
  (a) Orange and blue lines show the Fermi surface belonging to the glide-positive and glide-negative sector, respectively.
  (b) On $k_z=\pi$, the energy spectrum of each glide sector are doubly degenerate, and we show them by black lines. Thus, the number of Fermi surfaces counted for the glide $\ztwo$ invariants is odd on $k_z=0$, while it is even on $k_z=\pi$.
  }
  \label{fig:fs_model}
\end{figure}

To demonstrate the bulk-boundary correspondence and the emergence of Majorana states in CeRh$_2$As$_2$, we adopt the following parameters
\begin{equation}
    (t,t',\alpha,h,\mu)=(1.0,0.1,0.3,-0.1,-3.8),
\end{equation}
in which the number of Fermi surfaces, $\#{\rm FS}_{\Gamma_i \rightarrow \Gamma_j}^{\pm}$ modulo 2, is equivalent to the band structure calculation for CeRh$_2$As$_2$. 
Figure~\ref{fig:fs_model} shows the Fermi surfaces of the tight-binding model on the glide-invariant planes, $k_z=0, \pi$. 
The number of Fermi surfaces in each glide sector is odd at $k_z=0$, while it is even at $k_z=\pi$, consistent with the main text.

\subsection{Superconducting gap functions, BdG Hamiltonian, and glide sectors for each irreducible representations}
Now, we discuss the superconducting order parameter. We here ignore inter-sublattice pairing for simplicity.
Since the site symmetry of the Ce site is $C_{4v}$, spin-singlet component and spin-triplet component can locally mix, although the inversion parity can be globally defined.
In the BCS state, which is the conventional even-parity state expected as the low-field phase in CeRh$_2$As$_2$, the gap function has the form, 
\begin{align}
  \Delta_{\mathrm{BCS}}(\bk) &= \left(
  \begin{array}{cc}
    (\psi(\bk) + \bm{d}(\bk) \cdot \bm{s}) \, is_y & 0 \\
    0 & (\psi(\bk) - \bm{d}(\bk) \cdot \bm{s}) \, is_y \\
  \end{array}
  \right)_\sigma. 
\end{align}
The spin-triplet component of gap functions appear with the different sign between sublattices in accordance with the sign of ASOC.
As the inversion operation $\hat{I}$ interchanges sublattices and flips the wave number $\bk$, the space inversion parity is globally even.
On the other hand, in the PDW state~\cite{Supp_Yoshida2012}, the spin-singlet (spin-triplet) gap functions have different (same) sign, 
\begin{align}
  \Delta_{\mathrm{PDW}}(\bk) &= \left(
  \begin{array}{cc}
    (\psi(\bk) + \bm{d}(\bk) \cdot \bm{s}) \, is_y & 0 \\
    0 & (-\psi(\bk) + \bm{d}(\bk) \cdot \bm{s}) \, is_y \\
  \end{array}
  \right)_\sigma,
\end{align}
and therefore, the space inversion parity is globally odd even when the spin-singlet pairing is dominant. The even-odd parity transition between the BCS and PDW states has been theoretically proposed~\cite{Supp_Yoshida2012}. Because the odd-parity PDW state is more robust against the magnetic field than the BCS state, the two phases appear in the $H$-$T$ phase digaram. The low-field phase is the BCS state, while the high-field phase is the PDW state, as illustrated in the main text [Fig.~\ref{fig:crys_struc}(b)]. 

The gap functions are classified by the global point group symmetry $D_{4h}$, and 
the odd-parity PDW state belongs to one of the odd-parity irreducible representations. In this work we focus on the one-dimensional representations, $A_{1u}$, $A_{2u}$, $B_{1u}$, $B_{2u}$, as we mention in the main text. 
Here, we describe the gap functions with using the basis functions as $\psi(\bk) = \psi_0 \times (\mathrm{basis})$ and $\bm{d}(\bk) = d_0 \times (\mathrm{basis})$.
The basis are shown in Table.~\ref{tab:basis}.

\renewcommand{\arraystretch}{1.2}
\begin{table}[htb]
  \begin{center}
  \caption{Superconducting gap functions classified by the irreducible representations of $D_{4h}$ point group.}
\begin{tabular}{|c|c|c|}
\hline 
Irreducible representation & $\psi(\boldsymbol{k}) i \sigma_{y}/\psi_0$ & $[\bm{d}(\bm{k}) \cdot \bm{\sigma}] i \sigma_{y}/d_0$ \\
\hline \hline 
$A_{1 u}$ & $\sin k_{x} \sin k_{y}\left(\cos k_{x}-\cos k_{y}\right)$ & $\sin k_x\hat{\boldsymbol{x}}+\sin k_{y} \hat{\boldsymbol{y}}$\\
$A_{2 u}$ & $1$ & $\sin k_{y} \hat{\boldsymbol{x}}-\sin k_{x} \hat{\boldsymbol{y}}$\\
$B_{1 u}$ & $\sin k_{x}\sin k_{y}$ & $\sin k_{x} \hat{\boldsymbol{x}}-\sin k_{y} \hat{\boldsymbol{y}}$ \\
$B_{2 u}$ & $\cos k_{x}-\cos k_{y}$ & $\sin k_{y} \hat{\boldsymbol{x}}+\sin k_{x} \hat{\boldsymbol{y}}$\\
\hline  
  \end{tabular}
  \label{tab:basis}
  \end{center}
\end{table}
\renewcommand{\arraystretch}{1.0}

We perform the unitary transformation for the BdG Hamiltonian on the glide-invariant planes with
\begin{align}
  U_{\bdg} &= \frac{1}{\sqrt{2}}\left(
  \begin{array}{cccc}
    e^{-i(k_x+k_y)a/2} s_z & 0 & -e^{-i(k_x+k_y)a/2} s_z & 0 \\
    s_0 & 0 & s_0 & 0 \\
    0 & e^{-i(k_x+k_y)a/2} s_z & 0 & -e^{-i(k_x+k_y)a/2} s_z \\
    0 & s_0 & 0 & s_0 
  \end{array}
  \right),  \\
\end{align}
and obtain the block-diagonalized (glide sector) Hamiltonian $\mathcal{H}^{\mathfrak{g}^{\pm}}_{\bdg}(\bk)$ as in the normal state,
\begin{align}
  &U_{\bdg}^\dagger \mathcal{H}_{\bdg}(\bk) U_{\bdg} =\left(
  \begin{array}{cc}
    \mathcal{H}^{\mathfrak{g}^+}_{\bdg}(\bk) & 0  \\
    0 & \mathcal{H}^{\mathfrak{g}^-}_{\bdg}(\bk)  \\
  \end{array}
  \right)_{\sigma}, \\
  &\mathcal{H}^{\mathfrak{g}^{\pm}}_{\bdg}(\bk)=\left(
  \begin{array}{cc}
    \mathcal{H}^{\mathfrak{g}^{\pm}}(\bk) & \Delta^{\mathfrak{g}}(\bk)  \\
    (\Delta^{\mathfrak{g}}(\bk))^{\dagger} & \left(-\mathcal{H}^{\mathfrak{g}^{\pm}}(-\bk)\right)^{\top} \\
  \end{array}
  \right), \\
  &\mathcal{H}^{\mathfrak{g}^{\pm}}(\bk) =-2t(\cos k_x + \cos k_y)s_0 - \alpha \bm{g}(\bm{k}) \cdot \bm{s} -\left(\mu_B H \pm 4t'\cos \frac{k_x}{2} \cos \frac{k_y}{2} \, \delta_{k_z,0}\right)\sigma_z, \\
  &\Delta^{\mathfrak{g}}(\bk) = \left(-\psi(\bk)+\bm{d}(\bk)\cdot \bm{s}\right) is_y.
\end{align}

Since Majorana states appear at surfaces preserving the glide symmetry, which is generated by $\bm{a}+\bm{b}$ and $\bm{c}$, the $(\bar{1}10)$ surface is the host of Majorana surface states.
To calculate the energy spectrum in the open boundary condition, we retake primitive translation vector as $\bm{a}'= \bm{a}+\bm{b}=(a,a,0)$, $\bm{b}'=\bm{b}=(0,a,0)$, and $\bm{c}'=\bm{c}=(0,0,c)$.
The volume of a unit cell does not change under this operation, 
\begin{equation}
    V = |(\bm{a}\times\bm{b})\cdot\bm{c}| = |(\bm{a}'\times\bm{b}')\cdot\bm{c}'|.
\end{equation}
Correspondingly, the wave numbers are $k_{a'} = \bm{k}\cdot\bm{a}' = k_x+k_y$, $k_{b'} = \bm{k}\cdot\bm{b}' = k_y$, and $k_{c'} = \bm{k}\cdot\bm{c}' = k_z$.
We conduct calculations in the effective cubic Brillouin zone ($-\pi<k_{a'},k_{b'},k_{c'} \leq \pi$), with assuming the open boundary condition for $k_{b'}$ though periodic boundary condition for $k_{a'}$.
We take $L_{b'}=64$ and $L_{a'}=256$ for the system size.

\section{Majorana states on glide-symmetric surfaces}
\begin{figure}[htbp]
 \begin{center}
    \includegraphics[keepaspectratio, scale=0.38]{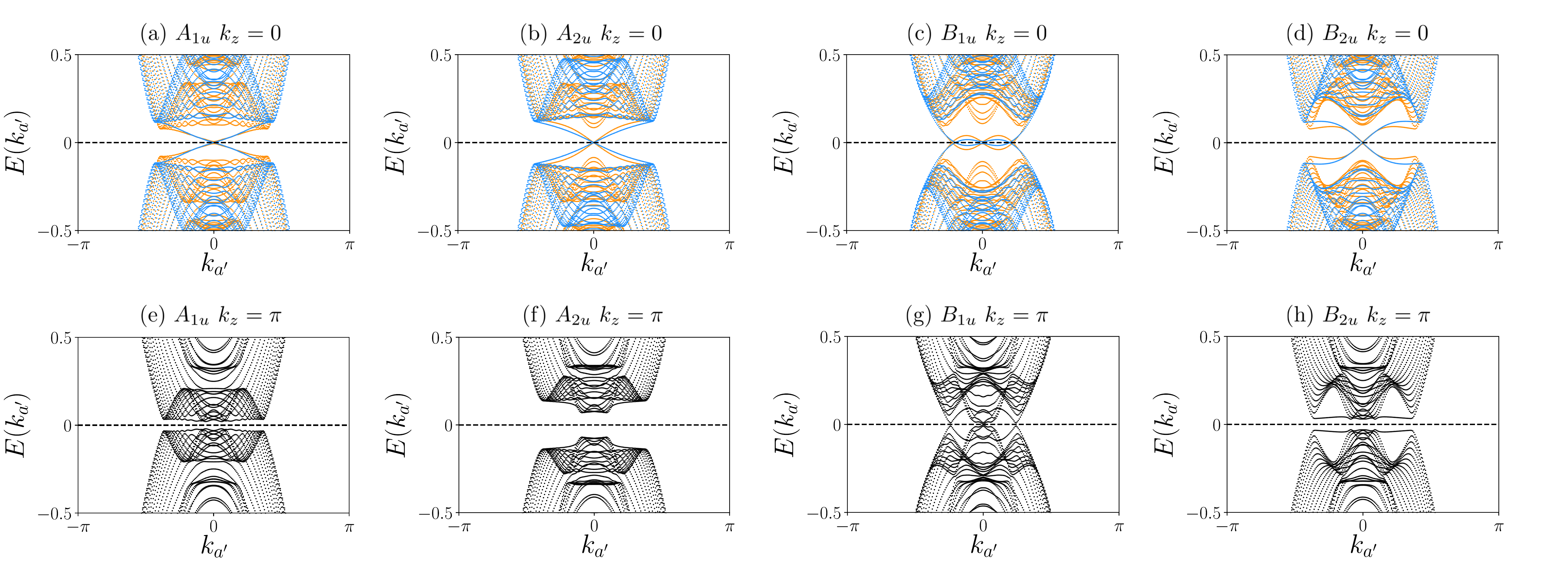}
  \end{center}
  \caption{$(\bar{1}10)$ and $(1\bar{1}0)$ surface states of the tight-binding BdG Hamiltonian on (a-d) $k_z=0$ and (e-f) $k_z=\pi$. Symmetry of superconducting gap functions is (a,e) $A_{1u}$, (b,f) $A_{2u}$, (c,g) $B_{1u}$, and (d,h) $B_{2u}$ representations. 
  We take $\psi_0=0.1$, $d_0=0.05$ for the $A_{2u}$ representation while $\psi_0=0.3$, $d_0=0.2$ for the others.
  The orange and blue lines represent the glide-positive and glide-negative sectors, respectively.  
  For $k_z=\pi$, only the spectrum of glide-positive sector is shown because it is equivalent to the glide-negative sector.
  }
  \label{fig:edge_supp}
\end{figure}

We show the surface states calculated by diagonalizing the BdG Hamiltonian introduced in the previous section with the open boundary condition. Figure~\ref{fig:edge_supp} reveals the  $(\bar{1}10)$ and $(1\bar{1}0)$ surface states on $k_z=0,\pi$ for all the odd-parity and glide-odd irreducible representations.
For the $k_z=0$ plane, we highlight glide sectors by color.
The orange lines and blue lines show the spectrum of the glide-positive and glide-negative sector, respectively.
On the other hand, for $k_z=\pi$, we show only the glide-positive sector because it is degenerate to the glide-negative sector.

All the results are consistent with the formula relating the Fermi surface topology with the glide $\ztwo$ invariants.
In Figs.~\ref{fig:edge_supp}(a)-(d), a non-degenerate Majorana state in each glide sector appears with zero energy at $k_{a'}=0$. This is consistent with the fact that $\#{\rm FS}_{\Gamma \rightarrow M}^{\pm}$ is odd [Fig.~\ref{fig:fs_model}(a)] and the $\ztwo$ invariants are nontrivial on the $k_z=0$ plane. 
On the other hand, in Figs.~\ref{fig:edge_supp}(e), (f) and (h), the energy gap exists in the surface spectrum at $k_z=\pi$, consistent with the trivial glide $\ztwo$ invariants concluded from the even number of $\#{\rm FS}_{Z \rightarrow A}^{\pm}$ [Fig.~\ref{fig:fs_model}(b)].
Although Fig.~\ref{fig:edge_supp}(g) shows zero modes at $k_{a'}=0$,
This is also consistent with the trivial glide $\ztwo$ invariants.

From the results in this section, our statements on the topological superconductivity and Majorana states have been verified based on the minimal tight-binding model study. We stress that the arguments  
do not essentially depend on the details of superconducting order parameter as well as the electronic structure, unless the topology of Fermi surfaces is altered.

%

\begin{thebibliography}{105}%
\makeatletter
\providecommand \@ifxundefined [1]{%
 \@ifx{#1\undefined}
}%
\providecommand \@ifnum [1]{%
 \ifnum #1\expandafter \@firstoftwo
 \else \expandafter \@secondoftwo
 \fi
}%
\providecommand \@ifx [1]{%
 \ifx #1\expandafter \@firstoftwo
 \else \expandafter \@secondoftwo
 \fi
}%
\providecommand \natexlab [1]{#1}%
\providecommand \enquote  [1]{``#1''}%
\providecommand \bibnamefont  [1]{#1}%
\providecommand \bibfnamefont [1]{#1}%
\providecommand \citenamefont [1]{#1}%
\providecommand \href@noop [0]{\@secondoftwo}%
\providecommand \href [0]{\begingroup \@sanitize@url \@href}%
\providecommand \@href[1]{\@@startlink{#1}\@@href}%
\providecommand \@@href[1]{\endgroup#1\@@endlink}%
\providecommand \@sanitize@url [0]{\catcode `\\12\catcode `\$12\catcode
  `\&12\catcode `\#12\catcode `\^12\catcode `\_12\catcode `\%12\relax}%
\providecommand \@@startlink[1]{}%
\providecommand \@@endlink[0]{}%
\providecommand \url  [0]{\begingroup\@sanitize@url \@url }%
\providecommand \@url [1]{\endgroup\@href {#1}{\urlprefix }}%
\providecommand \urlprefix  [0]{URL }%
\providecommand \Eprint [0]{\href }%
\providecommand \doibase [0]{http://dx.doi.org/}%
\providecommand \selectlanguage [0]{\@gobble}%
\providecommand \bibinfo  [0]{\@secondoftwo}%
\providecommand \bibfield  [0]{\@secondoftwo}%
\providecommand \translation [1]{[#1]}%
\providecommand \BibitemOpen [0]{}%
\providecommand \bibitemStop [0]{}%
\providecommand \bibitemNoStop [0]{.\EOS\space}%
\providecommand \EOS [0]{\spacefactor3000\relax}%
\providecommand \BibitemShut  [1]{\csname bibitem#1\endcsname}%
\let\auto@bib@innerbib\@empty
\bibitem [{\citenamefont {Landau}\ and\ \citenamefont
  {Lifshitz}(2013)}]{Landau2013statistical}%
  \BibitemOpen
  \bibfield  {author} {\bibinfo {author} {\bibfnamefont {L.}~\bibnamefont
  {Landau}}\ and\ \bibinfo {author} {\bibfnamefont {E.}~\bibnamefont
  {Lifshitz}},\ }\href {https://books.google.co.jp/books?id=VzgJN-XPTRsC}
  {\emph {\bibinfo {title} {Statistical Physics: Volume 5}}}\ (\bibinfo
  {publisher} {Elsevier Science},\ \bibinfo {year} {2013})\BibitemShut
  {NoStop}%
\bibitem [{\citenamefont {Qi}\ and\ \citenamefont {Zhang}(2011)}]{Qi2011}%
  \BibitemOpen
  \bibfield  {author} {\bibinfo {author} {\bibfnamefont {X.-L.}\ \bibnamefont
  {Qi}}\ and\ \bibinfo {author} {\bibfnamefont {S.-C.}\ \bibnamefont {Zhang}},\
  }\href {\doibase 10.1103/RevModPhys.83.1057} {\bibfield  {journal} {\bibinfo
  {journal} {Rev. Mod. Phys.}\ }\textbf {\bibinfo {volume} {83}},\ \bibinfo
  {pages} {1057} (\bibinfo {year} {2011})}\BibitemShut {NoStop}%
\bibitem [{\citenamefont {Tanaka}\ \emph {et~al.}(2012)\citenamefont {Tanaka},
  \citenamefont {Sato},\ and\ \citenamefont {Nagaosa}}]{Tanaka2012}%
  \BibitemOpen
  \bibfield  {author} {\bibinfo {author} {\bibfnamefont {Y.}~\bibnamefont
  {Tanaka}}, \bibinfo {author} {\bibfnamefont {M.}~\bibnamefont {Sato}}, \ and\
  \bibinfo {author} {\bibfnamefont {N.}~\bibnamefont {Nagaosa}},\ }\href
  {\doibase 10.1143/JPSJ.81.011013} {\bibfield  {journal} {\bibinfo  {journal}
  {Journal of the Physical Society of Japan}\ }\textbf {\bibinfo {volume}
  {81}},\ \bibinfo {pages} {011013} (\bibinfo {year} {2012})}\BibitemShut
  {NoStop}%
\bibitem [{\citenamefont {Sato}\ and\ \citenamefont
  {Fujimoto}(2016)}]{Sato2016}%
  \BibitemOpen
  \bibfield  {author} {\bibinfo {author} {\bibfnamefont {M.}~\bibnamefont
  {Sato}}\ and\ \bibinfo {author} {\bibfnamefont {S.}~\bibnamefont
  {Fujimoto}},\ }\href {\doibase 10.7566/JPSJ.85.072001} {\bibfield  {journal}
  {\bibinfo  {journal} {Journal of the Physical Society of Japan}\ }\textbf
  {\bibinfo {volume} {85}},\ \bibinfo {pages} {072001} (\bibinfo {year}
  {2016})}\BibitemShut {NoStop}%
\bibitem [{\citenamefont {Sato}\ and\ \citenamefont {Ando}(2017)}]{Sato2017}%
  \BibitemOpen
  \bibfield  {author} {\bibinfo {author} {\bibfnamefont {M.}~\bibnamefont
  {Sato}}\ and\ \bibinfo {author} {\bibfnamefont {Y.}~\bibnamefont {Ando}},\
  }\href {\doibase 10.1088/1361-6633/aa6ac7} {\bibfield  {journal} {\bibinfo
  {journal} {Reports on Progress in Physics}\ }\textbf {\bibinfo {volume}
  {80}},\ \bibinfo {pages} {076501} (\bibinfo {year} {2017})}\BibitemShut
  {NoStop}%
\bibitem [{\citenamefont {Kitaev}(2001)}]{Kitaev2001}%
  \BibitemOpen
  \bibfield  {author} {\bibinfo {author} {\bibfnamefont {A.~Y.}\ \bibnamefont
  {Kitaev}},\ }\href {\doibase 10.1070/1063-7869/44/10s/s29} {\bibfield
  {journal} {\bibinfo  {journal} {Physics-Uspekhi}\ }\textbf {\bibinfo {volume}
  {44}},\ \bibinfo {pages} {131} (\bibinfo {year} {2001})}\BibitemShut
  {NoStop}%
\bibitem [{\citenamefont {Nayak}\ \emph {et~al.}(2008)\citenamefont {Nayak},
  \citenamefont {Simon}, \citenamefont {Stern}, \citenamefont {Freedman},\ and\
  \citenamefont {Das~Sarma}}]{Nayak2008}%
  \BibitemOpen
  \bibfield  {author} {\bibinfo {author} {\bibfnamefont {C.}~\bibnamefont
  {Nayak}}, \bibinfo {author} {\bibfnamefont {S.~H.}\ \bibnamefont {Simon}},
  \bibinfo {author} {\bibfnamefont {A.}~\bibnamefont {Stern}}, \bibinfo
  {author} {\bibfnamefont {M.}~\bibnamefont {Freedman}}, \ and\ \bibinfo
  {author} {\bibfnamefont {S.}~\bibnamefont {Das~Sarma}},\ }\href {\doibase
  10.1103/RevModPhys.80.1083} {\bibfield  {journal} {\bibinfo  {journal} {Rev.
  Mod. Phys.}\ }\textbf {\bibinfo {volume} {80}},\ \bibinfo {pages} {1083}
  (\bibinfo {year} {2008})}\BibitemShut {NoStop}%
\bibitem [{\citenamefont {Fu}\ and\ \citenamefont {Kane}(2008)}]{Fu2008}%
  \BibitemOpen
  \bibfield  {author} {\bibinfo {author} {\bibfnamefont {L.}~\bibnamefont
  {Fu}}\ and\ \bibinfo {author} {\bibfnamefont {C.~L.}\ \bibnamefont {Kane}},\
  }\href {\doibase 10.1103/PhysRevLett.100.096407} {\bibfield  {journal}
  {\bibinfo  {journal} {Phys. Rev. Lett.}\ }\textbf {\bibinfo {volume} {100}},\
  \bibinfo {pages} {096407} (\bibinfo {year} {2008})}\BibitemShut {NoStop}%
\bibitem [{\citenamefont {Sato}\ \emph {et~al.}(2009)\citenamefont {Sato},
  \citenamefont {Takahashi},\ and\ \citenamefont {Fujimoto}}]{Sato2009}%
  \BibitemOpen
  \bibfield  {author} {\bibinfo {author} {\bibfnamefont {M.}~\bibnamefont
  {Sato}}, \bibinfo {author} {\bibfnamefont {Y.}~\bibnamefont {Takahashi}}, \
  and\ \bibinfo {author} {\bibfnamefont {S.}~\bibnamefont {Fujimoto}},\ }\href
  {\doibase 10.1103/PhysRevLett.103.020401} {\bibfield  {journal} {\bibinfo
  {journal} {Phys. Rev. Lett.}\ }\textbf {\bibinfo {volume} {103}},\ \bibinfo
  {pages} {020401} (\bibinfo {year} {2009})}\BibitemShut {NoStop}%
\bibitem [{\citenamefont {Sau}\ \emph {et~al.}(2010)\citenamefont {Sau},
  \citenamefont {Lutchyn}, \citenamefont {Tewari},\ and\ \citenamefont
  {Das~Sarma}}]{Sau2010}%
  \BibitemOpen
  \bibfield  {author} {\bibinfo {author} {\bibfnamefont {J.~D.}\ \bibnamefont
  {Sau}}, \bibinfo {author} {\bibfnamefont {R.~M.}\ \bibnamefont {Lutchyn}},
  \bibinfo {author} {\bibfnamefont {S.}~\bibnamefont {Tewari}}, \ and\ \bibinfo
  {author} {\bibfnamefont {S.}~\bibnamefont {Das~Sarma}},\ }\href {\doibase
  10.1103/PhysRevLett.104.040502} {\bibfield  {journal} {\bibinfo  {journal}
  {Phys. Rev. Lett.}\ }\textbf {\bibinfo {volume} {104}},\ \bibinfo {pages}
  {040502} (\bibinfo {year} {2010})}\BibitemShut {NoStop}%
\bibitem [{\citenamefont {Lutchyn}\ \emph {et~al.}(2010)\citenamefont
  {Lutchyn}, \citenamefont {Sau},\ and\ \citenamefont
  {Das~Sarma}}]{Lutchyn2010}%
  \BibitemOpen
  \bibfield  {author} {\bibinfo {author} {\bibfnamefont {R.~M.}\ \bibnamefont
  {Lutchyn}}, \bibinfo {author} {\bibfnamefont {J.~D.}\ \bibnamefont {Sau}}, \
  and\ \bibinfo {author} {\bibfnamefont {S.}~\bibnamefont {Das~Sarma}},\ }\href
  {\doibase 10.1103/PhysRevLett.105.077001} {\bibfield  {journal} {\bibinfo
  {journal} {Phys. Rev. Lett.}\ }\textbf {\bibinfo {volume} {105}},\ \bibinfo
  {pages} {077001} (\bibinfo {year} {2010})}\BibitemShut {NoStop}%
\bibitem [{\citenamefont {Oreg}\ \emph {et~al.}(2010)\citenamefont {Oreg},
  \citenamefont {Refael},\ and\ \citenamefont {von Oppen}}]{Oreg2010}%
  \BibitemOpen
  \bibfield  {author} {\bibinfo {author} {\bibfnamefont {Y.}~\bibnamefont
  {Oreg}}, \bibinfo {author} {\bibfnamefont {G.}~\bibnamefont {Refael}}, \ and\
  \bibinfo {author} {\bibfnamefont {F.}~\bibnamefont {von Oppen}},\ }\href
  {\doibase 10.1103/PhysRevLett.105.177002} {\bibfield  {journal} {\bibinfo
  {journal} {Phys. Rev. Lett.}\ }\textbf {\bibinfo {volume} {105}},\ \bibinfo
  {pages} {177002} (\bibinfo {year} {2010})}\BibitemShut {NoStop}%
\bibitem [{\citenamefont {Alicea}(2010)}]{Alicea2010}%
  \BibitemOpen
  \bibfield  {author} {\bibinfo {author} {\bibfnamefont {J.}~\bibnamefont
  {Alicea}},\ }\href {\doibase 10.1103/PhysRevB.81.125318} {\bibfield
  {journal} {\bibinfo  {journal} {Phys. Rev. B}\ }\textbf {\bibinfo {volume}
  {81}},\ \bibinfo {pages} {125318} (\bibinfo {year} {2010})}\BibitemShut
  {NoStop}%
\bibitem [{\citenamefont {Qi}\ \emph {et~al.}(2010)\citenamefont {Qi},
  \citenamefont {Hughes},\ and\ \citenamefont {Zhang}}]{Qi2010}%
  \BibitemOpen
  \bibfield  {author} {\bibinfo {author} {\bibfnamefont {X.-L.}\ \bibnamefont
  {Qi}}, \bibinfo {author} {\bibfnamefont {T.~L.}\ \bibnamefont {Hughes}}, \
  and\ \bibinfo {author} {\bibfnamefont {S.-C.}\ \bibnamefont {Zhang}},\ }\href
  {\doibase 10.1103/PhysRevB.82.184516} {\bibfield  {journal} {\bibinfo
  {journal} {Phys. Rev. B}\ }\textbf {\bibinfo {volume} {82}},\ \bibinfo
  {pages} {184516} (\bibinfo {year} {2010})}\BibitemShut {NoStop}%
\bibitem [{\citenamefont {Chung}\ \emph {et~al.}(2011)\citenamefont {Chung},
  \citenamefont {Qi}, \citenamefont {Maciejko},\ and\ \citenamefont
  {Zhang}}]{Chung2011}%
  \BibitemOpen
  \bibfield  {author} {\bibinfo {author} {\bibfnamefont {S.~B.}\ \bibnamefont
  {Chung}}, \bibinfo {author} {\bibfnamefont {X.-L.}\ \bibnamefont {Qi}},
  \bibinfo {author} {\bibfnamefont {J.}~\bibnamefont {Maciejko}}, \ and\
  \bibinfo {author} {\bibfnamefont {S.-C.}\ \bibnamefont {Zhang}},\ }\href
  {\doibase 10.1103/PhysRevB.83.100512} {\bibfield  {journal} {\bibinfo
  {journal} {Phys. Rev. B}\ }\textbf {\bibinfo {volume} {83}},\ \bibinfo
  {pages} {100512(R)} (\bibinfo {year} {2011})}\BibitemShut {NoStop}%
\bibitem [{\citenamefont {Mourik}\ \emph {et~al.}(2012)\citenamefont {Mourik},
  \citenamefont {Zuo}, \citenamefont {Frolov}, \citenamefont {Plissard},
  \citenamefont {Bakkers},\ and\ \citenamefont {Kouwenhoven}}]{Mourik2012}%
  \BibitemOpen
  \bibfield  {author} {\bibinfo {author} {\bibfnamefont {V.}~\bibnamefont
  {Mourik}}, \bibinfo {author} {\bibfnamefont {K.}~\bibnamefont {Zuo}},
  \bibinfo {author} {\bibfnamefont {S.~M.}\ \bibnamefont {Frolov}}, \bibinfo
  {author} {\bibfnamefont {S.~R.}\ \bibnamefont {Plissard}}, \bibinfo {author}
  {\bibfnamefont {E.~P. A.~M.}\ \bibnamefont {Bakkers}}, \ and\ \bibinfo
  {author} {\bibfnamefont {L.~P.}\ \bibnamefont {Kouwenhoven}},\ }\href
  {\doibase 10.1126/science.1222360} {\bibfield  {journal} {\bibinfo  {journal}
  {Science}\ }\textbf {\bibinfo {volume} {336}},\ \bibinfo {pages} {1003}
  (\bibinfo {year} {2012})}\BibitemShut {NoStop}%
\bibitem [{\citenamefont {Das}\ \emph {et~al.}(2012)\citenamefont {Das},
  \citenamefont {Ronen}, \citenamefont {Most}, \citenamefont {Oreg},
  \citenamefont {Heiblum},\ and\ \citenamefont {Shtrikman}}]{Das2012}%
  \BibitemOpen
  \bibfield  {author} {\bibinfo {author} {\bibfnamefont {A.}~\bibnamefont
  {Das}}, \bibinfo {author} {\bibfnamefont {Y.}~\bibnamefont {Ronen}}, \bibinfo
  {author} {\bibfnamefont {Y.}~\bibnamefont {Most}}, \bibinfo {author}
  {\bibfnamefont {Y.}~\bibnamefont {Oreg}}, \bibinfo {author} {\bibfnamefont
  {M.}~\bibnamefont {Heiblum}}, \ and\ \bibinfo {author} {\bibfnamefont
  {H.}~\bibnamefont {Shtrikman}},\ }\href {\doibase 10.1038/nphys2479}
  {\bibfield  {journal} {\bibinfo  {journal} {Nature Physics}\ }\textbf
  {\bibinfo {volume} {8}},\ \bibinfo {pages} {887} (\bibinfo {year}
  {2012})}\BibitemShut {NoStop}%
\bibitem [{\citenamefont {Deng}\ \emph {et~al.}(2012)\citenamefont {Deng},
  \citenamefont {Yu}, \citenamefont {Huang}, \citenamefont {Larsson},
  \citenamefont {Caroff},\ and\ \citenamefont {Xu}}]{Deng2012}%
  \BibitemOpen
  \bibfield  {author} {\bibinfo {author} {\bibfnamefont {M.~T.}\ \bibnamefont
  {Deng}}, \bibinfo {author} {\bibfnamefont {C.~L.}\ \bibnamefont {Yu}},
  \bibinfo {author} {\bibfnamefont {G.~Y.}\ \bibnamefont {Huang}}, \bibinfo
  {author} {\bibfnamefont {M.}~\bibnamefont {Larsson}}, \bibinfo {author}
  {\bibfnamefont {P.}~\bibnamefont {Caroff}}, \ and\ \bibinfo {author}
  {\bibfnamefont {H.~Q.}\ \bibnamefont {Xu}},\ }\href {\doibase
  10.1021/nl303758w} {\bibfield  {journal} {\bibinfo  {journal} {Nano Letters}\
  }\textbf {\bibinfo {volume} {12}},\ \bibinfo {pages} {6414} (\bibinfo {year}
  {2012})}\BibitemShut {NoStop}%
\bibitem [{\citenamefont {Wang}\ \emph {et~al.}(2012)\citenamefont {Wang},
  \citenamefont {Liu}, \citenamefont {Xu}, \citenamefont {Yang}, \citenamefont
  {Miao}, \citenamefont {Yao}, \citenamefont {Gao}, \citenamefont {Shen},
  \citenamefont {Ma}, \citenamefont {Chen}, \citenamefont {Xu}, \citenamefont
  {Liu}, \citenamefont {Zhang}, \citenamefont {Qian}, \citenamefont {Jia},\
  and\ \citenamefont {Xue}}]{Wang2012}%
  \BibitemOpen
  \bibfield  {author} {\bibinfo {author} {\bibfnamefont {M.-X.}\ \bibnamefont
  {Wang}}, \bibinfo {author} {\bibfnamefont {C.}~\bibnamefont {Liu}}, \bibinfo
  {author} {\bibfnamefont {J.-P.}\ \bibnamefont {Xu}}, \bibinfo {author}
  {\bibfnamefont {F.}~\bibnamefont {Yang}}, \bibinfo {author} {\bibfnamefont
  {L.}~\bibnamefont {Miao}}, \bibinfo {author} {\bibfnamefont {M.-Y.}\
  \bibnamefont {Yao}}, \bibinfo {author} {\bibfnamefont {C.~L.}\ \bibnamefont
  {Gao}}, \bibinfo {author} {\bibfnamefont {C.}~\bibnamefont {Shen}}, \bibinfo
  {author} {\bibfnamefont {X.}~\bibnamefont {Ma}}, \bibinfo {author}
  {\bibfnamefont {X.}~\bibnamefont {Chen}}, \bibinfo {author} {\bibfnamefont
  {Z.-A.}\ \bibnamefont {Xu}}, \bibinfo {author} {\bibfnamefont
  {Y.}~\bibnamefont {Liu}}, \bibinfo {author} {\bibfnamefont {S.-C.}\
  \bibnamefont {Zhang}}, \bibinfo {author} {\bibfnamefont {D.}~\bibnamefont
  {Qian}}, \bibinfo {author} {\bibfnamefont {J.-F.}\ \bibnamefont {Jia}}, \
  and\ \bibinfo {author} {\bibfnamefont {Q.-K.}\ \bibnamefont {Xue}},\ }\href
  {\doibase 10.1126/science.1216466} {\bibfield  {journal} {\bibinfo  {journal}
  {Science}\ }\textbf {\bibinfo {volume} {336}},\ \bibinfo {pages} {52}
  (\bibinfo {year} {2012})}\BibitemShut {NoStop}%
\bibitem [{\citenamefont {Nadj-Perge}\ \emph {et~al.}(2014)\citenamefont
  {Nadj-Perge}, \citenamefont {Drozdov}, \citenamefont {Li}, \citenamefont
  {Chen}, \citenamefont {Jeon}, \citenamefont {Seo}, \citenamefont {MacDonald},
  \citenamefont {Bernevig},\ and\ \citenamefont {Yazdani}}]{Nadj-Perge2014}%
  \BibitemOpen
  \bibfield  {author} {\bibinfo {author} {\bibfnamefont {S.}~\bibnamefont
  {Nadj-Perge}}, \bibinfo {author} {\bibfnamefont {I.~K.}\ \bibnamefont
  {Drozdov}}, \bibinfo {author} {\bibfnamefont {J.}~\bibnamefont {Li}},
  \bibinfo {author} {\bibfnamefont {H.}~\bibnamefont {Chen}}, \bibinfo {author}
  {\bibfnamefont {S.}~\bibnamefont {Jeon}}, \bibinfo {author} {\bibfnamefont
  {J.}~\bibnamefont {Seo}}, \bibinfo {author} {\bibfnamefont {A.~H.}\
  \bibnamefont {MacDonald}}, \bibinfo {author} {\bibfnamefont {B.~A.}\
  \bibnamefont {Bernevig}}, \ and\ \bibinfo {author} {\bibfnamefont
  {A.}~\bibnamefont {Yazdani}},\ }\href {\doibase 10.1126/science.1259327}
  {\bibfield  {journal} {\bibinfo  {journal} {Science}\ }\textbf {\bibinfo
  {volume} {346}},\ \bibinfo {pages} {602} (\bibinfo {year}
  {2014})}\BibitemShut {NoStop}%
\bibitem [{\citenamefont {Xu}\ \emph {et~al.}(2014)\citenamefont {Xu},
  \citenamefont {Liu}, \citenamefont {Wang}, \citenamefont {Ge}, \citenamefont
  {Liu}, \citenamefont {Yang}, \citenamefont {Chen}, \citenamefont {Liu},
  \citenamefont {Xu}, \citenamefont {Gao}, \citenamefont {Qian}, \citenamefont
  {Zhang},\ and\ \citenamefont {Jia}}]{Xu2014}%
  \BibitemOpen
  \bibfield  {author} {\bibinfo {author} {\bibfnamefont {J.-P.}\ \bibnamefont
  {Xu}}, \bibinfo {author} {\bibfnamefont {C.}~\bibnamefont {Liu}}, \bibinfo
  {author} {\bibfnamefont {M.-X.}\ \bibnamefont {Wang}}, \bibinfo {author}
  {\bibfnamefont {J.}~\bibnamefont {Ge}}, \bibinfo {author} {\bibfnamefont
  {Z.-L.}\ \bibnamefont {Liu}}, \bibinfo {author} {\bibfnamefont
  {X.}~\bibnamefont {Yang}}, \bibinfo {author} {\bibfnamefont {Y.}~\bibnamefont
  {Chen}}, \bibinfo {author} {\bibfnamefont {Y.}~\bibnamefont {Liu}}, \bibinfo
  {author} {\bibfnamefont {Z.-A.}\ \bibnamefont {Xu}}, \bibinfo {author}
  {\bibfnamefont {C.-L.}\ \bibnamefont {Gao}}, \bibinfo {author} {\bibfnamefont
  {D.}~\bibnamefont {Qian}}, \bibinfo {author} {\bibfnamefont {F.-C.}\
  \bibnamefont {Zhang}}, \ and\ \bibinfo {author} {\bibfnamefont {J.-F.}\
  \bibnamefont {Jia}},\ }\href {\doibase 10.1103/PhysRevLett.112.217001}
  {\bibfield  {journal} {\bibinfo  {journal} {Phys. Rev. Lett.}\ }\textbf
  {\bibinfo {volume} {112}},\ \bibinfo {pages} {217001} (\bibinfo {year}
  {2014})}\BibitemShut {NoStop}%
\bibitem [{\citenamefont {Xu}\ \emph {et~al.}(2015)\citenamefont {Xu},
  \citenamefont {Wang}, \citenamefont {Liu}, \citenamefont {Ge}, \citenamefont
  {Yang}, \citenamefont {Liu}, \citenamefont {Xu}, \citenamefont {Guan},
  \citenamefont {Gao}, \citenamefont {Qian}, \citenamefont {Liu}, \citenamefont
  {Wang}, \citenamefont {Zhang}, \citenamefont {Xue},\ and\ \citenamefont
  {Jia}}]{Xu2015}%
  \BibitemOpen
  \bibfield  {author} {\bibinfo {author} {\bibfnamefont {J.-P.}\ \bibnamefont
  {Xu}}, \bibinfo {author} {\bibfnamefont {M.-X.}\ \bibnamefont {Wang}},
  \bibinfo {author} {\bibfnamefont {Z.~L.}\ \bibnamefont {Liu}}, \bibinfo
  {author} {\bibfnamefont {J.-F.}\ \bibnamefont {Ge}}, \bibinfo {author}
  {\bibfnamefont {X.}~\bibnamefont {Yang}}, \bibinfo {author} {\bibfnamefont
  {C.}~\bibnamefont {Liu}}, \bibinfo {author} {\bibfnamefont {Z.~A.}\
  \bibnamefont {Xu}}, \bibinfo {author} {\bibfnamefont {D.}~\bibnamefont
  {Guan}}, \bibinfo {author} {\bibfnamefont {C.~L.}\ \bibnamefont {Gao}},
  \bibinfo {author} {\bibfnamefont {D.}~\bibnamefont {Qian}}, \bibinfo {author}
  {\bibfnamefont {Y.}~\bibnamefont {Liu}}, \bibinfo {author} {\bibfnamefont
  {Q.-H.}\ \bibnamefont {Wang}}, \bibinfo {author} {\bibfnamefont {F.-C.}\
  \bibnamefont {Zhang}}, \bibinfo {author} {\bibfnamefont {Q.-K.}\ \bibnamefont
  {Xue}}, \ and\ \bibinfo {author} {\bibfnamefont {J.-F.}\ \bibnamefont
  {Jia}},\ }\href {\doibase 10.1103/PhysRevLett.114.017001} {\bibfield
  {journal} {\bibinfo  {journal} {Phys. Rev. Lett.}\ }\textbf {\bibinfo
  {volume} {114}},\ \bibinfo {pages} {017001} (\bibinfo {year}
  {2015})}\BibitemShut {NoStop}%
\bibitem [{\citenamefont {Wang}\ \emph
  {et~al.}(2015{\natexlab{a}})\citenamefont {Wang}, \citenamefont {Zhou},
  \citenamefont {Lian},\ and\ \citenamefont {Zhang}}]{Wang2015}%
  \BibitemOpen
  \bibfield  {author} {\bibinfo {author} {\bibfnamefont {J.}~\bibnamefont
  {Wang}}, \bibinfo {author} {\bibfnamefont {Q.}~\bibnamefont {Zhou}}, \bibinfo
  {author} {\bibfnamefont {B.}~\bibnamefont {Lian}}, \ and\ \bibinfo {author}
  {\bibfnamefont {S.-C.}\ \bibnamefont {Zhang}},\ }\href {\doibase
  10.1103/PhysRevB.92.064520} {\bibfield  {journal} {\bibinfo  {journal} {Phys.
  Rev. B}\ }\textbf {\bibinfo {volume} {92}},\ \bibinfo {pages} {064520}
  (\bibinfo {year} {2015}{\natexlab{a}})}\BibitemShut {NoStop}%
\bibitem [{\citenamefont {Sun}\ \emph {et~al.}(2016)\citenamefont {Sun},
  \citenamefont {Zhang}, \citenamefont {Hu}, \citenamefont {Li}, \citenamefont
  {Wang}, \citenamefont {Ma}, \citenamefont {Xu}, \citenamefont {Gao},
  \citenamefont {Guan}, \citenamefont {Li}, \citenamefont {Liu}, \citenamefont
  {Qian}, \citenamefont {Zhou}, \citenamefont {Fu}, \citenamefont {Li},
  \citenamefont {Zhang},\ and\ \citenamefont {Jia}}]{Sun2016}%
  \BibitemOpen
  \bibfield  {author} {\bibinfo {author} {\bibfnamefont {H.-H.}\ \bibnamefont
  {Sun}}, \bibinfo {author} {\bibfnamefont {K.-W.}\ \bibnamefont {Zhang}},
  \bibinfo {author} {\bibfnamefont {L.-H.}\ \bibnamefont {Hu}}, \bibinfo
  {author} {\bibfnamefont {C.}~\bibnamefont {Li}}, \bibinfo {author}
  {\bibfnamefont {G.-Y.}\ \bibnamefont {Wang}}, \bibinfo {author}
  {\bibfnamefont {H.-Y.}\ \bibnamefont {Ma}}, \bibinfo {author} {\bibfnamefont
  {Z.-A.}\ \bibnamefont {Xu}}, \bibinfo {author} {\bibfnamefont {C.-L.}\
  \bibnamefont {Gao}}, \bibinfo {author} {\bibfnamefont {D.-D.}\ \bibnamefont
  {Guan}}, \bibinfo {author} {\bibfnamefont {Y.-Y.}\ \bibnamefont {Li}},
  \bibinfo {author} {\bibfnamefont {C.}~\bibnamefont {Liu}}, \bibinfo {author}
  {\bibfnamefont {D.}~\bibnamefont {Qian}}, \bibinfo {author} {\bibfnamefont
  {Y.}~\bibnamefont {Zhou}}, \bibinfo {author} {\bibfnamefont {L.}~\bibnamefont
  {Fu}}, \bibinfo {author} {\bibfnamefont {S.-C.}\ \bibnamefont {Li}}, \bibinfo
  {author} {\bibfnamefont {F.-C.}\ \bibnamefont {Zhang}}, \ and\ \bibinfo
  {author} {\bibfnamefont {J.-F.}\ \bibnamefont {Jia}},\ }\href {\doibase
  10.1103/PhysRevLett.116.257003} {\bibfield  {journal} {\bibinfo  {journal}
  {Phys. Rev. Lett.}\ }\textbf {\bibinfo {volume} {116}},\ \bibinfo {pages}
  {257003} (\bibinfo {year} {2016})}\BibitemShut {NoStop}%
\bibitem [{\citenamefont {He}\ \emph {et~al.}(2017)\citenamefont {He},
  \citenamefont {Pan}, \citenamefont {Stern}, \citenamefont {Burks},
  \citenamefont {Che}, \citenamefont {Yin}, \citenamefont {Wang}, \citenamefont
  {Lian}, \citenamefont {Zhou}, \citenamefont {Choi}, \citenamefont {Murata},
  \citenamefont {Kou}, \citenamefont {Chen}, \citenamefont {Nie}, \citenamefont
  {Shao}, \citenamefont {Fan}, \citenamefont {Zhang}, \citenamefont {Liu},
  \citenamefont {Xia},\ and\ \citenamefont {Wang}}]{He2017}%
  \BibitemOpen
  \bibfield  {author} {\bibinfo {author} {\bibfnamefont {Q.~L.}\ \bibnamefont
  {He}}, \bibinfo {author} {\bibfnamefont {L.}~\bibnamefont {Pan}}, \bibinfo
  {author} {\bibfnamefont {A.~L.}\ \bibnamefont {Stern}}, \bibinfo {author}
  {\bibfnamefont {E.~C.}\ \bibnamefont {Burks}}, \bibinfo {author}
  {\bibfnamefont {X.}~\bibnamefont {Che}}, \bibinfo {author} {\bibfnamefont
  {G.}~\bibnamefont {Yin}}, \bibinfo {author} {\bibfnamefont {J.}~\bibnamefont
  {Wang}}, \bibinfo {author} {\bibfnamefont {B.}~\bibnamefont {Lian}}, \bibinfo
  {author} {\bibfnamefont {Q.}~\bibnamefont {Zhou}}, \bibinfo {author}
  {\bibfnamefont {E.~S.}\ \bibnamefont {Choi}}, \bibinfo {author}
  {\bibfnamefont {K.}~\bibnamefont {Murata}}, \bibinfo {author} {\bibfnamefont
  {X.}~\bibnamefont {Kou}}, \bibinfo {author} {\bibfnamefont {Z.}~\bibnamefont
  {Chen}}, \bibinfo {author} {\bibfnamefont {T.}~\bibnamefont {Nie}}, \bibinfo
  {author} {\bibfnamefont {Q.}~\bibnamefont {Shao}}, \bibinfo {author}
  {\bibfnamefont {Y.}~\bibnamefont {Fan}}, \bibinfo {author} {\bibfnamefont
  {S.-C.}\ \bibnamefont {Zhang}}, \bibinfo {author} {\bibfnamefont
  {K.}~\bibnamefont {Liu}}, \bibinfo {author} {\bibfnamefont {J.}~\bibnamefont
  {Xia}}, \ and\ \bibinfo {author} {\bibfnamefont {K.~L.}\ \bibnamefont
  {Wang}},\ }\href {\doibase 10.1126/science.aag2792} {\bibfield  {journal}
  {\bibinfo  {journal} {Science}\ }\textbf {\bibinfo {volume} {357}},\ \bibinfo
  {pages} {294} (\bibinfo {year} {2017})}\BibitemShut {NoStop}%
\bibitem [{\citenamefont {M{\'e}nard}\ \emph {et~al.}(2017)\citenamefont
  {M{\'e}nard}, \citenamefont {Guissart}, \citenamefont {Brun}, \citenamefont
  {Leriche}, \citenamefont {Trif}, \citenamefont {Debontridder}, \citenamefont
  {Demaille}, \citenamefont {Roditchev}, \citenamefont {Simon},\ and\
  \citenamefont {Cren}}]{Menard2017}%
  \BibitemOpen
  \bibfield  {author} {\bibinfo {author} {\bibfnamefont {G.~C.}\ \bibnamefont
  {M{\'e}nard}}, \bibinfo {author} {\bibfnamefont {S.}~\bibnamefont
  {Guissart}}, \bibinfo {author} {\bibfnamefont {C.}~\bibnamefont {Brun}},
  \bibinfo {author} {\bibfnamefont {R.~T.}\ \bibnamefont {Leriche}}, \bibinfo
  {author} {\bibfnamefont {M.}~\bibnamefont {Trif}}, \bibinfo {author}
  {\bibfnamefont {F.}~\bibnamefont {Debontridder}}, \bibinfo {author}
  {\bibfnamefont {D.}~\bibnamefont {Demaille}}, \bibinfo {author}
  {\bibfnamefont {D.}~\bibnamefont {Roditchev}}, \bibinfo {author}
  {\bibfnamefont {P.}~\bibnamefont {Simon}}, \ and\ \bibinfo {author}
  {\bibfnamefont {T.}~\bibnamefont {Cren}},\ }\href {\doibase
  10.1038/s41467-017-02192-x} {\bibfield  {journal} {\bibinfo  {journal}
  {Nature Communications}\ }\textbf {\bibinfo {volume} {8}},\ \bibinfo {pages}
  {2040} (\bibinfo {year} {2017})}\BibitemShut {NoStop}%
\bibitem [{\citenamefont {Zhang}\ \emph
  {et~al.}(2018{\natexlab{a}})\citenamefont {Zhang}, \citenamefont {Liu},
  \citenamefont {Gazibegovic}, \citenamefont {Xu}, \citenamefont {Logan},
  \citenamefont {Wang}, \citenamefont {van Loo}, \citenamefont {Bommer},
  \citenamefont {de~Moor}, \citenamefont {Car}, \citenamefont {Op~het Veld},
  \citenamefont {van Veldhoven}, \citenamefont {Koelling}, \citenamefont
  {Verheijen}, \citenamefont {Pendharkar}, \citenamefont {Pennachio},
  \citenamefont {Shojaei}, \citenamefont {Lee}, \citenamefont {Palmstr{\o}m},
  \citenamefont {Bakkers}, \citenamefont {Sarma},\ and\ \citenamefont
  {Kouwenhoven}}]{Zhang2018}%
  \BibitemOpen
  \bibfield  {author} {\bibinfo {author} {\bibfnamefont {H.}~\bibnamefont
  {Zhang}}, \bibinfo {author} {\bibfnamefont {C.-X.}\ \bibnamefont {Liu}},
  \bibinfo {author} {\bibfnamefont {S.}~\bibnamefont {Gazibegovic}}, \bibinfo
  {author} {\bibfnamefont {D.}~\bibnamefont {Xu}}, \bibinfo {author}
  {\bibfnamefont {J.~A.}\ \bibnamefont {Logan}}, \bibinfo {author}
  {\bibfnamefont {G.}~\bibnamefont {Wang}}, \bibinfo {author} {\bibfnamefont
  {N.}~\bibnamefont {van Loo}}, \bibinfo {author} {\bibfnamefont {J.~D.~S.}\
  \bibnamefont {Bommer}}, \bibinfo {author} {\bibfnamefont {M.~W.~A.}\
  \bibnamefont {de~Moor}}, \bibinfo {author} {\bibfnamefont {D.}~\bibnamefont
  {Car}}, \bibinfo {author} {\bibfnamefont {R.~L.~M.}\ \bibnamefont {Op~het
  Veld}}, \bibinfo {author} {\bibfnamefont {P.~J.}\ \bibnamefont {van
  Veldhoven}}, \bibinfo {author} {\bibfnamefont {S.}~\bibnamefont {Koelling}},
  \bibinfo {author} {\bibfnamefont {M.~A.}\ \bibnamefont {Verheijen}}, \bibinfo
  {author} {\bibfnamefont {M.}~\bibnamefont {Pendharkar}}, \bibinfo {author}
  {\bibfnamefont {D.~J.}\ \bibnamefont {Pennachio}}, \bibinfo {author}
  {\bibfnamefont {B.}~\bibnamefont {Shojaei}}, \bibinfo {author} {\bibfnamefont
  {J.~S.}\ \bibnamefont {Lee}}, \bibinfo {author} {\bibfnamefont {C.~J.}\
  \bibnamefont {Palmstr{\o}m}}, \bibinfo {author} {\bibfnamefont {E.~P. A.~M.}\
  \bibnamefont {Bakkers}}, \bibinfo {author} {\bibfnamefont {S.~D.}\
  \bibnamefont {Sarma}}, \ and\ \bibinfo {author} {\bibfnamefont {L.~P.}\
  \bibnamefont {Kouwenhoven}},\ }\href {\doibase 10.1038/nature26142}
  {\bibfield  {journal} {\bibinfo  {journal} {Nature}\ }\textbf {\bibinfo
  {volume} {556}},\ \bibinfo {pages} {74} (\bibinfo {year}
  {2018}{\natexlab{a}})}\BibitemShut {NoStop}%
\bibitem [{\citenamefont {Wang}\ \emph {et~al.}(2018)\citenamefont {Wang},
  \citenamefont {Kong}, \citenamefont {Fan}, \citenamefont {Chen},
  \citenamefont {Zhu}, \citenamefont {Liu}, \citenamefont {Cao}, \citenamefont
  {Sun}, \citenamefont {Du}, \citenamefont {Schneeloch} \emph
  {et~al.}}]{Wang2018}%
  \BibitemOpen
  \bibfield  {author} {\bibinfo {author} {\bibfnamefont {D.}~\bibnamefont
  {Wang}}, \bibinfo {author} {\bibfnamefont {L.}~\bibnamefont {Kong}}, \bibinfo
  {author} {\bibfnamefont {P.}~\bibnamefont {Fan}}, \bibinfo {author}
  {\bibfnamefont {H.}~\bibnamefont {Chen}}, \bibinfo {author} {\bibfnamefont
  {S.}~\bibnamefont {Zhu}}, \bibinfo {author} {\bibfnamefont {W.}~\bibnamefont
  {Liu}}, \bibinfo {author} {\bibfnamefont {L.}~\bibnamefont {Cao}}, \bibinfo
  {author} {\bibfnamefont {Y.}~\bibnamefont {Sun}}, \bibinfo {author}
  {\bibfnamefont {S.}~\bibnamefont {Du}}, \bibinfo {author} {\bibfnamefont
  {J.}~\bibnamefont {Schneeloch}},  \emph {et~al.},\ }\href@noop {} {\bibfield
  {journal} {\bibinfo  {journal} {Science}\ }\textbf {\bibinfo {volume}
  {362}},\ \bibinfo {pages} {333} (\bibinfo {year} {2018})}\BibitemShut
  {NoStop}%
\bibitem [{\citenamefont {Machida}\ \emph {et~al.}(2019)\citenamefont
  {Machida}, \citenamefont {Sun}, \citenamefont {Pyon}, \citenamefont {Takeda},
  \citenamefont {Kohsaka}, \citenamefont {Hanaguri}, \citenamefont {Sasagawa},\
  and\ \citenamefont {Tamegai}}]{Machida2019}%
  \BibitemOpen
  \bibfield  {author} {\bibinfo {author} {\bibfnamefont {T.}~\bibnamefont
  {Machida}}, \bibinfo {author} {\bibfnamefont {Y.}~\bibnamefont {Sun}},
  \bibinfo {author} {\bibfnamefont {S.}~\bibnamefont {Pyon}}, \bibinfo {author}
  {\bibfnamefont {S.}~\bibnamefont {Takeda}}, \bibinfo {author} {\bibfnamefont
  {Y.}~\bibnamefont {Kohsaka}}, \bibinfo {author} {\bibfnamefont
  {T.}~\bibnamefont {Hanaguri}}, \bibinfo {author} {\bibfnamefont
  {T.}~\bibnamefont {Sasagawa}}, \ and\ \bibinfo {author} {\bibfnamefont
  {T.}~\bibnamefont {Tamegai}},\ }\href@noop {} {\bibfield  {journal} {\bibinfo
   {journal} {Nature materials}\ }\textbf {\bibinfo {volume} {18}},\ \bibinfo
  {pages} {811} (\bibinfo {year} {2019})}\BibitemShut {NoStop}%
\bibitem [{\citenamefont {Wang}\ \emph {et~al.}(2020)\citenamefont {Wang},
  \citenamefont {Rodriguez}, \citenamefont {Jiao}, \citenamefont {Howard},
  \citenamefont {Graham}, \citenamefont {Gu}, \citenamefont {Hughes},
  \citenamefont {Morr},\ and\ \citenamefont {Madhavan}}]{Wang2020}%
  \BibitemOpen
  \bibfield  {author} {\bibinfo {author} {\bibfnamefont {Z.}~\bibnamefont
  {Wang}}, \bibinfo {author} {\bibfnamefont {J.~O.}\ \bibnamefont {Rodriguez}},
  \bibinfo {author} {\bibfnamefont {L.}~\bibnamefont {Jiao}}, \bibinfo {author}
  {\bibfnamefont {S.}~\bibnamefont {Howard}}, \bibinfo {author} {\bibfnamefont
  {M.}~\bibnamefont {Graham}}, \bibinfo {author} {\bibfnamefont {G.~D.}\
  \bibnamefont {Gu}}, \bibinfo {author} {\bibfnamefont {T.~L.}\ \bibnamefont
  {Hughes}}, \bibinfo {author} {\bibfnamefont {D.~K.}\ \bibnamefont {Morr}}, \
  and\ \bibinfo {author} {\bibfnamefont {V.}~\bibnamefont {Madhavan}},\ }\href
  {\doibase 10.1126/science.aaw8419} {\bibfield  {journal} {\bibinfo  {journal}
  {Science}\ }\textbf {\bibinfo {volume} {367}},\ \bibinfo {pages} {104}
  (\bibinfo {year} {2020})},\ \Eprint
  {http://arxiv.org/abs/https://science.sciencemag.org/content/367/6473/104.full.pdf}
  {https://science.sciencemag.org/content/367/6473/104.full.pdf} \BibitemShut
  {NoStop}%
\bibitem [{\citenamefont {Fu}\ and\ \citenamefont {Berg}(2010)}]{Fu2010}%
  \BibitemOpen
  \bibfield  {author} {\bibinfo {author} {\bibfnamefont {L.}~\bibnamefont
  {Fu}}\ and\ \bibinfo {author} {\bibfnamefont {E.}~\bibnamefont {Berg}},\
  }\href {\doibase 10.1103/PhysRevLett.105.097001} {\bibfield  {journal}
  {\bibinfo  {journal} {Phys. Rev. Lett.}\ }\textbf {\bibinfo {volume} {105}},\
  \bibinfo {pages} {097001} (\bibinfo {year} {2010})}\BibitemShut {NoStop}%
\bibitem [{\citenamefont {Sato}(2010)}]{Sato2010}%
  \BibitemOpen
  \bibfield  {author} {\bibinfo {author} {\bibfnamefont {M.}~\bibnamefont
  {Sato}},\ }\href {\doibase 10.1103/PhysRevB.81.220504} {\bibfield  {journal}
  {\bibinfo  {journal} {Phys. Rev. B}\ }\textbf {\bibinfo {volume} {81}},\
  \bibinfo {pages} {220504(R)} (\bibinfo {year} {2010})}\BibitemShut {NoStop}%
\bibitem [{\citenamefont {Hosur}\ \emph {et~al.}(2011)\citenamefont {Hosur},
  \citenamefont {Ghaemi}, \citenamefont {Mong},\ and\ \citenamefont
  {Vishwanath}}]{Hosuer2011}%
  \BibitemOpen
  \bibfield  {author} {\bibinfo {author} {\bibfnamefont {P.}~\bibnamefont
  {Hosur}}, \bibinfo {author} {\bibfnamefont {P.}~\bibnamefont {Ghaemi}},
  \bibinfo {author} {\bibfnamefont {R.~S.~K.}\ \bibnamefont {Mong}}, \ and\
  \bibinfo {author} {\bibfnamefont {A.}~\bibnamefont {Vishwanath}},\ }\href
  {\doibase 10.1103/PhysRevLett.107.097001} {\bibfield  {journal} {\bibinfo
  {journal} {Phys. Rev. Lett.}\ }\textbf {\bibinfo {volume} {107}},\ \bibinfo
  {pages} {097001} (\bibinfo {year} {2011})}\BibitemShut {NoStop}%
\bibitem [{\citenamefont {Fu}(2014)}]{Fu2014}%
  \BibitemOpen
  \bibfield  {author} {\bibinfo {author} {\bibfnamefont {L.}~\bibnamefont
  {Fu}},\ }\href {\doibase 10.1103/PhysRevB.90.100509} {\bibfield  {journal}
  {\bibinfo  {journal} {Phys. Rev. B}\ }\textbf {\bibinfo {volume} {90}},\
  \bibinfo {pages} {100509(R)} (\bibinfo {year} {2014})}\BibitemShut {NoStop}%
\bibitem [{\citenamefont {Hosur}\ \emph {et~al.}(2014)\citenamefont {Hosur},
  \citenamefont {Dai}, \citenamefont {Fang},\ and\ \citenamefont
  {Qi}}]{Hosuer2014}%
  \BibitemOpen
  \bibfield  {author} {\bibinfo {author} {\bibfnamefont {P.}~\bibnamefont
  {Hosur}}, \bibinfo {author} {\bibfnamefont {X.}~\bibnamefont {Dai}}, \bibinfo
  {author} {\bibfnamefont {Z.}~\bibnamefont {Fang}}, \ and\ \bibinfo {author}
  {\bibfnamefont {X.-L.}\ \bibnamefont {Qi}},\ }\href {\doibase
  10.1103/PhysRevB.90.045130} {\bibfield  {journal} {\bibinfo  {journal} {Phys.
  Rev. B}\ }\textbf {\bibinfo {volume} {90}},\ \bibinfo {pages} {045130}
  (\bibinfo {year} {2014})}\BibitemShut {NoStop}%
\bibitem [{\citenamefont {Kobayashi}\ and\ \citenamefont
  {Sato}(2015)}]{Kobayashi2015}%
  \BibitemOpen
  \bibfield  {author} {\bibinfo {author} {\bibfnamefont {S.}~\bibnamefont
  {Kobayashi}}\ and\ \bibinfo {author} {\bibfnamefont {M.}~\bibnamefont
  {Sato}},\ }\href {\doibase 10.1103/PhysRevLett.115.187001} {\bibfield
  {journal} {\bibinfo  {journal} {Phys. Rev. Lett.}\ }\textbf {\bibinfo
  {volume} {115}},\ \bibinfo {pages} {187001} (\bibinfo {year}
  {2015})}\BibitemShut {NoStop}%
\bibitem [{\citenamefont {Wang}\ \emph
  {et~al.}(2015{\natexlab{b}})\citenamefont {Wang}, \citenamefont {Zhang},
  \citenamefont {Xu}, \citenamefont {Zeng}, \citenamefont {Miao}, \citenamefont
  {Xu}, \citenamefont {Qian}, \citenamefont {Weng}, \citenamefont {Richard},
  \citenamefont {Fedorov}, \citenamefont {Ding}, \citenamefont {Dai},\ and\
  \citenamefont {Fang}}]{Wang2015_2}%
  \BibitemOpen
  \bibfield  {author} {\bibinfo {author} {\bibfnamefont {Z.}~\bibnamefont
  {Wang}}, \bibinfo {author} {\bibfnamefont {P.}~\bibnamefont {Zhang}},
  \bibinfo {author} {\bibfnamefont {G.}~\bibnamefont {Xu}}, \bibinfo {author}
  {\bibfnamefont {L.~K.}\ \bibnamefont {Zeng}}, \bibinfo {author}
  {\bibfnamefont {H.}~\bibnamefont {Miao}}, \bibinfo {author} {\bibfnamefont
  {X.}~\bibnamefont {Xu}}, \bibinfo {author} {\bibfnamefont {T.}~\bibnamefont
  {Qian}}, \bibinfo {author} {\bibfnamefont {H.}~\bibnamefont {Weng}}, \bibinfo
  {author} {\bibfnamefont {P.}~\bibnamefont {Richard}}, \bibinfo {author}
  {\bibfnamefont {A.~V.}\ \bibnamefont {Fedorov}}, \bibinfo {author}
  {\bibfnamefont {H.}~\bibnamefont {Ding}}, \bibinfo {author} {\bibfnamefont
  {X.}~\bibnamefont {Dai}}, \ and\ \bibinfo {author} {\bibfnamefont
  {Z.}~\bibnamefont {Fang}},\ }\href {\doibase 10.1103/PhysRevB.92.115119}
  {\bibfield  {journal} {\bibinfo  {journal} {Phys. Rev. B}\ }\textbf {\bibinfo
  {volume} {92}},\ \bibinfo {pages} {115119} (\bibinfo {year}
  {2015}{\natexlab{b}})}\BibitemShut {NoStop}%
\bibitem [{\citenamefont {Xu}\ \emph {et~al.}(2016)\citenamefont {Xu},
  \citenamefont {Lian}, \citenamefont {Tang}, \citenamefont {Qi},\ and\
  \citenamefont {Zhang}}]{Xu2016}%
  \BibitemOpen
  \bibfield  {author} {\bibinfo {author} {\bibfnamefont {G.}~\bibnamefont
  {Xu}}, \bibinfo {author} {\bibfnamefont {B.}~\bibnamefont {Lian}}, \bibinfo
  {author} {\bibfnamefont {P.}~\bibnamefont {Tang}}, \bibinfo {author}
  {\bibfnamefont {X.-L.}\ \bibnamefont {Qi}}, \ and\ \bibinfo {author}
  {\bibfnamefont {S.-C.}\ \bibnamefont {Zhang}},\ }\href {\doibase
  10.1103/PhysRevLett.117.047001} {\bibfield  {journal} {\bibinfo  {journal}
  {Phys. Rev. Lett.}\ }\textbf {\bibinfo {volume} {117}},\ \bibinfo {pages}
  {047001} (\bibinfo {year} {2016})}\BibitemShut {NoStop}%
\bibitem [{\citenamefont {Pan}\ \emph {et~al.}(2016)\citenamefont {Pan},
  \citenamefont {Nikitin}, \citenamefont {Araizi}, \citenamefont {Huang},
  \citenamefont {Matsushita}, \citenamefont {Naka},\ and\ \citenamefont
  {de~Visser}}]{Pan2016}%
  \BibitemOpen
  \bibfield  {author} {\bibinfo {author} {\bibfnamefont {Y.}~\bibnamefont
  {Pan}}, \bibinfo {author} {\bibfnamefont {A.~M.}\ \bibnamefont {Nikitin}},
  \bibinfo {author} {\bibfnamefont {G.~K.}\ \bibnamefont {Araizi}}, \bibinfo
  {author} {\bibfnamefont {Y.~K.}\ \bibnamefont {Huang}}, \bibinfo {author}
  {\bibfnamefont {Y.}~\bibnamefont {Matsushita}}, \bibinfo {author}
  {\bibfnamefont {T.}~\bibnamefont {Naka}}, \ and\ \bibinfo {author}
  {\bibfnamefont {A.}~\bibnamefont {de~Visser}},\ }\href {\doibase
  10.1038/srep28632} {\bibfield  {journal} {\bibinfo  {journal} {Scientific
  Reports}\ }\textbf {\bibinfo {volume} {6}},\ \bibinfo {pages} {28632}
  (\bibinfo {year} {2016})}\BibitemShut {NoStop}%
\bibitem [{\citenamefont {Matano}\ \emph {et~al.}(2016)\citenamefont {Matano},
  \citenamefont {Kriener}, \citenamefont {Segawa}, \citenamefont {Ando},\ and\
  \citenamefont {Zheng}}]{Matano2016}%
  \BibitemOpen
  \bibfield  {author} {\bibinfo {author} {\bibfnamefont {K.}~\bibnamefont
  {Matano}}, \bibinfo {author} {\bibfnamefont {M.}~\bibnamefont {Kriener}},
  \bibinfo {author} {\bibfnamefont {K.}~\bibnamefont {Segawa}}, \bibinfo
  {author} {\bibfnamefont {Y.}~\bibnamefont {Ando}}, \ and\ \bibinfo {author}
  {\bibfnamefont {G.-q.}\ \bibnamefont {Zheng}},\ }\href {\doibase
  10.1038/nphys3781} {\bibfield  {journal} {\bibinfo  {journal} {Nature
  Physics}\ }\textbf {\bibinfo {volume} {12}},\ \bibinfo {pages} {852}
  (\bibinfo {year} {2016})}\BibitemShut {NoStop}%
\bibitem [{\citenamefont {Yonezawa}\ \emph {et~al.}(2017)\citenamefont
  {Yonezawa}, \citenamefont {Tajiri}, \citenamefont {Nakata}, \citenamefont
  {Nagai}, \citenamefont {Wang}, \citenamefont {Segawa}, \citenamefont {Ando},\
  and\ \citenamefont {Maeno}}]{Yonezawa2017}%
  \BibitemOpen
  \bibfield  {author} {\bibinfo {author} {\bibfnamefont {S.}~\bibnamefont
  {Yonezawa}}, \bibinfo {author} {\bibfnamefont {K.}~\bibnamefont {Tajiri}},
  \bibinfo {author} {\bibfnamefont {S.}~\bibnamefont {Nakata}}, \bibinfo
  {author} {\bibfnamefont {Y.}~\bibnamefont {Nagai}}, \bibinfo {author}
  {\bibfnamefont {Z.}~\bibnamefont {Wang}}, \bibinfo {author} {\bibfnamefont
  {K.}~\bibnamefont {Segawa}}, \bibinfo {author} {\bibfnamefont
  {Y.}~\bibnamefont {Ando}}, \ and\ \bibinfo {author} {\bibfnamefont
  {Y.}~\bibnamefont {Maeno}},\ }\href {\doibase 10.1038/nphys3907} {\bibfield
  {journal} {\bibinfo  {journal} {Nature Physics}\ }\textbf {\bibinfo {volume}
  {13}},\ \bibinfo {pages} {123} (\bibinfo {year} {2017})}\BibitemShut
  {NoStop}%
\bibitem [{\citenamefont {Zhang}\ \emph
  {et~al.}(2018{\natexlab{b}})\citenamefont {Zhang}, \citenamefont {Yaji},
  \citenamefont {Hashimoto}, \citenamefont {Ota}, \citenamefont {Kondo},
  \citenamefont {Okazaki}, \citenamefont {Wang}, \citenamefont {Wen},
  \citenamefont {Gu}, \citenamefont {Ding},\ and\ \citenamefont
  {Shin}}]{Zhang2018_2}%
  \BibitemOpen
  \bibfield  {author} {\bibinfo {author} {\bibfnamefont {P.}~\bibnamefont
  {Zhang}}, \bibinfo {author} {\bibfnamefont {K.}~\bibnamefont {Yaji}},
  \bibinfo {author} {\bibfnamefont {T.}~\bibnamefont {Hashimoto}}, \bibinfo
  {author} {\bibfnamefont {Y.}~\bibnamefont {Ota}}, \bibinfo {author}
  {\bibfnamefont {T.}~\bibnamefont {Kondo}}, \bibinfo {author} {\bibfnamefont
  {K.}~\bibnamefont {Okazaki}}, \bibinfo {author} {\bibfnamefont
  {Z.}~\bibnamefont {Wang}}, \bibinfo {author} {\bibfnamefont {J.}~\bibnamefont
  {Wen}}, \bibinfo {author} {\bibfnamefont {G.~D.}\ \bibnamefont {Gu}},
  \bibinfo {author} {\bibfnamefont {H.}~\bibnamefont {Ding}}, \ and\ \bibinfo
  {author} {\bibfnamefont {S.}~\bibnamefont {Shin}},\ }\href {\doibase
  10.1126/science.aan4596} {\bibfield  {journal} {\bibinfo  {journal}
  {Science}\ }\textbf {\bibinfo {volume} {360}},\ \bibinfo {pages} {182}
  (\bibinfo {year} {2018}{\natexlab{b}})}\BibitemShut {NoStop}%
\bibitem [{\citenamefont {Yoshida}\ and\ \citenamefont
  {Yanase}(2016)}]{Yoshida2016}%
  \BibitemOpen
  \bibfield  {author} {\bibinfo {author} {\bibfnamefont {T.}~\bibnamefont
  {Yoshida}}\ and\ \bibinfo {author} {\bibfnamefont {Y.}~\bibnamefont
  {Yanase}},\ }\href {\doibase 10.1103/PhysRevB.93.054504} {\bibfield
  {journal} {\bibinfo  {journal} {Phys. Rev. B}\ }\textbf {\bibinfo {volume}
  {93}},\ \bibinfo {pages} {054504} (\bibinfo {year} {2016})}\BibitemShut
  {NoStop}%
\bibitem [{\citenamefont {Daido}\ and\ \citenamefont
  {Yanase}(2016)}]{Daido2016}%
  \BibitemOpen
  \bibfield  {author} {\bibinfo {author} {\bibfnamefont {A.}~\bibnamefont
  {Daido}}\ and\ \bibinfo {author} {\bibfnamefont {Y.}~\bibnamefont {Yanase}},\
  }\href {\doibase 10.1103/PhysRevB.94.054519} {\bibfield  {journal} {\bibinfo
  {journal} {Phys. Rev. B}\ }\textbf {\bibinfo {volume} {94}},\ \bibinfo
  {pages} {054519} (\bibinfo {year} {2016})}\BibitemShut {NoStop}%
\bibitem [{\citenamefont {Can}\ \emph {et~al.}(2021)\citenamefont {Can},
  \citenamefont {Tummuru}, \citenamefont {Day}, \citenamefont {Elfimov},
  \citenamefont {Damascelli},\ and\ \citenamefont {Franz}}]{Can2021}%
  \BibitemOpen
  \bibfield  {author} {\bibinfo {author} {\bibfnamefont {O.}~\bibnamefont
  {Can}}, \bibinfo {author} {\bibfnamefont {T.}~\bibnamefont {Tummuru}},
  \bibinfo {author} {\bibfnamefont {R.~P.}\ \bibnamefont {Day}}, \bibinfo
  {author} {\bibfnamefont {I.}~\bibnamefont {Elfimov}}, \bibinfo {author}
  {\bibfnamefont {A.}~\bibnamefont {Damascelli}}, \ and\ \bibinfo {author}
  {\bibfnamefont {M.}~\bibnamefont {Franz}},\ }\href {\doibase
  10.1038/s41567-020-01142-7} {\bibfield  {journal} {\bibinfo  {journal}
  {Nature Physics}\ } (\bibinfo {year} {2021}),\
  10.1038/s41567-020-01142-7}\BibitemShut {NoStop}%
\bibitem [{\citenamefont {Schnyder}\ \emph {et~al.}(2008)\citenamefont
  {Schnyder}, \citenamefont {Ryu}, \citenamefont {Furusaki},\ and\
  \citenamefont {Ludwig}}]{Schnyder2008}%
  \BibitemOpen
  \bibfield  {author} {\bibinfo {author} {\bibfnamefont {A.~P.}\ \bibnamefont
  {Schnyder}}, \bibinfo {author} {\bibfnamefont {S.}~\bibnamefont {Ryu}},
  \bibinfo {author} {\bibfnamefont {A.}~\bibnamefont {Furusaki}}, \ and\
  \bibinfo {author} {\bibfnamefont {A.~W.~W.}\ \bibnamefont {Ludwig}},\ }\href
  {\doibase 10.1103/PhysRevB.78.195125} {\bibfield  {journal} {\bibinfo
  {journal} {Phys. Rev. B}\ }\textbf {\bibinfo {volume} {78}},\ \bibinfo
  {pages} {195125} (\bibinfo {year} {2008})}\BibitemShut {NoStop}%
\bibitem [{\citenamefont {Kitaev}(2009)}]{Kitaev2009}%
  \BibitemOpen
  \bibfield  {author} {\bibinfo {author} {\bibfnamefont {A.}~\bibnamefont
  {Kitaev}},\ }\href {\doibase 10.1063/1.3149495} {\bibfield  {journal}
  {\bibinfo  {journal} {AIP Conf. Proc.}\ }\textbf {\bibinfo {volume} {1134}},\
  \bibinfo {pages} {22} (\bibinfo {year} {2009})}\BibitemShut {NoStop}%
\bibitem [{\citenamefont {Ryu}\ \emph {et~al.}(2010)\citenamefont {Ryu},
  \citenamefont {Schnyder}, \citenamefont {Furusaki},\ and\ \citenamefont
  {Ludwig}}]{Ryu2010}%
  \BibitemOpen
  \bibfield  {author} {\bibinfo {author} {\bibfnamefont {S.}~\bibnamefont
  {Ryu}}, \bibinfo {author} {\bibfnamefont {A.~P.}\ \bibnamefont {Schnyder}},
  \bibinfo {author} {\bibfnamefont {A.}~\bibnamefont {Furusaki}}, \ and\
  \bibinfo {author} {\bibfnamefont {A.~W.~W.}\ \bibnamefont {Ludwig}},\ }\href
  {http://stacks.iop.org/1367-2630/12/i=6/a=065010} {\bibfield  {journal}
  {\bibinfo  {journal} {New J. Phys.}\ }\textbf {\bibinfo {volume} {12}},\
  \bibinfo {pages} {065010} (\bibinfo {year} {2010})}\BibitemShut {NoStop}%
\bibitem [{\citenamefont {Fu}(2011)}]{Fu2011}%
  \BibitemOpen
  \bibfield  {author} {\bibinfo {author} {\bibfnamefont {L.}~\bibnamefont
  {Fu}},\ }\href {\doibase 10.1103/PhysRevLett.106.106802} {\bibfield
  {journal} {\bibinfo  {journal} {Phys. Rev. Lett.}\ }\textbf {\bibinfo
  {volume} {106}},\ \bibinfo {pages} {106802} (\bibinfo {year}
  {2011})}\BibitemShut {NoStop}%
\bibitem [{\citenamefont {Zhang}\ \emph {et~al.}(2013)\citenamefont {Zhang},
  \citenamefont {Kane},\ and\ \citenamefont {Mele}}]{Zhang2013}%
  \BibitemOpen
  \bibfield  {author} {\bibinfo {author} {\bibfnamefont {F.}~\bibnamefont
  {Zhang}}, \bibinfo {author} {\bibfnamefont {C.~L.}\ \bibnamefont {Kane}}, \
  and\ \bibinfo {author} {\bibfnamefont {E.~J.}\ \bibnamefont {Mele}},\ }\href
  {\doibase 10.1103/PhysRevLett.111.056403} {\bibfield  {journal} {\bibinfo
  {journal} {Phys. Rev. Lett.}\ }\textbf {\bibinfo {volume} {111}},\ \bibinfo
  {pages} {056403} (\bibinfo {year} {2013})}\BibitemShut {NoStop}%
\bibitem [{\citenamefont {Chiu}\ \emph {et~al.}(2013)\citenamefont {Chiu},
  \citenamefont {Yao},\ and\ \citenamefont {Ryu}}]{Chiu2013}%
  \BibitemOpen
  \bibfield  {author} {\bibinfo {author} {\bibfnamefont {C.-K.}\ \bibnamefont
  {Chiu}}, \bibinfo {author} {\bibfnamefont {H.}~\bibnamefont {Yao}}, \ and\
  \bibinfo {author} {\bibfnamefont {S.}~\bibnamefont {Ryu}},\ }\href {\doibase
  10.1103/PhysRevB.88.075142} {\bibfield  {journal} {\bibinfo  {journal} {Phys.
  Rev. B}\ }\textbf {\bibinfo {volume} {88}},\ \bibinfo {pages} {075142}
  (\bibinfo {year} {2013})}\BibitemShut {NoStop}%
\bibitem [{\citenamefont {Morimoto}\ and\ \citenamefont
  {Furusaki}(2013)}]{Morimoto2013}%
  \BibitemOpen
  \bibfield  {author} {\bibinfo {author} {\bibfnamefont {T.}~\bibnamefont
  {Morimoto}}\ and\ \bibinfo {author} {\bibfnamefont {A.}~\bibnamefont
  {Furusaki}},\ }\href {\doibase 10.1103/PhysRevB.88.125129} {\bibfield
  {journal} {\bibinfo  {journal} {Phys. Rev. B}\ }\textbf {\bibinfo {volume}
  {88}},\ \bibinfo {pages} {125129} (\bibinfo {year} {2013})}\BibitemShut
  {NoStop}%
\bibitem [{\citenamefont {Shiozaki}\ and\ \citenamefont
  {Sato}(2014)}]{Shiozaki2014}%
  \BibitemOpen
  \bibfield  {author} {\bibinfo {author} {\bibfnamefont {K.}~\bibnamefont
  {Shiozaki}}\ and\ \bibinfo {author} {\bibfnamefont {M.}~\bibnamefont
  {Sato}},\ }\href {\doibase 10.1103/PhysRevB.90.165114} {\bibfield  {journal}
  {\bibinfo  {journal} {Phys. Rev. B}\ }\textbf {\bibinfo {volume} {90}},\
  \bibinfo {pages} {165114} (\bibinfo {year} {2014})}\BibitemShut {NoStop}%
\bibitem [{\citenamefont {Chiu}\ and\ \citenamefont
  {Schnyder}(2014)}]{Chiu2014}%
  \BibitemOpen
  \bibfield  {author} {\bibinfo {author} {\bibfnamefont {C.-K.}\ \bibnamefont
  {Chiu}}\ and\ \bibinfo {author} {\bibfnamefont {A.~P.}\ \bibnamefont
  {Schnyder}},\ }\href {\doibase 10.1103/PhysRevB.90.205136} {\bibfield
  {journal} {\bibinfo  {journal} {Phys. Rev. B}\ }\textbf {\bibinfo {volume}
  {90}},\ \bibinfo {pages} {205136} (\bibinfo {year} {2014})}\BibitemShut
  {NoStop}%
\bibitem [{\citenamefont {Ueno}\ \emph {et~al.}(2013)\citenamefont {Ueno},
  \citenamefont {Yamakage}, \citenamefont {Tanaka},\ and\ \citenamefont
  {Sato}}]{Ueno2013}%
  \BibitemOpen
  \bibfield  {author} {\bibinfo {author} {\bibfnamefont {Y.}~\bibnamefont
  {Ueno}}, \bibinfo {author} {\bibfnamefont {A.}~\bibnamefont {Yamakage}},
  \bibinfo {author} {\bibfnamefont {Y.}~\bibnamefont {Tanaka}}, \ and\ \bibinfo
  {author} {\bibfnamefont {M.}~\bibnamefont {Sato}},\ }\href {\doibase
  10.1103/PhysRevLett.111.087002} {\bibfield  {journal} {\bibinfo  {journal}
  {Phys. Rev. Lett.}\ }\textbf {\bibinfo {volume} {111}},\ \bibinfo {pages}
  {087002} (\bibinfo {year} {2013})}\BibitemShut {NoStop}%
\bibitem [{\citenamefont {Tsutsumi}\ \emph {et~al.}(2013)\citenamefont
  {Tsutsumi}, \citenamefont {Ishikawa}, \citenamefont {Kawakami}, \citenamefont
  {Mizushima}, \citenamefont {Sato}, \citenamefont {Ichioka},\ and\
  \citenamefont {Machida}}]{Tsutsumi2013}%
  \BibitemOpen
  \bibfield  {author} {\bibinfo {author} {\bibfnamefont {Y.}~\bibnamefont
  {Tsutsumi}}, \bibinfo {author} {\bibfnamefont {M.}~\bibnamefont {Ishikawa}},
  \bibinfo {author} {\bibfnamefont {T.}~\bibnamefont {Kawakami}}, \bibinfo
  {author} {\bibfnamefont {T.}~\bibnamefont {Mizushima}}, \bibinfo {author}
  {\bibfnamefont {M.}~\bibnamefont {Sato}}, \bibinfo {author} {\bibfnamefont
  {M.}~\bibnamefont {Ichioka}}, \ and\ \bibinfo {author} {\bibfnamefont
  {K.}~\bibnamefont {Machida}},\ }\href {\doibase 10.7566/JPSJ.82.113707}
  {\bibfield  {journal} {\bibinfo  {journal} {Journal of the Physical Society
  of Japan}\ }\textbf {\bibinfo {volume} {82}},\ \bibinfo {pages} {113707}
  (\bibinfo {year} {2013})}\BibitemShut {NoStop}%
\bibitem [{\citenamefont {Yoshida}\ \emph {et~al.}(2015)\citenamefont
  {Yoshida}, \citenamefont {Sigrist},\ and\ \citenamefont
  {Yanase}}]{Yoshida2015}%
  \BibitemOpen
  \bibfield  {author} {\bibinfo {author} {\bibfnamefont {T.}~\bibnamefont
  {Yoshida}}, \bibinfo {author} {\bibfnamefont {M.}~\bibnamefont {Sigrist}}, \
  and\ \bibinfo {author} {\bibfnamefont {Y.}~\bibnamefont {Yanase}},\ }\href
  {\doibase 10.1103/PhysRevLett.115.027001} {\bibfield  {journal} {\bibinfo
  {journal} {Phys. Rev. Lett.}\ }\textbf {\bibinfo {volume} {115}},\ \bibinfo
  {pages} {027001} (\bibinfo {year} {2015})}\BibitemShut {NoStop}%
\bibitem [{\citenamefont {Fang}\ and\ \citenamefont {Fu}(2015)}]{Fang2015}%
  \BibitemOpen
  \bibfield  {author} {\bibinfo {author} {\bibfnamefont {C.}~\bibnamefont
  {Fang}}\ and\ \bibinfo {author} {\bibfnamefont {L.}~\bibnamefont {Fu}},\
  }\href {\doibase 10.1103/PhysRevB.91.161105} {\bibfield  {journal} {\bibinfo
  {journal} {Phys. Rev. B}\ }\textbf {\bibinfo {volume} {91}},\ \bibinfo
  {pages} {161105(R)} (\bibinfo {year} {2015})}\BibitemShut {NoStop}%
\bibitem [{\citenamefont {Shiozaki}\ \emph {et~al.}(2015)\citenamefont
  {Shiozaki}, \citenamefont {Sato},\ and\ \citenamefont {Gomi}}]{Shiozaki2015}%
  \BibitemOpen
  \bibfield  {author} {\bibinfo {author} {\bibfnamefont {K.}~\bibnamefont
  {Shiozaki}}, \bibinfo {author} {\bibfnamefont {M.}~\bibnamefont {Sato}}, \
  and\ \bibinfo {author} {\bibfnamefont {K.}~\bibnamefont {Gomi}},\ }\href
  {\doibase 10.1103/PhysRevB.91.155120} {\bibfield  {journal} {\bibinfo
  {journal} {Phys. Rev. B}\ }\textbf {\bibinfo {volume} {91}},\ \bibinfo
  {pages} {155120} (\bibinfo {year} {2015})}\BibitemShut {NoStop}%
\bibitem [{\citenamefont {Shiozaki}\ \emph {et~al.}(2016)\citenamefont
  {Shiozaki}, \citenamefont {Sato},\ and\ \citenamefont {Gomi}}]{Shiozaki2016}%
  \BibitemOpen
  \bibfield  {author} {\bibinfo {author} {\bibfnamefont {K.}~\bibnamefont
  {Shiozaki}}, \bibinfo {author} {\bibfnamefont {M.}~\bibnamefont {Sato}}, \
  and\ \bibinfo {author} {\bibfnamefont {K.}~\bibnamefont {Gomi}},\ }\href
  {\doibase 10.1103/PhysRevB.93.195413} {\bibfield  {journal} {\bibinfo
  {journal} {Phys. Rev. B}\ }\textbf {\bibinfo {volume} {93}},\ \bibinfo
  {pages} {195413} (\bibinfo {year} {2016})}\BibitemShut {NoStop}%
\bibitem [{\citenamefont {Shapourian}\ \emph {et~al.}(2018)\citenamefont
  {Shapourian}, \citenamefont {Wang},\ and\ \citenamefont
  {Ryu}}]{Shapourian2018}%
  \BibitemOpen
  \bibfield  {author} {\bibinfo {author} {\bibfnamefont {H.}~\bibnamefont
  {Shapourian}}, \bibinfo {author} {\bibfnamefont {Y.}~\bibnamefont {Wang}}, \
  and\ \bibinfo {author} {\bibfnamefont {S.}~\bibnamefont {Ryu}},\ }\href
  {\doibase 10.1103/PhysRevB.97.094508} {\bibfield  {journal} {\bibinfo
  {journal} {Phys. Rev. B}\ }\textbf {\bibinfo {volume} {97}},\ \bibinfo
  {pages} {094508} (\bibinfo {year} {2018})}\BibitemShut {NoStop}%
\bibitem [{\citenamefont {Yanase}\ and\ \citenamefont
  {Shiozaki}(2017)}]{Yanase2017}%
  \BibitemOpen
  \bibfield  {author} {\bibinfo {author} {\bibfnamefont {Y.}~\bibnamefont
  {Yanase}}\ and\ \bibinfo {author} {\bibfnamefont {K.}~\bibnamefont
  {Shiozaki}},\ }\href {\doibase 10.1103/PhysRevB.95.224514} {\bibfield
  {journal} {\bibinfo  {journal} {Phys. Rev. B}\ }\textbf {\bibinfo {volume}
  {95}},\ \bibinfo {pages} {224514} (\bibinfo {year} {2017})}\BibitemShut
  {NoStop}%
\bibitem [{\citenamefont {Daido}\ \emph {et~al.}(2019)\citenamefont {Daido},
  \citenamefont {Yoshida},\ and\ \citenamefont {Yanase}}]{Daido2019}%
  \BibitemOpen
  \bibfield  {author} {\bibinfo {author} {\bibfnamefont {A.}~\bibnamefont
  {Daido}}, \bibinfo {author} {\bibfnamefont {T.}~\bibnamefont {Yoshida}}, \
  and\ \bibinfo {author} {\bibfnamefont {Y.}~\bibnamefont {Yanase}},\ }\href
  {\doibase 10.1103/PhysRevLett.122.227001} {\bibfield  {journal} {\bibinfo
  {journal} {Phys. Rev. Lett.}\ }\textbf {\bibinfo {volume} {122}},\ \bibinfo
  {pages} {227001} (\bibinfo {year} {2019})}\BibitemShut {NoStop}%
\bibitem [{\citenamefont {Ono}\ \emph {et~al.}(2019)\citenamefont {Ono},
  \citenamefont {Yanase},\ and\ \citenamefont {Watanabe}}]{Ono2019}%
  \BibitemOpen
  \bibfield  {author} {\bibinfo {author} {\bibfnamefont {S.}~\bibnamefont
  {Ono}}, \bibinfo {author} {\bibfnamefont {Y.}~\bibnamefont {Yanase}}, \ and\
  \bibinfo {author} {\bibfnamefont {H.}~\bibnamefont {Watanabe}},\ }\href
  {\doibase 10.1103/PhysRevResearch.1.013012} {\bibfield  {journal} {\bibinfo
  {journal} {Phys. Rev. Research}\ }\textbf {\bibinfo {volume} {1}},\ \bibinfo
  {pages} {013012} (\bibinfo {year} {2019})}\BibitemShut {NoStop}%
\bibitem [{\citenamefont {Ono}\ \emph {et~al.}(2020{\natexlab{a}})\citenamefont
  {Ono}, \citenamefont {Po},\ and\ \citenamefont {Watanabe}}]{Ono2020}%
  \BibitemOpen
  \bibfield  {author} {\bibinfo {author} {\bibfnamefont {S.}~\bibnamefont
  {Ono}}, \bibinfo {author} {\bibfnamefont {H.~C.}\ \bibnamefont {Po}}, \ and\
  \bibinfo {author} {\bibfnamefont {H.}~\bibnamefont {Watanabe}},\ }\href
  {\doibase 10.1126/sciadv.aaz8367} {\bibfield  {journal} {\bibinfo  {journal}
  {Science Advances}\ }\textbf {\bibinfo {volume} {6}} (\bibinfo {year}
  {2020}{\natexlab{a}}),\ 10.1126/sciadv.aaz8367}\BibitemShut {NoStop}%
\bibitem [{\citenamefont {Ono}\ \emph {et~al.}(2020{\natexlab{b}})\citenamefont
  {Ono}, \citenamefont {Po},\ and\ \citenamefont {Shiozaki}}]{Ono2020_2}%
  \BibitemOpen
  \bibfield  {author} {\bibinfo {author} {\bibfnamefont {S.}~\bibnamefont
  {Ono}}, \bibinfo {author} {\bibfnamefont {H.~C.}\ \bibnamefont {Po}}, \ and\
  \bibinfo {author} {\bibfnamefont {K.}~\bibnamefont {Shiozaki}},\ }\href@noop
  {} {\enquote {\bibinfo {title} {$\mathbb{Z}_2$-enriched symmetry indicators
  for topological superconductors in the 1651 magnetic space groups},}\ }
  (\bibinfo {year} {2020}{\natexlab{b}}),\ \Eprint
  {http://arxiv.org/abs/2008.05499} {arXiv:2008.05499 [cond-mat.supr-con]}
  \BibitemShut {NoStop}%
\bibitem [{\citenamefont {Skurativska}\ \emph {et~al.}(2020)\citenamefont
  {Skurativska}, \citenamefont {Neupert},\ and\ \citenamefont
  {Fischer}}]{Skurativska2020}%
  \BibitemOpen
  \bibfield  {author} {\bibinfo {author} {\bibfnamefont {A.}~\bibnamefont
  {Skurativska}}, \bibinfo {author} {\bibfnamefont {T.}~\bibnamefont
  {Neupert}}, \ and\ \bibinfo {author} {\bibfnamefont {M.~H.}\ \bibnamefont
  {Fischer}},\ }\href {\doibase 10.1103/PhysRevResearch.2.013064} {\bibfield
  {journal} {\bibinfo  {journal} {Phys. Rev. Research}\ }\textbf {\bibinfo
  {volume} {2}},\ \bibinfo {pages} {013064} (\bibinfo {year}
  {2020})}\BibitemShut {NoStop}%
\bibitem [{\citenamefont {Geier}\ \emph {et~al.}(2020)\citenamefont {Geier},
  \citenamefont {Brouwer},\ and\ \citenamefont {Trifunovic}}]{Geier2020}%
  \BibitemOpen
  \bibfield  {author} {\bibinfo {author} {\bibfnamefont {M.}~\bibnamefont
  {Geier}}, \bibinfo {author} {\bibfnamefont {P.~W.}\ \bibnamefont {Brouwer}},
  \ and\ \bibinfo {author} {\bibfnamefont {L.}~\bibnamefont {Trifunovic}},\
  }\href {\doibase 10.1103/PhysRevB.101.245128} {\bibfield  {journal} {\bibinfo
   {journal} {Phys. Rev. B}\ }\textbf {\bibinfo {volume} {101}},\ \bibinfo
  {pages} {245128} (\bibinfo {year} {2020})}\BibitemShut {NoStop}%
\bibitem [{\citenamefont {Shiozaki}(2019)}]{Shiozaki2019}%
  \BibitemOpen
  \bibfield  {author} {\bibinfo {author} {\bibfnamefont {K.}~\bibnamefont
  {Shiozaki}},\ }\href@noop {} {\enquote {\bibinfo {title} {Variants of the
  symmetry-based indicator},}\ } (\bibinfo {year} {2019}),\ \Eprint
  {http://arxiv.org/abs/1907.13632} {arXiv:1907.13632 [cond-mat.mes-hall]}
  \BibitemShut {NoStop}%
\bibitem [{\citenamefont {Ahn}\ and\ \citenamefont {Yang}(2020)}]{Ahn2020}%
  \BibitemOpen
  \bibfield  {author} {\bibinfo {author} {\bibfnamefont {J.}~\bibnamefont
  {Ahn}}\ and\ \bibinfo {author} {\bibfnamefont {B.-J.}\ \bibnamefont {Yang}},\
  }\href {\doibase 10.1103/PhysRevResearch.2.012060} {\bibfield  {journal}
  {\bibinfo  {journal} {Phys. Rev. Research}\ }\textbf {\bibinfo {volume}
  {2}},\ \bibinfo {pages} {012060(R)} (\bibinfo {year} {2020})}\BibitemShut
  {NoStop}%
\bibitem [{\citenamefont {Sigrist}\ and\ \citenamefont
  {Ueda}(1991)}]{Sigrist-Ueda}%
  \BibitemOpen
  \bibfield  {author} {\bibinfo {author} {\bibfnamefont {M.}~\bibnamefont
  {Sigrist}}\ and\ \bibinfo {author} {\bibfnamefont {K.}~\bibnamefont {Ueda}},\
  }\href {\doibase 10.1103/RevModPhys.63.239} {\bibfield  {journal} {\bibinfo
  {journal} {Rev. Mod. Phys.}\ }\textbf {\bibinfo {volume} {63}},\ \bibinfo
  {pages} {239} (\bibinfo {year} {1991})}\BibitemShut {NoStop}%
\bibitem [{\citenamefont {Khim}\ \emph {et~al.}(2021)\citenamefont {Khim},
  \citenamefont {Landaeta}, \citenamefont {Banda}, \citenamefont {Bannor},
  \citenamefont {Brando}, \citenamefont {Brydon}, \citenamefont {Hafner},
  \citenamefont {Küchler}, \citenamefont {Cardoso-Gil}, \citenamefont
  {Stockert}, \citenamefont {Mackenzie}, \citenamefont {Agterberg},
  \citenamefont {Geibel},\ and\ \citenamefont {Hassinger}}]{Khim2021}%
  \BibitemOpen
  \bibfield  {author} {\bibinfo {author} {\bibfnamefont {S.}~\bibnamefont
  {Khim}}, \bibinfo {author} {\bibfnamefont {J.~F.}\ \bibnamefont {Landaeta}},
  \bibinfo {author} {\bibfnamefont {J.}~\bibnamefont {Banda}}, \bibinfo
  {author} {\bibfnamefont {N.}~\bibnamefont {Bannor}}, \bibinfo {author}
  {\bibfnamefont {M.}~\bibnamefont {Brando}}, \bibinfo {author} {\bibfnamefont
  {P.~M.~R.}\ \bibnamefont {Brydon}}, \bibinfo {author} {\bibfnamefont
  {D.}~\bibnamefont {Hafner}}, \bibinfo {author} {\bibfnamefont
  {R.}~\bibnamefont {Küchler}}, \bibinfo {author} {\bibfnamefont
  {R.}~\bibnamefont {Cardoso-Gil}}, \bibinfo {author} {\bibfnamefont
  {U.}~\bibnamefont {Stockert}}, \bibinfo {author} {\bibfnamefont {A.~P.}\
  \bibnamefont {Mackenzie}}, \bibinfo {author} {\bibfnamefont {D.~F.}\
  \bibnamefont {Agterberg}}, \bibinfo {author} {\bibfnamefont {C.}~\bibnamefont
  {Geibel}}, \ and\ \bibinfo {author} {\bibfnamefont {E.}~\bibnamefont
  {Hassinger}},\ }\href@noop {} {\enquote {\bibinfo {title} {Field-induced
  transition from even to odd parity superconductivity in cerh$_2$as$_2$},}\ }
  (\bibinfo {year} {2021}),\ \Eprint {http://arxiv.org/abs/2101.09522}
  {arXiv:2101.09522 [cond-mat.supr-con]} \BibitemShut {NoStop}%
\bibitem [{\citenamefont {Joynt}\ and\ \citenamefont
  {Taillefer}(2002)}]{Joynt_review}%
  \BibitemOpen
  \bibfield  {author} {\bibinfo {author} {\bibfnamefont {R.}~\bibnamefont
  {Joynt}}\ and\ \bibinfo {author} {\bibfnamefont {L.}~\bibnamefont
  {Taillefer}},\ }\href {\doibase 10.1103/RevModPhys.74.235} {\bibfield
  {journal} {\bibinfo  {journal} {Rev. Mod. Phys.}\ }\textbf {\bibinfo {volume}
  {74}},\ \bibinfo {pages} {235} (\bibinfo {year} {2002})}\BibitemShut
  {NoStop}%
\bibitem [{\citenamefont {Braithwaite}\ \emph {et~al.}(2019)\citenamefont
  {Braithwaite}, \citenamefont {Vali{\v{s}}ka}, \citenamefont {Knebel},
  \citenamefont {Lapertot}, \citenamefont {Brison}, \citenamefont {Pourret},
  \citenamefont {Zhitomirsky}, \citenamefont {Flouquet}, \citenamefont
  {Honda},\ and\ \citenamefont {Aoki}}]{Braithwaite_UTe2_2019}%
  \BibitemOpen
  \bibfield  {author} {\bibinfo {author} {\bibfnamefont {D.}~\bibnamefont
  {Braithwaite}}, \bibinfo {author} {\bibfnamefont {M.}~\bibnamefont
  {Vali{\v{s}}ka}}, \bibinfo {author} {\bibfnamefont {G.}~\bibnamefont
  {Knebel}}, \bibinfo {author} {\bibfnamefont {G.}~\bibnamefont {Lapertot}},
  \bibinfo {author} {\bibfnamefont {J.-P.}\ \bibnamefont {Brison}}, \bibinfo
  {author} {\bibfnamefont {A.}~\bibnamefont {Pourret}}, \bibinfo {author}
  {\bibfnamefont {M.~E.}\ \bibnamefont {Zhitomirsky}}, \bibinfo {author}
  {\bibfnamefont {J.}~\bibnamefont {Flouquet}}, \bibinfo {author}
  {\bibfnamefont {F.}~\bibnamefont {Honda}}, \ and\ \bibinfo {author}
  {\bibfnamefont {D.}~\bibnamefont {Aoki}},\ }\href {\doibase
  10.1038/s42005-019-0248-z} {\bibfield  {journal} {\bibinfo  {journal}
  {Communications Physics}\ }\textbf {\bibinfo {volume} {2}},\ \bibinfo {pages}
  {147} (\bibinfo {year} {2019})}\BibitemShut {NoStop}%
\bibitem [{\citenamefont {Ran}\ \emph {et~al.}(2020)\citenamefont {Ran},
  \citenamefont {Kim}, \citenamefont {Liu}, \citenamefont {Saha}, \citenamefont
  {Hayes}, \citenamefont {Metz}, \citenamefont {Eo}, \citenamefont {Paglione},\
  and\ \citenamefont {Butch}}]{Ran_UTe2_pressure}%
  \BibitemOpen
  \bibfield  {author} {\bibinfo {author} {\bibfnamefont {S.}~\bibnamefont
  {Ran}}, \bibinfo {author} {\bibfnamefont {H.}~\bibnamefont {Kim}}, \bibinfo
  {author} {\bibfnamefont {I.-L.}\ \bibnamefont {Liu}}, \bibinfo {author}
  {\bibfnamefont {S.~R.}\ \bibnamefont {Saha}}, \bibinfo {author}
  {\bibfnamefont {I.}~\bibnamefont {Hayes}}, \bibinfo {author} {\bibfnamefont
  {T.}~\bibnamefont {Metz}}, \bibinfo {author} {\bibfnamefont {Y.~S.}\
  \bibnamefont {Eo}}, \bibinfo {author} {\bibfnamefont {J.}~\bibnamefont
  {Paglione}}, \ and\ \bibinfo {author} {\bibfnamefont {N.~P.}\ \bibnamefont
  {Butch}},\ }\href {\doibase 10.1103/PhysRevB.101.140503} {\bibfield
  {journal} {\bibinfo  {journal} {Phys. Rev. B}\ }\textbf {\bibinfo {volume}
  {101}},\ \bibinfo {pages} {140503(R)} (\bibinfo {year} {2020})}\BibitemShut
  {NoStop}%
\bibitem [{\citenamefont {Aoki}\ \emph {et~al.}(2020)\citenamefont {Aoki},
  \citenamefont {Honda}, \citenamefont {Knebel}, \citenamefont {Braithwaite},
  \citenamefont {Nakamura}, \citenamefont {Li}, \citenamefont {Homma},
  \citenamefont {Shimizu}, \citenamefont {Sato}, \citenamefont {Brison},\ and\
  \citenamefont {Flouquet}}]{Aoki_UTe2_2020}%
  \BibitemOpen
  \bibfield  {author} {\bibinfo {author} {\bibfnamefont {D.}~\bibnamefont
  {Aoki}}, \bibinfo {author} {\bibfnamefont {F.}~\bibnamefont {Honda}},
  \bibinfo {author} {\bibfnamefont {G.}~\bibnamefont {Knebel}}, \bibinfo
  {author} {\bibfnamefont {D.}~\bibnamefont {Braithwaite}}, \bibinfo {author}
  {\bibfnamefont {A.}~\bibnamefont {Nakamura}}, \bibinfo {author}
  {\bibfnamefont {D.}~\bibnamefont {Li}}, \bibinfo {author} {\bibfnamefont
  {Y.}~\bibnamefont {Homma}}, \bibinfo {author} {\bibfnamefont
  {Y.}~\bibnamefont {Shimizu}}, \bibinfo {author} {\bibfnamefont {Y.~J.}\
  \bibnamefont {Sato}}, \bibinfo {author} {\bibfnamefont {J.-P.}\ \bibnamefont
  {Brison}}, \ and\ \bibinfo {author} {\bibfnamefont {J.}~\bibnamefont
  {Flouquet}},\ }\href {\doibase 10.7566/JPSJ.89.053705} {\bibfield  {journal}
  {\bibinfo  {journal} {J. Phys. Soc. Jpn.}\ }\textbf {\bibinfo {volume}
  {89}},\ \bibinfo {pages} {053705} (\bibinfo {year} {2020})}\BibitemShut
  {NoStop}%
\bibitem [{\citenamefont {Ishizuka}\ and\ \citenamefont
  {Yanase}(2020)}]{Ishizuka2020}%
  \BibitemOpen
  \bibfield  {author} {\bibinfo {author} {\bibfnamefont {J.}~\bibnamefont
  {Ishizuka}}\ and\ \bibinfo {author} {\bibfnamefont {Y.}~\bibnamefont
  {Yanase}},\ }\href@noop {} {\enquote {\bibinfo {title} {A periodic anderson
  model for magnetism and superconductivity in ute2},}\ } (\bibinfo {year}
  {2020}),\ \Eprint {http://arxiv.org/abs/2008.01945} {arXiv:2008.01945
  [cond-mat.supr-con]} \BibitemShut {NoStop}%
\bibitem [{\citenamefont {Yoshida}\ \emph {et~al.}(2012)\citenamefont
  {Yoshida}, \citenamefont {Sigrist},\ and\ \citenamefont
  {Yanase}}]{Yoshida2012}%
  \BibitemOpen
  \bibfield  {author} {\bibinfo {author} {\bibfnamefont {T.}~\bibnamefont
  {Yoshida}}, \bibinfo {author} {\bibfnamefont {M.}~\bibnamefont {Sigrist}}, \
  and\ \bibinfo {author} {\bibfnamefont {Y.}~\bibnamefont {Yanase}},\ }\href
  {\doibase 10.1103/PhysRevB.86.134514} {\bibfield  {journal} {\bibinfo
  {journal} {Phys. Rev. B}\ }\textbf {\bibinfo {volume} {86}},\ \bibinfo
  {pages} {134514} (\bibinfo {year} {2012})}\BibitemShut {NoStop}%
\bibitem [{\citenamefont {Schertenleib}\ \emph {et~al.}(2021)\citenamefont
  {Schertenleib}, \citenamefont {Fischer},\ and\ \citenamefont
  {Sigrist}}]{Schertenleib2021}%
  \BibitemOpen
  \bibfield  {author} {\bibinfo {author} {\bibfnamefont {E.~G.}\ \bibnamefont
  {Schertenleib}}, \bibinfo {author} {\bibfnamefont {M.~H.}\ \bibnamefont
  {Fischer}}, \ and\ \bibinfo {author} {\bibfnamefont {M.}~\bibnamefont
  {Sigrist}},\ }\href@noop {} {\enquote {\bibinfo {title} {Unusual $h$-$t$
  phase diagram of cerh$_2$as$_2$ -- the role of staggered
  non-centrosymmetricity},}\ } (\bibinfo {year} {2021}),\ \Eprint
  {http://arxiv.org/abs/2101.08821} {arXiv:2101.08821 [cond-mat.supr-con]}
  \BibitemShut {NoStop}%
\bibitem [{\citenamefont {Möckli}\ and\ \citenamefont
  {Ramires}(2021)}]{moeckli2021}%
  \BibitemOpen
  \bibfield  {author} {\bibinfo {author} {\bibfnamefont {D.}~\bibnamefont
  {Möckli}}\ and\ \bibinfo {author} {\bibfnamefont {A.}~\bibnamefont
  {Ramires}},\ }\href@noop {} {\enquote {\bibinfo {title} {Two scenarios for
  superconductivity in cerh$_2$as$_2$},}\ } (\bibinfo {year} {2021}),\ \Eprint
  {http://arxiv.org/abs/2102.09425} {arXiv:2102.09425 [cond-mat.supr-con]}
  \BibitemShut {NoStop}%
\bibitem [{\citenamefont {Edel'shtein}(1989)}]{Edelstein1989}%
  \BibitemOpen
  \bibfield  {author} {\bibinfo {author} {\bibfnamefont {V.~M.}\ \bibnamefont
  {Edel'shtein}},\ }\href
  {http://inis.iaea.org/search/search.aspx?orig_q=RN:22082571} {\bibfield
  {journal} {\bibinfo  {journal} {Soviet Physics - JETP (English Translation)}\
  }\textbf {\bibinfo {volume} {68}},\ \bibinfo {pages} {1244} (\bibinfo {year}
  {1989})}\BibitemShut {NoStop}%
\bibitem [{\citenamefont {Edelstein}(1995)}]{Edelstein1995}%
  \BibitemOpen
  \bibfield  {author} {\bibinfo {author} {\bibfnamefont {V.~M.}\ \bibnamefont
  {Edelstein}},\ }\href {\doibase 10.1103/PhysRevLett.75.2004} {\bibfield
  {journal} {\bibinfo  {journal} {Phys. Rev. Lett.}\ }\textbf {\bibinfo
  {volume} {75}},\ \bibinfo {pages} {2004} (\bibinfo {year}
  {1995})}\BibitemShut {NoStop}%
\bibitem [{\citenamefont {Bauer}\ \emph {et~al.}(2004)\citenamefont {Bauer},
  \citenamefont {Hilscher}, \citenamefont {Michor}, \citenamefont {Paul},
  \citenamefont {Scheidt}, \citenamefont {Gribanov}, \citenamefont {Seropegin},
  \citenamefont {No\"el}, \citenamefont {Sigrist},\ and\ \citenamefont
  {Rogl}}]{Bauer2004}%
  \BibitemOpen
  \bibfield  {author} {\bibinfo {author} {\bibfnamefont {E.}~\bibnamefont
  {Bauer}}, \bibinfo {author} {\bibfnamefont {G.}~\bibnamefont {Hilscher}},
  \bibinfo {author} {\bibfnamefont {H.}~\bibnamefont {Michor}}, \bibinfo
  {author} {\bibfnamefont {C.}~\bibnamefont {Paul}}, \bibinfo {author}
  {\bibfnamefont {E.~W.}\ \bibnamefont {Scheidt}}, \bibinfo {author}
  {\bibfnamefont {A.}~\bibnamefont {Gribanov}}, \bibinfo {author}
  {\bibfnamefont {Y.}~\bibnamefont {Seropegin}}, \bibinfo {author}
  {\bibfnamefont {H.}~\bibnamefont {No\"el}}, \bibinfo {author} {\bibfnamefont
  {M.}~\bibnamefont {Sigrist}}, \ and\ \bibinfo {author} {\bibfnamefont
  {P.}~\bibnamefont {Rogl}},\ }\href {\doibase 10.1103/PhysRevLett.92.027003}
  {\bibfield  {journal} {\bibinfo  {journal} {Phys. Rev. Lett.}\ }\textbf
  {\bibinfo {volume} {92}},\ \bibinfo {pages} {027003} (\bibinfo {year}
  {2004})}\BibitemShut {NoStop}%
\bibitem [{\citenamefont {Agterberg}\ and\ \citenamefont
  {Kaur}(2007)}]{Agterberg2007}%
  \BibitemOpen
  \bibfield  {author} {\bibinfo {author} {\bibfnamefont {D.~F.}\ \bibnamefont
  {Agterberg}}\ and\ \bibinfo {author} {\bibfnamefont {R.~P.}\ \bibnamefont
  {Kaur}},\ }\href {\doibase 10.1103/PhysRevB.75.064511} {\bibfield  {journal}
  {\bibinfo  {journal} {Phys. Rev. B}\ }\textbf {\bibinfo {volume} {75}},\
  \bibinfo {pages} {064511} (\bibinfo {year} {2007})}\BibitemShut {NoStop}%
\bibitem [{\citenamefont {Bauer}\ and\ \citenamefont
  {Sigrist}(2012)}]{Bauer2012}%
  \BibitemOpen
  \bibfield  {author} {\bibinfo {author} {\bibfnamefont {E.}~\bibnamefont
  {Bauer}}\ and\ \bibinfo {author} {\bibfnamefont {M.}~\bibnamefont
  {Sigrist}},\ }\href@noop {} {\emph {\bibinfo {title} {Non-centrosymmetric
  superconductors: introduction and overview}}},\ Vol.\ \bibinfo {volume}
  {847}\ (\bibinfo  {publisher} {Springer Science \& Business Media},\ \bibinfo
  {year} {2012})\BibitemShut {NoStop}%
\bibitem [{\citenamefont {Smidman}\ \emph {et~al.}(2017)\citenamefont
  {Smidman}, \citenamefont {Salamon}, \citenamefont {Yuan},\ and\ \citenamefont
  {Agterberg}}]{Smidman_2017}%
  \BibitemOpen
  \bibfield  {author} {\bibinfo {author} {\bibfnamefont {M.}~\bibnamefont
  {Smidman}}, \bibinfo {author} {\bibfnamefont {M.~B.}\ \bibnamefont
  {Salamon}}, \bibinfo {author} {\bibfnamefont {H.~Q.}\ \bibnamefont {Yuan}}, \
  and\ \bibinfo {author} {\bibfnamefont {D.~F.}\ \bibnamefont {Agterberg}},\
  }\href {\doibase 10.1088/1361-6633/80/3/036501} {\bibfield  {journal}
  {\bibinfo  {journal} {Reports on Progress in Physics}\ }\textbf {\bibinfo
  {volume} {80}},\ \bibinfo {pages} {036501} (\bibinfo {year}
  {2017})}\BibitemShut {NoStop}%
\bibitem [{\citenamefont {Saito}\ \emph {et~al.}(2016)\citenamefont {Saito},
  \citenamefont {Nakamura}, \citenamefont {Bahramy}, \citenamefont {Kohama},
  \citenamefont {Ye}, \citenamefont {Kasahara}, \citenamefont {Nakagawa},
  \citenamefont {Onga}, \citenamefont {Tokunaga}, \citenamefont {Nojima},
  \citenamefont {Yanase},\ and\ \citenamefont {Iwasa}}]{Saito2016}%
  \BibitemOpen
  \bibfield  {author} {\bibinfo {author} {\bibfnamefont {Y.}~\bibnamefont
  {Saito}}, \bibinfo {author} {\bibfnamefont {Y.}~\bibnamefont {Nakamura}},
  \bibinfo {author} {\bibfnamefont {M.~S.}\ \bibnamefont {Bahramy}}, \bibinfo
  {author} {\bibfnamefont {Y.}~\bibnamefont {Kohama}}, \bibinfo {author}
  {\bibfnamefont {J.}~\bibnamefont {Ye}}, \bibinfo {author} {\bibfnamefont
  {Y.}~\bibnamefont {Kasahara}}, \bibinfo {author} {\bibfnamefont
  {Y.}~\bibnamefont {Nakagawa}}, \bibinfo {author} {\bibfnamefont
  {M.}~\bibnamefont {Onga}}, \bibinfo {author} {\bibfnamefont {M.}~\bibnamefont
  {Tokunaga}}, \bibinfo {author} {\bibfnamefont {T.}~\bibnamefont {Nojima}},
  \bibinfo {author} {\bibfnamefont {Y.}~\bibnamefont {Yanase}}, \ and\ \bibinfo
  {author} {\bibfnamefont {Y.}~\bibnamefont {Iwasa}},\ }\href {\doibase
  10.1038/nphys3580} {\bibfield  {journal} {\bibinfo  {journal} {Nature
  Physics}\ }\textbf {\bibinfo {volume} {12}},\ \bibinfo {pages} {144}
  (\bibinfo {year} {2016})}\BibitemShut {NoStop}%
\bibitem [{\citenamefont {Wakatsuki}\ and\ \citenamefont
  {Nagaosa}(2018)}]{Wakatsuki2018}%
  \BibitemOpen
  \bibfield  {author} {\bibinfo {author} {\bibfnamefont {R.}~\bibnamefont
  {Wakatsuki}}\ and\ \bibinfo {author} {\bibfnamefont {N.}~\bibnamefont
  {Nagaosa}},\ }\href {\doibase 10.1103/PhysRevLett.121.026601} {\bibfield
  {journal} {\bibinfo  {journal} {Phys. Rev. Lett.}\ }\textbf {\bibinfo
  {volume} {121}},\ \bibinfo {pages} {026601} (\bibinfo {year}
  {2018})}\BibitemShut {NoStop}%
\bibitem [{\citenamefont {Ando}\ \emph {et~al.}(2020)\citenamefont {Ando},
  \citenamefont {Miyasaka}, \citenamefont {Li}, \citenamefont {Ishizuka},
  \citenamefont {Arakawa}, \citenamefont {Shiota}, \citenamefont {Moriyama},
  \citenamefont {Yanase},\ and\ \citenamefont {Ono}}]{Ando2020}%
  \BibitemOpen
  \bibfield  {author} {\bibinfo {author} {\bibfnamefont {F.}~\bibnamefont
  {Ando}}, \bibinfo {author} {\bibfnamefont {Y.}~\bibnamefont {Miyasaka}},
  \bibinfo {author} {\bibfnamefont {T.}~\bibnamefont {Li}}, \bibinfo {author}
  {\bibfnamefont {J.}~\bibnamefont {Ishizuka}}, \bibinfo {author}
  {\bibfnamefont {T.}~\bibnamefont {Arakawa}}, \bibinfo {author} {\bibfnamefont
  {Y.}~\bibnamefont {Shiota}}, \bibinfo {author} {\bibfnamefont
  {T.}~\bibnamefont {Moriyama}}, \bibinfo {author} {\bibfnamefont
  {Y.}~\bibnamefont {Yanase}}, \ and\ \bibinfo {author} {\bibfnamefont
  {T.}~\bibnamefont {Ono}},\ }\href {\doibase 10.1038/s41586-020-2590-4}
  {\bibfield  {journal} {\bibinfo  {journal} {Nature}\ }\textbf {\bibinfo
  {volume} {584}},\ \bibinfo {pages} {373} (\bibinfo {year}
  {2020})}\BibitemShut {NoStop}%
\bibitem [{\citenamefont {Nogaki}\ and\ \citenamefont
  {Yanase}(2020)}]{Nogaki2020}%
  \BibitemOpen
  \bibfield  {author} {\bibinfo {author} {\bibfnamefont {K.}~\bibnamefont
  {Nogaki}}\ and\ \bibinfo {author} {\bibfnamefont {Y.}~\bibnamefont
  {Yanase}},\ }\href {\doibase 10.1103/PhysRevB.102.165114} {\bibfield
  {journal} {\bibinfo  {journal} {Phys. Rev. B}\ }\textbf {\bibinfo {volume}
  {102}},\ \bibinfo {pages} {165114} (\bibinfo {year} {2020})}\BibitemShut
  {NoStop}%
\bibitem [{\citenamefont {Fischer}\ \emph {et~al.}(2011)\citenamefont
  {Fischer}, \citenamefont {Loder},\ and\ \citenamefont
  {Sigrist}}]{Fischer2011}%
  \BibitemOpen
  \bibfield  {author} {\bibinfo {author} {\bibfnamefont {M.~H.}\ \bibnamefont
  {Fischer}}, \bibinfo {author} {\bibfnamefont {F.}~\bibnamefont {Loder}}, \
  and\ \bibinfo {author} {\bibfnamefont {M.}~\bibnamefont {Sigrist}},\ }\href
  {\doibase 10.1103/PhysRevB.84.184533} {\bibfield  {journal} {\bibinfo
  {journal} {Phys. Rev. B}\ }\textbf {\bibinfo {volume} {84}},\ \bibinfo
  {pages} {184533} (\bibinfo {year} {2011})}\BibitemShut {NoStop}%
\bibitem [{\citenamefont {Maruyama}\ \emph {et~al.}(2012)\citenamefont
  {Maruyama}, \citenamefont {Sigrist},\ and\ \citenamefont
  {Yanase}}]{Maruyama2012}%
  \BibitemOpen
  \bibfield  {author} {\bibinfo {author} {\bibfnamefont {D.}~\bibnamefont
  {Maruyama}}, \bibinfo {author} {\bibfnamefont {M.}~\bibnamefont {Sigrist}}, \
  and\ \bibinfo {author} {\bibfnamefont {Y.}~\bibnamefont {Yanase}},\ }\href
  {\doibase 10.1143/JPSJ.81.034702} {\bibfield  {journal} {\bibinfo  {journal}
  {Journal of the Physical Society of Japan}\ }\textbf {\bibinfo {volume}
  {81}},\ \bibinfo {pages} {034702} (\bibinfo {year} {2012})}\BibitemShut
  {NoStop}%
\bibitem [{\citenamefont {Maruyama}\ \emph {et~al.}(2013)\citenamefont
  {Maruyama}, \citenamefont {Sigrist},\ and\ \citenamefont
  {Yanase}}]{Maruyama2013}%
  \BibitemOpen
  \bibfield  {author} {\bibinfo {author} {\bibfnamefont {D.}~\bibnamefont
  {Maruyama}}, \bibinfo {author} {\bibfnamefont {M.}~\bibnamefont {Sigrist}}, \
  and\ \bibinfo {author} {\bibfnamefont {Y.}~\bibnamefont {Yanase}},\ }\href
  {\doibase 10.7566/JPSJ.82.043703} {\bibfield  {journal} {\bibinfo  {journal}
  {Journal of the Physical Society of Japan}\ }\textbf {\bibinfo {volume}
  {82}},\ \bibinfo {pages} {043703} (\bibinfo {year} {2013})}\BibitemShut
  {NoStop}%
\bibitem [{\citenamefont {Yoshida}\ \emph {et~al.}(2013)\citenamefont
  {Yoshida}, \citenamefont {Sigrist},\ and\ \citenamefont
  {Yanase}}]{Yoshida2013}%
  \BibitemOpen
  \bibfield  {author} {\bibinfo {author} {\bibfnamefont {T.}~\bibnamefont
  {Yoshida}}, \bibinfo {author} {\bibfnamefont {M.}~\bibnamefont {Sigrist}}, \
  and\ \bibinfo {author} {\bibfnamefont {Y.}~\bibnamefont {Yanase}},\ }\href
  {\doibase 10.7566/JPSJ.82.074714} {\bibfield  {journal} {\bibinfo  {journal}
  {Journal of the Physical Society of Japan}\ }\textbf {\bibinfo {volume}
  {82}},\ \bibinfo {pages} {074714} (\bibinfo {year} {2013})}\BibitemShut
  {NoStop}%
\bibitem [{\citenamefont {Yoshida}\ \emph {et~al.}(2014)\citenamefont
  {Yoshida}, \citenamefont {Sigrist},\ and\ \citenamefont
  {Yanase}}]{Yoshida2014}%
  \BibitemOpen
  \bibfield  {author} {\bibinfo {author} {\bibfnamefont {T.}~\bibnamefont
  {Yoshida}}, \bibinfo {author} {\bibfnamefont {M.}~\bibnamefont {Sigrist}}, \
  and\ \bibinfo {author} {\bibfnamefont {Y.}~\bibnamefont {Yanase}},\ }\href
  {\doibase 10.7566/JPSJ.83.013703} {\bibfield  {journal} {\bibinfo  {journal}
  {Journal of the Physical Society of Japan}\ }\textbf {\bibinfo {volume}
  {83}},\ \bibinfo {pages} {013703} (\bibinfo {year} {2014})}\BibitemShut
  {NoStop}%
\bibitem [{\citenamefont {Shimozawa}\ \emph {et~al.}(2016)\citenamefont
  {Shimozawa}, \citenamefont {Goh}, \citenamefont {Shibauchi},\ and\
  \citenamefont {Matsuda}}]{Shimozawa_2016}%
  \BibitemOpen
  \bibfield  {author} {\bibinfo {author} {\bibfnamefont {M.}~\bibnamefont
  {Shimozawa}}, \bibinfo {author} {\bibfnamefont {S.~K.}\ \bibnamefont {Goh}},
  \bibinfo {author} {\bibfnamefont {T.}~\bibnamefont {Shibauchi}}, \ and\
  \bibinfo {author} {\bibfnamefont {Y.}~\bibnamefont {Matsuda}},\ }\href
  {\doibase 10.1088/0034-4885/79/7/074503} {\bibfield  {journal} {\bibinfo
  {journal} {Reports on Progress in Physics}\ }\textbf {\bibinfo {volume}
  {79}},\ \bibinfo {pages} {074503} (\bibinfo {year} {2016})}\BibitemShut
  {NoStop}%
\bibitem [{\citenamefont {M\"ockli}\ \emph {et~al.}(2018)\citenamefont
  {M\"ockli}, \citenamefont {Yanase},\ and\ \citenamefont
  {Sigrist}}]{Mockli2018}%
  \BibitemOpen
  \bibfield  {author} {\bibinfo {author} {\bibfnamefont {D.}~\bibnamefont
  {M\"ockli}}, \bibinfo {author} {\bibfnamefont {Y.}~\bibnamefont {Yanase}}, \
  and\ \bibinfo {author} {\bibfnamefont {M.}~\bibnamefont {Sigrist}},\ }\href
  {\doibase 10.1103/PhysRevB.97.144508} {\bibfield  {journal} {\bibinfo
  {journal} {Phys. Rev. B}\ }\textbf {\bibinfo {volume} {97}},\ \bibinfo
  {pages} {144508} (\bibinfo {year} {2018})}\BibitemShut {NoStop}%
\bibitem [{sup()}]{supplement}%
  \BibitemOpen
  \href@noop {} {}\bibinfo {note} {See Supplemental Materials.}\BibitemShut
  {Stop}%
\bibitem [{\citenamefont {Ishizuka}\ \emph {et~al.}(2019)\citenamefont
  {Ishizuka}, \citenamefont {Sumita}, \citenamefont {Daido},\ and\
  \citenamefont {Yanase}}]{Ishizuka2019}%
  \BibitemOpen
  \bibfield  {author} {\bibinfo {author} {\bibfnamefont {J.}~\bibnamefont
  {Ishizuka}}, \bibinfo {author} {\bibfnamefont {S.}~\bibnamefont {Sumita}},
  \bibinfo {author} {\bibfnamefont {A.}~\bibnamefont {Daido}}, \ and\ \bibinfo
  {author} {\bibfnamefont {Y.}~\bibnamefont {Yanase}},\ }\href {\doibase
  10.1103/PhysRevLett.123.217001} {\bibfield  {journal} {\bibinfo  {journal}
  {Phys. Rev. Lett.}\ }\textbf {\bibinfo {volume} {123}},\ \bibinfo {pages}
  {217001} (\bibinfo {year} {2019})}\BibitemShut {NoStop}%
\bibitem [{\citenamefont {Momma}\ and\ \citenamefont
  {Izumi}(2011)}]{Momma2011}%
  \BibitemOpen
  \bibfield  {author} {\bibinfo {author} {\bibfnamefont {K.}~\bibnamefont
  {Momma}}\ and\ \bibinfo {author} {\bibfnamefont {F.}~\bibnamefont {Izumi}},\
  }\href {\doibase 10.1107/S0021889811038970} {\bibfield  {journal} {\bibinfo
  {journal} {Journal of Applied Crystallography}\ }\textbf {\bibinfo {volume}
  {44}},\ \bibinfo {pages} {1272} (\bibinfo {year} {2011})}\BibitemShut
  {NoStop}%
\bibitem [{\citenamefont {Kokalj}(1999)}]{Kokaji1999}%
  \BibitemOpen
  \bibfield  {author} {\bibinfo {author} {\bibfnamefont {A.}~\bibnamefont
  {Kokalj}},\ }\href {\doibase https://doi.org/10.1016/S1093-3263(99)00028-5}
  {\bibfield  {journal} {\bibinfo  {journal} {Journal of Molecular Graphics and
  Modelling}\ }\textbf {\bibinfo {volume} {17}},\ \bibinfo {pages} {176}
  (\bibinfo {year} {1999})}\BibitemShut {NoStop}%
\bibitem [{\citenamefont {Blaha}\ \emph {et~al.}(2019)\citenamefont {Blaha},
  \citenamefont {Schwarz}, \citenamefont {Madsen}, \citenamefont {Kvasnicka},
  \citenamefont {Luitz}, \citenamefont {Laskowsk}, \citenamefont {Tran},
  \citenamefont {Marks},\ and\ \citenamefont {Marks}}]{Blaha2019}%
  \BibitemOpen
  \bibfield  {author} {\bibinfo {author} {\bibfnamefont {P.}~\bibnamefont
  {Blaha}}, \bibinfo {author} {\bibfnamefont {K.}~\bibnamefont {Schwarz}},
  \bibinfo {author} {\bibfnamefont {G.}~\bibnamefont {Madsen}}, \bibinfo
  {author} {\bibfnamefont {D.}~\bibnamefont {Kvasnicka}}, \bibinfo {author}
  {\bibfnamefont {J.}~\bibnamefont {Luitz}}, \bibinfo {author} {\bibfnamefont
  {R.}~\bibnamefont {Laskowsk}}, \bibinfo {author} {\bibfnamefont
  {F.}~\bibnamefont {Tran}}, \bibinfo {author} {\bibfnamefont {L.}~\bibnamefont
  {Marks}}, \ and\ \bibinfo {author} {\bibfnamefont {L.}~\bibnamefont
  {Marks}},\ }\href
  {http://susi.theochem.tuwien.ac.at/reg_user/textbooks/usersguide.pdf} {\emph
  {\bibinfo {title} {WIEN2k: An Augmented Plane Wave Plus Local Orbitals
  Program for Calculating Crystal Properties}}}\ (\bibinfo  {publisher} {Techn.
  Universitat},\ \bibinfo {year} {2019})\BibitemShut {NoStop}%
\bibitem [{\citenamefont {Ptok}\ \emph {et~al.}(2021)\citenamefont {Ptok},
  \citenamefont {Kapcia}, \citenamefont {Jochym}, \citenamefont {Łażewski},
  \citenamefont {Oleś},\ and\ \citenamefont {Piekarz}}]{Ptok2021}%
  \BibitemOpen
  \bibfield  {author} {\bibinfo {author} {\bibfnamefont {A.}~\bibnamefont
  {Ptok}}, \bibinfo {author} {\bibfnamefont {K.~J.}\ \bibnamefont {Kapcia}},
  \bibinfo {author} {\bibfnamefont {P.~T.}\ \bibnamefont {Jochym}}, \bibinfo
  {author} {\bibfnamefont {J.}~\bibnamefont {Łażewski}}, \bibinfo {author}
  {\bibfnamefont {A.~M.}\ \bibnamefont {Oleś}}, \ and\ \bibinfo {author}
  {\bibfnamefont {P.}~\bibnamefont {Piekarz}},\ }\href@noop {} {\enquote
  {\bibinfo {title} {Electronic and dynamical properties of cerh$_{2}$as$_{2}$:
  Role of rh$_{2}$as$_{2}$ layers and expected hidden orbital order},}\ }
  (\bibinfo {year} {2021}),\ \Eprint {http://arxiv.org/abs/2102.02735}
  {arXiv:2102.02735 [cond-mat.mtrl-sci]} \BibitemShut {NoStop}%
\bibitem [{\citenamefont {Ishida}()}]{Ishida2021}%
  \BibitemOpen
  \bibfield  {author} {\bibinfo {author} {\bibfnamefont {K.}~\bibnamefont
  {Ishida}},\ }\href@noop {} {}\bibinfo {howpublished} {private
  communication}\BibitemShut {NoStop}%
\bibitem [{\citenamefont {Yanase}(2016)}]{Yanase2016}%
  \BibitemOpen
  \bibfield  {author} {\bibinfo {author} {\bibfnamefont {Y.}~\bibnamefont
  {Yanase}},\ }\href {\doibase 10.1103/PhysRevB.94.174502} {\bibfield
  {journal} {\bibinfo  {journal} {Phys. Rev. B}\ }\textbf {\bibinfo {volume}
  {94}},\ \bibinfo {pages} {174502} (\bibinfo {year} {2016})}\BibitemShut
  {NoStop}%
\end{thebibliography}

\begin{thebibliography}{5}%
\makeatletter
\providecommand \@ifxundefined [1]{%
 \@ifx{#1\undefined}
}%
\providecommand \@ifnum [1]{%
 \ifnum #1\expandafter \@firstoftwo
 \else \expandafter \@secondoftwo
 \fi
}%
\providecommand \@ifx [1]{%
 \ifx #1\expandafter \@firstoftwo
 \else \expandafter \@secondoftwo
 \fi
}%
\providecommand \natexlab [1]{#1}%
\providecommand \enquote  [1]{``#1''}%
\providecommand \bibnamefont  [1]{#1}%
\providecommand \bibfnamefont [1]{#1}%
\providecommand \citenamefont [1]{#1}%
\providecommand \href@noop [0]{\@secondoftwo}%
\providecommand \href [0]{\begingroup \@sanitize@url \@href}%
\providecommand \@href[1]{\@@startlink{#1}\@@href}%
\providecommand \@@href[1]{\endgroup#1\@@endlink}%
\providecommand \@sanitize@url [0]{\catcode `\\12\catcode `\$12\catcode
  `\&12\catcode `\#12\catcode `\^12\catcode `\_12\catcode `\%12\relax}%
\providecommand \@@startlink[1]{}%
\providecommand \@@endlink[0]{}%
\providecommand \url  [0]{\begingroup\@sanitize@url \@url }%
\providecommand \@url [1]{\endgroup\@href {#1}{\urlprefix }}%
\providecommand \urlprefix  [0]{URL }%
\providecommand \Eprint [0]{\href }%
\providecommand \doibase [0]{http://dx.doi.org/}%
\providecommand \selectlanguage [0]{\@gobble}%
\providecommand \bibinfo  [0]{\@secondoftwo}%
\providecommand \bibfield  [0]{\@secondoftwo}%
\providecommand \translation [1]{[#1]}%
\providecommand \BibitemOpen [0]{}%
\providecommand \bibitemStop [0]{}%
\providecommand \bibitemNoStop [0]{.\EOS\space}%
\providecommand \EOS [0]{\spacefactor3000\relax}%
\providecommand \BibitemShut  [1]{\csname bibitem#1\endcsname}%
\let\auto@bib@innerbib\@empty
\bibitem [{\citenamefont {Blaha}\ \emph {et~al.}(2019)\citenamefont {Blaha},
  \citenamefont {Schwarz}, \citenamefont {Madsen}, \citenamefont {Kvasnicka},
  \citenamefont {Luitz}, \citenamefont {Laskowsk}, \citenamefont {Tran},
  \citenamefont {Marks},\ and\ \citenamefont {Marks}}]{Supp_Blaha2019}%
  \BibitemOpen
  \bibfield  {author} {\bibinfo {author} {\bibfnamefont {P.}~\bibnamefont
  {Blaha}}, \bibinfo {author} {\bibfnamefont {K.}~\bibnamefont {Schwarz}},
  \bibinfo {author} {\bibfnamefont {G.}~\bibnamefont {Madsen}}, \bibinfo
  {author} {\bibfnamefont {D.}~\bibnamefont {Kvasnicka}}, \bibinfo {author}
  {\bibfnamefont {J.}~\bibnamefont {Luitz}}, \bibinfo {author} {\bibfnamefont
  {R.}~\bibnamefont {Laskowsk}}, \bibinfo {author} {\bibfnamefont
  {F.}~\bibnamefont {Tran}}, \bibinfo {author} {\bibfnamefont {L.}~\bibnamefont
  {Marks}}, \ and\ \bibinfo {author} {\bibfnamefont {L.}~\bibnamefont
  {Marks}},\ }\href
  {http://susi.theochem.tuwien.ac.at/reg_user/textbooks/usersguide.pdf} {\emph
  {\bibinfo {title} {WIEN2k: An Augmented Plane Wave Plus Local Orbitals
  Program for Calculating Crystal Properties}}}\ (\bibinfo  {publisher} {Techn.
  Universitat},\ \bibinfo {year} {2019})\BibitemShut {NoStop}%
\bibitem [{\citenamefont {Khim}\ \emph {et~al.}(2021)\citenamefont {Khim},
  \citenamefont {Landaeta}, \citenamefont {Banda}, \citenamefont {Bannor},
  \citenamefont {Brando}, \citenamefont {Brydon}, \citenamefont {Hafner},
  \citenamefont {Küchler}, \citenamefont {Cardoso-Gil}, \citenamefont
  {Stockert}, \citenamefont {Mackenzie}, \citenamefont {Agterberg},
  \citenamefont {Geibel},\ and\ \citenamefont {Hassinger}}]{Supp_Khim2021}%
  \BibitemOpen
  \bibfield  {author} {\bibinfo {author} {\bibfnamefont {S.}~\bibnamefont
  {Khim}}, \bibinfo {author} {\bibfnamefont {J.~F.}\ \bibnamefont {Landaeta}},
  \bibinfo {author} {\bibfnamefont {J.}~\bibnamefont {Banda}}, \bibinfo
  {author} {\bibfnamefont {N.}~\bibnamefont {Bannor}}, \bibinfo {author}
  {\bibfnamefont {M.}~\bibnamefont {Brando}}, \bibinfo {author} {\bibfnamefont
  {P.~M.~R.}\ \bibnamefont {Brydon}}, \bibinfo {author} {\bibfnamefont
  {D.}~\bibnamefont {Hafner}}, \bibinfo {author} {\bibfnamefont
  {R.}~\bibnamefont {Küchler}}, \bibinfo {author} {\bibfnamefont
  {R.}~\bibnamefont {Cardoso-Gil}}, \bibinfo {author} {\bibfnamefont
  {U.}~\bibnamefont {Stockert}}, \bibinfo {author} {\bibfnamefont {A.~P.}\
  \bibnamefont {Mackenzie}}, \bibinfo {author} {\bibfnamefont {D.~F.}\
  \bibnamefont {Agterberg}}, \bibinfo {author} {\bibfnamefont {C.}~\bibnamefont
  {Geibel}}, \ and\ \bibinfo {author} {\bibfnamefont {E.}~\bibnamefont
  {Hassinger}},\ }\href@noop {} {\enquote {\bibinfo {title} {Field-induced
  transition from even to odd parity superconductivity in cerh$_2$as$_2$},}\ }
  (\bibinfo {year} {2021}),\ \Eprint {http://arxiv.org/abs/2101.09522}
  {arXiv:2101.09522 [cond-mat.supr-con]} \BibitemShut {NoStop}%
\bibitem [{\citenamefont {Ptok}\ \emph {et~al.}(2021)\citenamefont {Ptok},
  \citenamefont {Kapcia}, \citenamefont {Jochym}, \citenamefont {Łażewski},
  \citenamefont {Oleś},\ and\ \citenamefont {Piekarz}}]{Supp_Ptok2021}%
  \BibitemOpen
  \bibfield  {author} {\bibinfo {author} {\bibfnamefont {A.}~\bibnamefont
  {Ptok}}, \bibinfo {author} {\bibfnamefont {K.~J.}\ \bibnamefont {Kapcia}},
  \bibinfo {author} {\bibfnamefont {P.~T.}\ \bibnamefont {Jochym}}, \bibinfo
  {author} {\bibfnamefont {J.}~\bibnamefont {Łażewski}}, \bibinfo {author}
  {\bibfnamefont {A.~M.}\ \bibnamefont {Oleś}}, \ and\ \bibinfo {author}
  {\bibfnamefont {P.}~\bibnamefont {Piekarz}},\ }\href@noop {} {\enquote
  {\bibinfo {title} {Electronic and dynamical properties of cerh$_{2}$as$_{2}$:
  Role of rh$_{2}$as$_{2}$ layers and expected hidden orbital order},}\ }
  (\bibinfo {year} {2021}),\ \Eprint {http://arxiv.org/abs/2102.02735}
  {arXiv:2102.02735 [cond-mat.mtrl-sci]} \BibitemShut {NoStop}%
\bibitem [{\citenamefont {Ishida}()}]{Supp_Ishida2021}%
  \BibitemOpen
  \bibfield  {author} {\bibinfo {author} {\bibfnamefont {K.}~\bibnamefont
  {Ishida}},\ }\href@noop {} {}\bibinfo {howpublished} {private
  communication}\BibitemShut {NoStop}%
\bibitem [{\citenamefont {Yoshida}\ \emph {et~al.}(2012)\citenamefont
  {Yoshida}, \citenamefont {Sigrist},\ and\ \citenamefont
  {Yanase}}]{Supp_Yoshida2012}%
  \BibitemOpen
  \bibfield  {author} {\bibinfo {author} {\bibfnamefont {T.}~\bibnamefont
  {Yoshida}}, \bibinfo {author} {\bibfnamefont {M.}~\bibnamefont {Sigrist}}, \
  and\ \bibinfo {author} {\bibfnamefont {Y.}~\bibnamefont {Yanase}},\ }\href
  {\doibase 10.1103/PhysRevB.86.134514} {\bibfield  {journal} {\bibinfo
  {journal} {Phys. Rev. B}\ }\textbf {\bibinfo {volume} {86}},\ \bibinfo
  {pages} {134514} (\bibinfo {year} {2012})}\BibitemShut {NoStop}%
\end{thebibliography}

\end{document}